\documentclass[12pt]{article}
\pdfoutput=1

\usepackage{putex}
\usepackage{graphicx}
\usepackage{caption}
\usepackage{amsmath}
\usepackage{array}
\usepackage{subcaption}
\usepackage{epstopdf}
\usepackage{enumerate}
\usepackage{cite}
\usepackage{youngtab}
\usepackage{tensor}
\usepackage{slashed}
\usepackage[aligntableaux=center]{ytableau}
\usepackage[utf8]{inputenc}
\usepackage[
      colorlinks=true,
      linkcolor=blue,
      urlcolor=blue,
      filecolor=black,
      citecolor=red,
      ]{hyperref}
\usepackage{braket}
\usepackage{mathtools}
\usepackage{rotating}
\usepackage{tabularx}
\newcolumntype{Y}{>{\centering\arraybackslash}X}
\newcolumntype{C}[1]{>{\centering\arraybackslash}p{#1}}
\usepackage{color, colortbl}
\definecolor{LightCyan}{rgb}{0.7,1,1}
\definecolor{Gray}{gray}{0.9}
\usepackage{tikz}
\usetikzlibrary{decorations.pathreplacing,decorations.markings,shapes.misc,arrows}
\usepackage{siunitx}

\newcommand {\be} {\begin {equation}}
\newcommand {\ee} {\end {equation}}

\newcommand {\bes} {\begin {equation*}}
\newcommand {\ees} {\end {equation*}}

\newcommand{\es}[2] {\begin{equation} \label{#1} \begin{split} #2 \end{split} \end{equation}}

\newcommand{\Z}{\mathbb{Z}}

\newcommand{\R}{\mathbb{R}}

\newcommand{\cT}{{\mathcal T}}

\newcommand{\beq}{\begin{equation}}
\newcommand{\eeq}{\end{equation}}

\def\p{\partial}

\def\ie{\begin{equation}\begin{aligned}}
\def\fe{\end{aligned}\end{equation}}

\newcommand{\m}{\mu}

\numberwithin{equation}{section}

\def\<{\langle}
\def\>{\rangle}

\newcommand{\quark}{v}
\newcommand{\Quark}{q}

\def\hc{{\rm h.c.}}
\DeclareMathOperator{\Tr}{Tr}
\DeclareMathOperator{\SU}{SU}
\DeclareMathOperator{\SO}{SO}
\DeclareMathOperator{\U}{U}

\begin{document}

\preprint{PUPT-2623}

\institution{PU}{Joseph Henry Laboratories, Princeton University, Princeton, NJ 08544, USA}
\institution{PCTS}{Princeton Center for Theoretical Science, Princeton University, Princeton, NJ 08544}

\title{Exact Symmetries and Threshold States \\ in Two-Dimensional Models for QCD
}

\authors{Ross Dempsey,\worksat{\PU} Igor R.~Klebanov,\worksat{\PU, \PCTS} and Silviu S.~Pufu\worksat{\PU}}

\abstract{
Two-dimensional $\SU(N)$ gauge theory coupled to a Majorana fermion in the adjoint representation is a nice toy model for higher-dimensional gauge dynamics. 
It possesses a multitude of ``gluinoball" bound states whose spectrum has been studied using numerical diagonalizations of the light-cone Hamiltonian. 
We extend this model by coupling it to $N_f$ flavors of fundamental Dirac fermions (quarks). The extended model also contains meson-like bound states, both bosonic and fermionic, which in the large-$N$ limit decouple from the gluinoballs. We study the large-$N$ meson spectrum using the Discretized Light-Cone Quantization (DLCQ)\@. When all the fermions are massless,  
we exhibit an exact $\mathfrak{osp}(1|4)$ symmetry algebra that leads to an infinite number of degeneracies in the DLCQ approach. More generally, we show that many single-trace states in the theory are threshold bound states that are
degenerate with multi-trace states. These exact degeneracies can be explained using the Kac-Moody algebra of the $\SU(N)$ current. We also present strong numerical evidence that additional threshold states appear in the continuum limit. Finally, we make the quarks massive while keeping the adjoint fermion massless.  In this case too, we observe some exact degeneracies that show that the spectrum of mesons becomes continuous above a certain threshold. This demonstrates quantitatively that the fundamental string tension vanishes in the massless adjoint QCD$_2$.
}
\date{June 2021}

\maketitle

\tableofcontents

\section{Introduction and Summary}

Soon after the emergence of Quantum Chromodynamics as the $\SU(3)$ Yang-Mills theory of strong interactions \cite{Gross:1973id,Politzer:1973fx,Fritzsch:1973pi}, 
't Hooft introduced its generalization to gauge group $\SU(N)$ and the large $N$ limit where $g_{\rm YM}^2 N$ is held fixed \cite{tHooft:1973alw}. 
To demonstrate the power of this approach, he applied it to the theory now known as the 't Hooft model \cite{tHooft:1974pnl}: the $1+1$ dimensional $\SU(N)$ gauge theory coupled to massive Dirac fermions in the fundamental
representation.  Using light-cone quantization, he derived an equation for the meson spectrum.  This equation is exact in the large $N$ limit. A crucial simplification in this model is that the meson light-cone wave functions describe the bound states of only two quanta, a quark and an antiquark. 

It is interesting to generalize the 't Hooft model by replacing the fermions in the fundamental representation by those in two-index representations; such $1+1$ dimensional models contain discrete analogues of $\theta$-vacua \cite{Witten:1978ka}. A minimal model of this type, the $\SU(N)$ gauge theory coupled to one adjoint Majorana fermion of mass $m_\text{adj}$
\cite{Dalley:1992yy} has turned out to be an interesting playground for studying various non-perturbative phenomena in gauge theory. 
Instead of the meson spectrum, its large $N$ spectrum consists of glueball-like bound states that may be viewed as closed strings. Since the adjoint fermion is akin to a gluino, we will refer to them as gluinoballs.\footnote{It is also possible to study models with adjoint scalars \cite{Dalley:1992yy,Demeterfi:1993rs} and supersymmetric models containing both adjoint scalars and fermions
  \cite{Matsumura:1995kw,Gadde:2014ppa}.}
They are bosonic when the number of adjoint quanta is even and fermionic when it is odd. The light-cone bound state equations are not separable since they involve superpositions of states with different numbers of quanta. Nevertheless, some features of the spectra can be studied with good precision using the Discretized Light-Cone Quantization (DLCQ)
\cite{Pauli:1985ps,Hornbostel:1988fb,Hornbostel:1988ne,Brodsky:1997de}, where one of the light-cone coordinates, which we take to be $x^-$, 
 is formally compactified on a circle of radius $L$, thus identifying $x^-\sim x^- +2 \pi L$. As $m_\text{adj}\rightarrow 0$, all the bound states of the theory remain massive \cite{Dalley:1992yy,Kutasov:1993gq,Bhanot:1993xp}, but the vacua of the model have interesting 
topological properties \cite{Witten:1978ka,Smilga:1994hc,Lenz:1994du,Kutasov:1994xq,Gross:1995bp,Smilga:1996dn,Cherman:2019hbq,Komargodski:2020mxz}. 
They are described by the topological coset model ${  \SO(N^2-1)_1\over \SU(N)_{N} }$ which has a vanishing central charge.
Thus, for any $N$, the $\SU(N)$ gauge theory coupled to a massless adjoint fermion serves as a non-trivial example of a gapped topological phase.  

In another sign of interesting physics, it was  argued in \cite{Gross:1995bp} that as $m_\text{adj}\rightarrow 0$ the model makes a transition from the confining to screening behavior of the Wilson loop in the fundamental representation of $\SU(N)$. In recent literature there have been renewed discussions of the $m_\text{adj}\rightarrow 0$ limit of the model \cite{Cherman:2019hbq,Komargodski:2020mxz,Smilga:2021zrw}. In a very interesting development \cite{Komargodski:2020mxz}, it was demonstrated that, in addition to the $N$ discrete $\theta$-vacua \cite{Witten:1978ka}, the model possesses a large number of additional superselection sectors whose number grows exponentially in $N$. This has led to new arguments for the screening of the massless adjoint model. In Section~\ref{sec:MASSIVE} we will provide additional quantitative evidence for the screening behavior of the theory with $m_\text{adj}= 0$.

As noted in \cite{Kutasov:1994xq}, the massless limit of adjoint QCD$_2$ exhibits an important simplification because the DLCQ Hilbert space breaks up into separate Kac-Moody current algebra blocks \cite{Knizhnik:1984nr}.   At large $N$, this separation of the DLCQ spectrum leads to exact degeneracies between the value of $P^-$ for certain single-trace states at resolution $K$ and sums of the values of $P^-$ for certain fermionic single-trace states at resolutions $m$ and $K-m$ \cite{Gross:1997mx}. As a result, some single-particle states may be interpreted as threshold bound states. 
It therefore appears that the spectrum of the massless large $N$ model is built of a sequence of basic states, sometimes called ``single-particle states," which were investigated in
\cite{Kutasov:1994xq,Gross:1997mx,Trittmann:2000uj,Trittmann:2001dk,Trittmann:2015oka} using DLCQ\@. In Section~\ref{KMA} we will review the implications of the Kac-Moody structure of the massless model \cite{Kutasov:1994xq} and examine the exact DLCQ degeneracies more fully.
The massless model was also studied numerically using a conformal truncation approach (for a review, see \cite{Anand:2020gnn}), which provides another basis for light-cone wave functions
\cite{Katz:2013qua}.\footnote{Let us also mention that for $m_\text{adj}^2= \frac{g_{\rm YM}^2 N}{\pi}$ the model becomes supersymmetric 
\cite{Kutasov:1993gq,Bhanot:1993xp,Antonuccio:1998zp}. This leads to very interesting effects in non-trivial $\theta$-vacua, where the supersymmetry is broken \cite{Dubovsky:2018dlk}.}

While the model with a single adjoint Majorana fermion contains rich physics, it is of obvious interest to consider its generalizations. For example, the $\SU(N)$ gauge theory coupled to two adjoint Majorana fermions, or equivalently one Dirac adjoint, was studied in \cite{Gopakumar:2012gd}. As $m_\text{adj}\rightarrow 0$, this model is not gapped but is rather described by a gauged 
${\SO(2N^2-2)_1\over \SU(N)_{2N}}$
Wess-Zumino-Witten (WZW) model with central charge $\frac{N^2-1}{3}$. Surprisingly, this CFT turns out to have ${\cal N}=2$ supersymmetry 
\cite{Gopakumar:2012gd,Isachenkov:2014zua,Isachenkov:2014osa}. The $\U(1)_R$ symmetry of this CFT is traced back to the $\U(1)$ phase symmetry of the adjoint Dirac fermion.    

In this paper we study another extension of the 't Hooft model by coupling the $1+1$ dimensional $\SU(N)$ gauge theory to 
$N_f$ fundamental Dirac fermions (quarks) $\Quark_\alpha$ of mass $m_\text{fund}$ {\it and} an adjoint Majorana fermion $\Psi$ of mass $m_\text{adj}$: 
\es{STotal}{
    S  &= \int d^2x\,\left\lbrack \tr\left(-\frac{1}{4 g^2}F_{\mu\nu}F^{\mu\nu} + \frac{i}{2}
\overline{\Psi}\slashed{D}\Psi - \frac{m_\text{adj}}{2}  \overline{\Psi}\Psi\right) + \sum_{\alpha=1}^{N_f} \left ( i
\overline{\Quark}_\alpha\slashed{D}\Quark_\alpha - m_\text{fund} \overline{\Quark}_\alpha \Quark_\alpha\right )  \right\rbrack \,.
} 
 Then, similarly to the higher-dimensional large $N$ QCD, the spectrum contains both the glueball-like bound states akin to closed strings, and the meson-like bound states akin to open strings.\footnote{There are also baryons which consist of $N$ quarks and can include some number of adjoint quanta. The baryon masses grow as $N$ in the large $N$ limit \cite{Witten:1979kh}. We will not discuss them further in this paper.} The $N_f^2$ meson states transform in the adjoint representation of the $\U(N_f)$ global symmetry, while the gluinoballs are
$\U(N_f)$ singlets. We will consider the large $N$ limit where $N_f$ is held fixed.\footnote{The large $N$ limit of meson masses does not depend on $N_f$, and in some parts of the paper we will restrict to $N_f=1$.} 
The structure of the mesons is much more intricate than in the original 't Hooft model \cite{tHooft:1974pnl} because there can be an arbitrary number of adjoint quanta forming a string which connects the quark and anti-quark at its endpoints.
When this number is even, the mesons are bosonic, while when it is odd they are fermionic. Far in the IR, the mesonic sector of the massless theory is described by a gauged 
${  \SO(N^2-1+2N N_f)_1\over \SU(N)_{N+N_f} }$
WZW model with central charge $\frac{N_f( 3N^2 +2N_f N +1)}{2(2N+N_f)}$.  Therefore, in contrast to the pure adjoint model where the corresponding central charge vanishes, the IR limit of the theory (\ref{STotal}) is complicated and dynamical. 

The bound state equations in the mesonic sector involve mixing of states with different numbers of quanta. As a result, the problem of determining the meson masses is much more complicated than in  \cite{tHooft:1974pnl}. 
When we apply the DLCQ to the mesonic sector, we find new surprises in the limit where both $m_\text{adj}$ and $m_\text{fund}$ are sent to zero. The values of $P^-$ for certain massive meson states does not change as the resolution parameter is increased and is the same as the sum of the values of $P^-$ for one or more fermionic gluinoball states. We trace this exact result to the $\mathfrak{osp}(1|4)$ symmetry of the DLCQ system with antiperiodic boundary conditions around the circle in $x^-$ direction. This symmetry helps us prove that in the continuum limit, there is a large amount of degeneracy in the spectrum of the meson states at the same value of masses as possessed by certain gluinoball states; the first of them occurs at $M_0^2\approx 5.72  \frac{g^2 N}{\pi}$. 
This fascinating structure of the spectrum is similar to the threshold bound states in the pure gluinoball sector. Some of new degeneracies we observe (see Figures~\ref{fig:pminus_degeneracies} and~\ref{fig:m2_convergence}) suggest that the meson states with growing multiplicities 
may be thought of as threshold bound states of the singlet gluinoball of mass $M_0$ and states from the CFT sector. 
After we turn on the quark mass $m_\text{fund}$, we continue to find that some meson states are threshold bound states of gluinoballs and lighter mesons; see Section~\ref{sec:MASSIVE}.
This structure of some meson states 
in our model (\ref{STotal}) is also reminiscent of the exotic, possibly molecular, $XYZ$ mesons observed in the real world (for reviews, see \cite{Lebed:2016hpi,Esposito:2016noz}). In particular, the mass of the $X(3872)$ charmonium state is extremely close to the sum of the masses of $D$ and $D^*$ mesons. 
Clearly, the intricate and surprising structure of the bound state spectrum in model (\ref{STotal}) deserves further studies and a deeper understanding.

The rest of this paper is organized as follows.  In Section~\ref{ACTION} we set up our conventions and derive the expressions for the light-cone momenta in the continuum limit.  In Section~\ref{DISCRETE}, we describe the implementation of the DLCQ procedure and present numerical results for the spectrum of $P^-$ and the spectrum of masses for both gluinoballs and mesons at leading order in the large $N$ limit when $m_\text{adj} = m_\text{fund} = 0$.  Section~\ref{KMA} contains a review of the Kac-Moody approach developed in \cite{Kutasov:1994xq} as well as its implications for the model we study.  As we will show, many of the degeneracies we observe can be anticipated from this approach, and many are expected to survive at finite $N$.  In Section~\ref{sec:glue_spectrum} we present the numerical gluinoball spectrum obtained using DLCQ and show that introduction of the adjoint mass lifts the degeneracies. In Section~\ref{sec:MESON} we similarly present the numerical meson spectrum.  In Section~\ref{sec:symmetry} we explain many of the degeneracies in the $P^-$ meson spectrum by constructing the symmetry generators that form the $\mathfrak{osp}(1|4)$ algebra.  Lastly, in Section~\ref{sec:MASSIVE} we examine how the spectrum changes when we introduce non-zero 
fundamental mass $m_\text{fund}$ and show that some of the degeneracies are not lifted.  We provide additional evidence for the screening behavior of the theory with massless adjoints. It follows the approach in
\cite{Kutasov:1994xq,Gross:1995bp} and relies on the quantitative properties of meson spectra in theory (\ref{STotal}) with massive quarks.   We end with a discussion of our results in Section~\ref{sec:DISCUSSION}.

\section{Action and mode expansion}
\label{ACTION}

\subsection{Action}

When working with fermions in $1+1$ dimensions, we will use the gamma matrices $\gamma^0 = \sigma_2$, $\gamma^1 = i\sigma_1$, obeying the Clifford algebra $\{ \gamma^\mu, \gamma^\nu\} = 2 \eta^{\mu\nu}$ with $\eta^{\mu\nu} = \diag\{ 1, -1\}$.  The action for our model for an $\SU(N)$ gauge theory with gauge field $A_\mu$ as well as an adjoint Majorana fermion $\Psi$ and fundamental Dirac fermions $\Quark_\alpha$ was given in \eqref{STotal}, where the $\SU(N)$ gauge covariant derivative acts as
\begin{equation}
    D_\mu\Psi_{ij} = \partial_\mu\Psi_{ij} + i[A_\mu, \Psi]_{ij},\qquad D_\mu \Quark_{i \alpha } = \partial_\mu \Quark_{i \alpha } + i(A_\mu)_{ij} \Quark_{j\alpha } \,,
\end{equation}
where $(A_{\mu})_{ij}$ is hermitian and traceless. 

To study this action in light-cone quantization, we define the light-cone coordinates $x^\pm = (x^0 \pm x^1) / \sqrt{2}$ as well as the light-cone components of the gauge and matter fields.  For the gauge field, we have $A_\pm = (A_0 \pm A_1) / \sqrt{2}$, while for the fermions we define $\Psi_{ij} = 2^{-1/4}\begin{pmatrix} \psi_{ij} \\ \chi_{ij}\end{pmatrix}$ and $\Quark_{i \alpha} = 2^{-1/4}\begin{pmatrix} \quark_{i \alpha} \\ \phi_{i \alpha}\end{pmatrix}$.   In the gauge $A_- = 0$, the action \eqref{STotal} takes the form
 \es{Action}{
    S = \int d^2x\,\Bigg\lbrack&\tr\left(\frac{1}{2g^2}\left(\partial_- A_+\right)^2 + \frac{1}{2} i\psi \partial_+ \psi + \frac{1}{2} i\chi \partial_- \chi + A_+ J^+ - \frac{i}{\sqrt{2}}m_\text{adj} \chi \psi\right) \\
    &+ i\quark_{\alpha k}^\dagger \partial_+ \quark_{k \alpha} + i \phi_{\alpha k}^\dagger \partial_- \phi_k -\frac{i}{\sqrt{2}}m_\text{fund}\left(\phi_{\alpha k}^\dagger \quark_{k \alpha}+\phi_{k \alpha} \quark_{\alpha k}^\dagger\right)\Bigg\rbrack,
 }
where $J^+ = J^+_\text{adj} + J^+_\text{fund}$ is the right-moving component of the gauged $\SU(N)$ current, with contributions from the adjoint and fundamental fermions given by\footnote{Equivalently, we can also write the adjoint contribution more simply as the normal-ordered expression $J^+_{ij, \text{adj}} = :\psi^{ik} \psi^{kj}:$.  However, if we subtract the $\SU(N)$ trace, as in \eqref{JDefs}, then $(J^+_\text{adj})_{ij}$ is manifestly an $\SU(N)$ adjoint and we do not have to worry about the normal ordering prescription.  Indeed, the normal ordering ambiguity is a $c$-number, which represents a mixing of the current with the identity operator.  Since the identity operator and the current transform in different representations of $\SU(N)$, there can be no mixing.}
 \es{JDefs}{
  (J^+_\text{adj})_{ij} \equiv \psi_{ik}\psi_{kj} - \frac{1}{N} \delta_{ij} \psi_{kl} \psi_{lk} \,, \qquad
   (J^+_\text{fund})_{ij} = \quark_{i \alpha} \quark_{\alpha j}^\dagger - \frac{1}{N}\delta_{ij}\quark_{k \alpha} \quark_{\alpha k}^\dagger \,.
 }

If we treat $x^+$ as the time coordinate, then we see that $A_+$, $\chi$, $\phi$ and $\phi^\dagger$ are non-dynamical and can be eliminated using their equations of motion, which are
 \es{eoms}{
    J^+ =  \frac{1}{g^2}\partial_-^2 A_+ \,, \qquad 
     \partial_-\chi = \frac{m_\text{adj}}{\sqrt{2}} \psi \,, \qquad
    \partial_- \phi = \frac{m_\text{fund}}{\sqrt{2}}\quark \,, \qquad
      \partial_- \phi^\dagger = \frac{m_\text{fund}}{\sqrt{2}}\quark^\dagger \,.
 }
Eliminating each of the non-dynamical fields, we can write \eqref{Action} as
\begin{align}
    S = \int d^2x\left\lbrack \tr\left(\frac{g^2}{2} J^+ \frac{1}{\partial_-^2}J^+ + \frac{i}{2}\psi\partial_+\psi + \frac{im_\text{adj}^2}{4}\psi\frac{1}{\partial_-}\psi\right) + i\quark_{\alpha k}^\dagger \partial_+ \quark_{k \alpha} + \frac{im_\text{fund}^2}{2} \quark_{\alpha k}^\dagger \frac{1}{\partial_-}\quark_{k \alpha} \right\rbrack.
\end{align}
From this expression, it follows that the light-cone momentum operators are
 \es{Ppm}{
    P^+ &= \int dx^- \big\lbrack \frac{1}{2}\tr\left(i\psi \partial_- \psi\right) + i\quark_{\alpha k}^\dagger \partial_- \quark_{k \alpha} \big\rbrack\,, \\
    P^- &= \int dx^-\left\lbrack -\tr\left(\frac{g^2}{2} J^+ \frac{1}{\partial_-^2}J^+ + \frac{im_\text{adj}^2}{4}\psi\frac{1}{\partial_-}\psi\right) - \frac{im_\text{fund}^2}{2} \quark_{\alpha k}^\dagger \frac{1}{\partial_-}\quark_{k \alpha}\right\rbrack \,.
 }
The light-cone momenta $P^+$ and $P^-$ commute and can be simultaneously diagonalized.  To find the spectrum of the theory, we will be interested in finding the eigenvalues of the mass squared operator $M^2 = 2 P^+ P^-$.

In addition, both $P^+$ and $P^-$ commute with the charge conjugation operator $\mathcal{C}$ defined by
 \es{eq:charge_conjugation}{
	\mathcal{C}\psi_{ij}\mathcal{C}^{-1} = \psi_{ji} \,, \qquad  \mathcal{C}\quark_{k\alpha}\mathcal{C}^{-1} = \quark^\dagger_{\alpha k} \,.
 }
The operator $\mathcal{C}$ generates a $\mathbb{Z}_2$ symmetry which we sometimes use to classify states.

\subsection{Mode expansion and canonical quantization}

One can analyze this theory in canonical quantization.  The canonical anti-commutation relations are
  \es{psivAnticommRelns}{
  \{ \psi_{ij}(x^-), \psi_{kl}(y^-)\} &= \delta(x^--y^-) \left( \delta_{il} \delta_{kj} - \frac 1N \delta_{ij} \delta_{kl} \right) \,,\\
   \{v^\dagger_{\alpha i}(x^-), \quark_{j \beta}(y^-)\} &= \delta(x^--y^-) \delta_{ji} \delta_{\alpha \beta} \,.
 } 
To construct the states, we first expand $\psi$ and $v$ in Fourier modes in the $x^-$ direction
 \es{ModeExpansions}{
    \psi_{ij}(x^-) &= \frac{1}{\sqrt{2\pi}}\int_0^\infty dk^+\,\left(b_{ij}(k^+)e^{-ik^+ x^-} + b^\dagger_{ji}(k^+) e^{ik^+ x^-}\right) \,, \\
    \quark_{i \alpha }(x^-) &= \frac{1}{\sqrt{2\pi}}\int_0^\infty dk^+\,\left(d_{i \alpha} (k^+)e^{-ik^+ x^-} + c_{\alpha i}^\dagger(k^+) e^{ik^+ x^-}\right) \,,
 }
and then one can check that the canonical commutation relations \eqref{psivAnticommRelns} imply 
 \es{AnticommCreation}{
    \left\{c^\dagger_{\alpha i}(k), c_{\beta j}(k')\right\} &= \delta_{ij} \delta_{\alpha \beta} \delta(k-k')\,, \qquad \left\{d^\dagger_{i \alpha}(k) , d_{j \beta}(k')\right\} = \delta_{ij} \delta_{\alpha \beta} \delta(k-k') \,, \\
     \left\{b^\dagger_{ij}(k),b_{kl}(k')\right\} &= \delta(k-k')\left(\delta_{ik}\delta_{jl}-\frac{1}{N}\delta_{ij}\delta_{kl}\right) \, ,
 }
where we have omitted the superscript $+$ on the momenta. 
We treat $c_{\alpha i}(k)$, $d_{i \alpha}(k)$, and $b_{ij}(k)$ as annihilation operators and $c^\dagger_{\alpha i}(k)$, $d^\dagger_{i \alpha}(k)$, and $b^\dagger_{ji}(k)$ as creation operators, and we assume that all annihilation operators annihilate the vacuum $|0 \rangle$.  The space of states is obtained by acting with creation operators on this vacuum. The charge conjugation operator acts on the creation operators according to
\begin{align}
	\mathcal{C} b^\dagger_{ij}(k)\mathcal{C}^{-1} = b^\dagger_{ji}(k) \,, \qquad  \mathcal{C} c^\dagger_{\alpha i}(k)\mathcal{C}^{-1} = d^\dagger_{i\alpha}(k) \,,  \qquad \mathcal{C}d^\dagger_{i\alpha}(k)\mathcal{C}^{-1} = c^\dagger_{\alpha i}(k) \,.
\end{align}

At leading order in large $N$, the single-trace states decouple, in the sense that the matrix elements of $M^2 = 2 P^+ P^-$ between single-trace and multi-trace states are suppressed in $1/N$.  Our main interest here will be in the single-trace mesonic states.  For a fixed $P^+$ component of the momentum, the most general such single-trace mesonic state can be written as 
 \es{MesonState}{
    \ket{\{g\}_{\alpha \beta}; P^+} = \frac{(P^+)^{(n-1)/2}}{N^{(n-1)/2}} &\sum_{n} \int_0^{1} dx_1\,\,\cdots\,dx_{n}\, \delta\left(\sum_{i=1}^{n} x_i - 1 \right)\\
    &{}\times g_{n}(x_1,\ldots,x_{n})c_\alpha^\dagger(k_1)b^\dagger(k_2)\cdots b^\dagger(k_{n-1})d_\beta^\dagger(k_{n})\ket{0} \,,
 }
where $x_i = k_i / P^+$ are the momentum fractions, $g_n$ is the $n$-bit component of the wavefunction, and $\{g \}$ denotes the set of all $g_n$ with $n \geq 2$.  The overall powers of $P^+$ and $N$ are such that, at leading order in $1/N$, the inner product on the states \eqref{MesonState} takes the form
 \es{InnerProdMeson}{
  \langle \{g'\}_{\alpha \beta}; P^{\prime + } | \{ g\}_{\gamma \delta}; P^+ \rangle
   &= \sum_n \int_0^1 dx_1 \cdots dx_n \, g'_n(x_1, \ldots, x_n)^* g_n(x_1, \ldots, x_n) \\
   &{}\times \delta(P^{ \prime + } - P^+)  \delta_{\alpha \gamma} \delta_{\beta \delta} \,,
 } 
and thus is of order $N^0$ at large $N$.   The analogous single-trace gluinoball states previously studied in \cite{Bhanot:1993xp} are of the form
 \es{WaveFnGlueball}{
  |\{f\}; P^+ \rangle =   \sum_n  \frac{(P^+)^{(n-1)/2}}{N^{n/2}} \int_0^1 &dx_1\, \cdots dx_n\,  \delta  \left(\sum_{i=1}^n x_i - 1 \right) \\
    &{}\times f_n(x_1, \ldots, x_n) \tr \left[b^\dagger(k_1) \cdots b^\dagger(k_n)  \right] |0 \rangle  \,,
 }
where now, due to the cyclic property of the trace, the $n$-bit component functions $f_n(x_1, \ldots, x_n)$ obey $f_n(x_2, \ldots, x_n, x_1) = (-1)^{n-1}f_n(x_1, \ldots, x_n)$.  At leading order in $1/N$, the inner product on these states takes the form
 \es{InnerProdGlue}{
  \langle \{f'\}; P^{\prime + } | \{ f\}; P^+ \rangle
   = \delta(P^{ \prime + } - P^+) \sum_n n \int_0^1 dx_1 \cdots dx_n \, f'_n(x_1, \ldots, x_n)^* f_n(x_1, \ldots, x_n) \,.
 } 
Note that a different power of $N$ is needed in \eqref{MesonState} and \eqref{WaveFnGlueball} in order to achieve an inner product that does not scale with $N$.  The single-trace meson and gluionoball states in \eqref{MesonState} and \eqref{WaveFnGlueball} are orthogonal.

\subsection{Light-cone momentum in canonical quantization}\label{sec:pminus}

Inserting the mode expansions \eqref{ModeExpansions} into the expression for $P^+$ in \eqref{Ppm} gives
 \es{PpOscillators}{
    P^+ = \int_0^\infty dk\, k\left(b_{ij}^\dagger(k) b_{ij}(k) + c_{\alpha i}^\dagger(k) c_{\alpha i}(k) + d_{i \alpha}^\dagger(k) d_{i \alpha}(k)\right).
 }
The expression for $P^-$ is more complicated since it involves mass terms that are quadratic in the creation operators as well as interaction terms arising from the first term in \eqref{Ppm}.  We thus split up $P^-$ as 
 \es{PmGen}{
  P^- = P^-_\text{mass} + P^-_\text{int} 
 }
where
\begin{equation}\label{eq:pminus_mass}
    P^-_\text{mass}  = \frac{m^2_\text{adj}}{2}\int_0^\infty dk\,\frac{1}{k}b_{ij}^\dagger(k)b_{ij}(k) + \frac{m^2_\text{fund}}{2}\int_0^\infty dk\,\frac{1}{k}\left(c_{\alpha i}^\dagger(k)c_{\alpha i}(k)+d_{i \alpha}^\dagger(k)d_{i \alpha}(k)\right) \,,
\end{equation}
and 
 \es{Pmint}{
  P^-_\text{int} = - \frac{g^2}{2}\int dx^-\,\tr\left(J^+ \frac{1}{\partial_-^2}J^+\right) \,.
 }
To evaluate this term, note that the current is
 \es{GotJpExpansion}{
    J^+_{ij}(x^-) &= \frac{1}{2\pi}\int_0^\infty dk\,dk'\,\Bigg\lbrack \left(b_{ik}(k)e^{-ik x_-} + b^\dagger_{ki}(k) e^{ik x_-}\right)\left(b_{kj}(k')e^{-ik' x_-} + b^\dagger_{jk}(k') e^{ik' x_-}\right) \\
    &{}+ \left(d_{i \alpha}(k)e^{-ik x_-} + c_{\alpha i}^\dagger(k) e^{ik x_-}\right)\left(d_{j \alpha}^\dagger(k')e^{ik' x_-} + c_{\alpha j}(k') e^{-ik' x_-}\right)\Bigg\rbrack \,.
 }
When this is expanded out, $J^+_{ij}$ has 8 terms, and so $\tr \left( J^+ \frac{1}{\partial_-^2}J^+\right)$ has 64 terms.  A careful calculation gives
 \es{GotPminusInt}{
  P^-_\text{int} &= \frac{g^2}{2\pi}\int_0^\infty d\vec{k}\,\biggl[  \delta(k_1+k_2-k_3-k_4) \left(\frac{1}{(k_1-k_3)^2}-\frac{1}{(k_1+k_2)^2}\right)b_{ik}^\dagger(k_1)b_{kj}^\dagger(k_2)b_{il}(k_3)b_{lj}(k_4) \\
  &{}+ \delta(k_1+k_2+k_3-k_4)\left(\frac{1}{(k_1+k_2)^2}-\frac{1}{(k_2+k_3)^2}\right)b^\dagger_{ik}(k_1)b^\dagger_{kl}(k_2)b^\dagger_{lj}(k_3)b_{ij}(k_4)\\
  &{}+ \delta(k_1+k_2+k_3-k_4)\left(\frac{1}{(k_2+k_3)^2}-\frac{1}{(k_1+k_2)^2}\right)b^\dagger_{ij}(k_4)b_{ik}(k_1)b_{kl}(k_2)b_{lj}(k_3)\\
  &{}+ \delta(k_1+k_2-k_3-k_4) \frac{1}{(k_2-k_4)^2} c^\dagger_{\alpha i}(k_1)d^\dagger_{i \beta}(k_2)c_{\alpha j}(k_3)d_{j \beta}(k_4) \\
  &{}+ \delta(k_1-k_2-k_3-k_4) \frac{c^\dagger_{\alpha j}(k_1)c_{\alpha i}(k_2)b_{ik}(k_3)b_{kj}(k_4) - c^\dagger_{\alpha j}(k_2)b^\dagger_{jk}(k_3)b^\dagger_{ki}(k_4)c_{\alpha i}(k_1)}{(k_1-k_2)^2} \\
  &{}+ \delta(k_1-k_2-k_3-k_4)\frac{b^\dagger_{jk}(k_2)b^\dagger_{ki}(k_3)d_{i \alpha}^\dagger(k_4)d_{j \alpha}(k_1)- d^\dagger_{i \alpha}(k_1)b_{ik}(k_2)b_{kj}(k_3)d_{j \alpha}(k_4)}{(k_1-k_4)^2} \\
  &{}+ \delta(k_1+k_2-k_3-k_4) \frac{c^\dagger_{\alpha j}(k_1)b^\dagger_{jk}(k_2)c_{\alpha i}(k_3)b_{ik}(k_4) + b^\dagger_{ki}(k_1)d^\dagger_{i \alpha}(k_2)b_{kj}(k_3)d_{j \alpha}(k_4)}{(k_1-k_3)^2} \\
  &{}+ \frac{g^2N}{\pi}\int_0^\infty dk\, \biggl( b^\dagger_{ij}(k)b_{ij}(k)
    + c_{\alpha i}^\dagger(k)c_{\alpha i}(k) + d_{i \alpha}^\dagger(k)d_{i \alpha}(k) \biggr) \int_0^k \frac{dp}{(p-k)^2}  \,,
 }
where we ignored terms that are suppressed in $1/N$ when acting on the single-trace meson and gluinoball states.  More precisely, we ignored terms whose matrix elements between the states \eqref{MesonState} and \eqref{WaveFnGlueball} vanish as $N \to \infty$.

\section{Discretized eigenvalue problem}
\label{DISCRETE}

To treat the problem numerically, we compactify the $x^-$ direction into a circle of circumference $2 \pi L$ and impose antiperiodic boundary conditions for the fermions, namely
\begin{equation}
    \psi_{ij}(x^-) = -\psi_{ij}(x^-+2\pi L) \,, \qquad
     \quark_i(x^-) = - \quark_i(x^- + 2 \pi L) \,.
\end{equation}
Then the allowed momenta are $k_n = \frac{n}{2L}$, with odd $n$.  All the formulas in the previous section can be easily rewritten in the discretized case by limiting the range of the $x$ integrals to $[0, 2\pi L]$ and replacing $\delta(k - k') \mapsto L \delta_{k, k'}$ and $\int dk \mapsto \frac{1}{L} \sum_k$, where $\delta_{k, k'}$ is the Kronecker delta symbol.  Thus, in the discrete case the anti-commutation relations \eqref{AnticommCreation} become
  \es{AntiCommbCompact}{
  \left\{ b_{ij}(k_1), b^\dagger_{lk}(k_2) \right\} &= L \delta_{k_1, k_2} \left( \delta_{il} \delta_{jk} - \frac 1N \delta_{ij} \delta_{kl} \right) \,, \\
   \left\{ c_{\alpha i}(k_1), c^\dagger_{\beta j}(k_2) \right\}  &= \left\{ d_{i \alpha}(k_1), d^\dagger_{j \beta}(k_2) \right\}=  L\delta_{k_1, k_2} \delta_{ij} \delta_{\alpha \beta} \,.
 } 
To simplify the notation and to exhibit the fact that $L$ drops out from the formula for the masses, we define the dimensionless annihilation and creation operators.  The annihilation operators are defined by
 \es{BCDDefs}{
  B_{ij}(n) = \frac{1}{\sqrt{L}} b_{ij}\left( \frac{n}{2L} \right) \,, \qquad C_{\alpha i}(n) =  \frac{1}{\sqrt{L}} c_{\alpha i} \left( \frac{n}{2L} \right)\,, \qquad 
  D_{i \alpha}(n) =  \frac{1}{\sqrt{L}} d_{i \alpha}\left( \frac{n}{2L} \right)\,,
 }
with $n>0$ odd, and the corresponding creation operators are defined analogously.   These dimensionless operators obey
   \es{AntiCommDimless}{
  \left\{ B_{ij}(n_1), B^\dagger_{lk}(n_2) \right\} &= \delta_{n_1, n_2} \left( \delta_{il} \delta_{jk} - \frac 1N \delta_{ij} \delta_{kl} \right) \,, \\
   \left\{ C_{\alpha i}(n_1), C^\dagger_{\beta j}(n_2) \right\}  &= \left\{ D_{i \alpha}(n_1), D^\dagger_{j \beta}(n_2) \right\}=  \delta_{n_1, n_2} \delta_{ij} \delta_{\alpha \beta} \,.
 }

In terms of the dimensionless oscillators, the mode decompositions in \eqref{ModeExpansions} become
 \es{psiTobCompact}{
  \psi_{ij}(x) &= \frac{1}{\sqrt{2\pi L} } \sum_{\text{odd }n>0} \left( B_{ij}(n) e^{-i n \frac{x}{2L} } + B^\dagger_{ji}(n) e^{i n \frac{x}{2L}} \right) \,, \\
  \quark_{i \alpha}(x) &= \frac{1}{ \sqrt{2\pi L} }\sum_{\text{odd }n>0} \left( D_{i\alpha}(n) e^{-i n \frac{x}{2L}} + C^\dagger_{\alpha i}(n) e^{in \frac{x}{2L}} \right) \,,
 }
the operator $P^+$ is
  \es{PpOscillatorsDiscrete}{
    P^+ = \frac{1}{2L}  \sum_{\text{odd }n>0} \,n \left(B_{ij}^\dagger(n) B_{ij}(n) + C_{\alpha i}^\dagger(n) C_{\alpha i}(n) + D_{i \alpha}^\dagger(n) D_{i  \alpha}(n)\right) \,,
 }
and the operator $P^-$ can be written as
 \es{PmOscillatorsDiscrete}{
  P^- =  \frac{g^2 L}{\pi}\left(y_\text{adj}V_\text{adj} + y_\text{fund} V_\text{fund} + T\right) \,,
 }
where we have defined $y_\text{adj} = \frac{m^2_\text{adj} \pi}{g^2 N}$ and $y_\text{fund} = \frac{m^2_\text{fund} \pi}{g^2 N}$, with 
 \es{VDefs}{
    V_\text{adj} &=\sum_{\text{odd }n>0} \frac{1}{n}B^\dagger_{ij}(n)B_{ij}(n) \,, \qquad
       V_\text{fund} = \sum_{\text{odd }n>0} \frac{1}{n}\left(C^\dagger_{\alpha i}(n)C_{\alpha i}(n) + D^\dagger_{i \alpha}(n)D_{i \alpha}(n)\right) \,,
 }
and
 \es{TDef}{
    T =N \sum_{\text{odd }n>0} \Big(4B^\dagger_{ij}(n) B_{ij}(n)&+2C^\dagger_{\alpha i}(n)C_{\alpha i}(n) + 2D^\dagger_{i \alpha}(n) D_{i \alpha}(n)\Big) \sum_m^{n-2} \frac{1}{(n-m)^2} \\ 
      + 2\sum_{\text{odd }n_i>0}\Bigg\lbrace \delta_{n_1+n_2,n_3+n_4}\Bigg\lbrack &\left(\frac{1}{(n_1-n_3)^2}-\frac{1}{(n_1+n_2)^2}\right)B_{ik}^\dagger(n_1)B_{kj}^\dagger(n_2)B_{il}(n_3)B_{lj}(n_4) \\
    &+\frac{1}{(n_1-n_3)^2}C^\dagger_{\alpha i}(n_1)B^\dagger_{ik}(n_2)C_{\alpha j}(n_3)B_{jk}(n_4) \\
    &+\frac{1}{(n_1-n_3)^2}B^\dagger_{ki}(n_1)D^\dagger_{i \alpha}(n_2)B_{kj}(n_3)D_{j \alpha}(n_4) \\
    &+\frac{1}{(n_1-n_3)^2}C^\dagger_{\alpha i}(n_1)D^\dagger_{i \beta}(n_2)C_{\alpha j}(n_3)D_{j \beta}(n_4) \Bigg\rbrack \\
    +\delta_{n_1,n_2+n_3+n_4}\Bigg\lbrack &\left(\frac{1}{(n_3+n_4)^2}-\frac{1}{(n_2+n_3)^2}\right)B^\dagger_{ij}(n_1)B_{ik}(n_2)B_{kl}(n_3)B_{lj}(n_4) \\
    &+\frac{1}{(n_3+n_4)^2}C^\dagger_{\alpha j}(n_1)C_{\alpha i}(n_2)B_{ik}(n_3)B_{kj}(n_4) \\
    &-\frac{1}{(n_2+n_3)^2}D^\dagger_{i \alpha}(n_1)B_{ik}(n_2)B_{kj}(n_3)D_{j \alpha}(n_4)+\hc\Bigg\rbrack\Bigg\rbrace \,.
}

Since $P^+$ and $P^-$ commute, we can work at fixed $P^+ = K/(2L)$ for some integer $K$.  This means we consider single-trace meson states of the form 
 \es{MesonStates}{
    \frac{1}{N^{(p-1)/2}} C_\alpha^\dagger(n_1) B^\dagger(n_2)\cdots B^\dagger(n_{p-1}) D_\beta^\dagger(n_p) \ket{0} \,, 
 }
with $\sum_{i=1}^p n_i = K$, as well as single-trace gluinoball states
 \es{GlueballStates}{
    \frac{1}{N^{p/2}} \tr\left(B^\dagger(n_1)\cdots B^\dagger(n_p)\right)\ket{0} \,, 
}
with the same condition on the sum of the $n_i$.  At leading order in large $N$, one can choose an orthonormal basis of such states.   Just as in \eqref{MesonState} and \eqref{WaveFnGlueball}, the overall powers of $N$ in \eqref{MesonStates}--\eqref{GlueballStates} are such that the states have finite norm as $N \to \infty$, and from the expression for $P^-$ we dropped all terms whose matrix elements are suppressed in $1/N$ in the large $N$ limit.

The mass squared operator becomes
 \es{GotM2}{
    M^2 = 2P^+P^- = \frac{g^2 K}{\pi}\left(y_\text{adj}V_\text{adj} + y_\text{fund} V_\text{fund} + T\right),
 }
Notice that the factors of $L$ canceled from this expression.  Removing the cutoff corresponds to taking $L \to \infty$ and $K\to\infty$ with $P^+$ fixed.  In Sections~\ref{sec:glue_spectrum}, \ref{sec:MESON}, and~\ref{sec:MASSIVE} we will use the expression \eqref{GotM2} to find the spectrum of gluinoballs and mesons numerically at leading order in large $N$.

\section{Kac-Moody algebra in DLCQ and exact degeneracies of the spectrum}
\label{KMA}

When $m_\text{fund} = m_\text{adj} = 0$, the discretized problem of finding the spectrum of $M^2$ presented in the previous section can be somewhat simplified upon using the $\SU(N)$ gauge current algebra and its representations \cite{Kutasov:1994xq}.  While we will not use this simplification, let us now review this approach because it will be useful for interpreting the numerical results of Sections~\ref{sec:glue_spectrum} and \ref{sec:MESON}.

\subsection{Current algebra}

The simplification mentioned above is based on the fact that the operator $P^-$ in \eqref{PmOscillatorsDiscrete}--\eqref{TDef} can be expressed solely in terms of the Fourier modes of the gauge current $J^+_{ij}$.  Let us define the $n$th Fourier mode $J_{ij}(n)$ (with $n \in 2 \Z$) through
 \es{JModes}{
  J^+_{ij}(x^-) = \frac{1}{ 2\pi L } \sum_{\text{even }n} J_{ij}(n) e^{-i n \frac{x^-}{2 L} } \,,  \qquad J_{ij}(n)  = J_{\text{adj}, ij}(n) + J_{\text{fund}, ij}(n)
 }
with the adjoint and fundamental contributions $J_{\text{adj}, ij}(n)$ and $J_{\text{fund}, ij}(n)$ being
 \es{JnOscillator}{
  J_{\text{adj}, ij}(n) &\equiv \sum_{n_1 + n_2 = n} \left(  B_{ik}(n_1) B_{kj} (n_2) - \frac{1}{N} \delta_{ij} B_{kl} (n_1) B_{lk} (n_2) \right)  \\
  J_{\text{fund}, ij}(n) &\equiv  \sum_{n_1 + n_2 = n} \left( D_{i \alpha}(n_1) C_{\alpha j} (n_2) - \frac{1}{N} \delta_{ij} D_{k\alpha} (n_1) C_{\alpha k} (n_2) 
   \right) \,,
 }
where for simplicity we denoted $B_{ij}(-n) = B^\dagger_{ji}(n)$, $D_{i \alpha}(-n) = C^\dagger_{\alpha i}(n)$, and $C_{\alpha i}(-n) = D^\dagger_{i \alpha}(n)$ for $n>0$.  For $n>0$ we have that $J_{ij}(n) \ket{0} = 0$ because each term in $J_{ij}(n)$ involves at least one annihilation operator.  We also have that $J_{ij}(0) \ket{0} = 0$ because the $J_{ij}(0)$ are the conserved $\SU(N)$ charge operators that annihilate the Fock vacuum.  In fact, $J_{ij}(0) \ket{\psi}  = 0$ for any gauge-invariant state $\ket{\psi}$.  When $n<0$, $J_{ij}(n) \ket{0} \neq 0$ because in this case $J_{ij}(n)$ contains at least one term with only fermionic creation operators.

Using the commutation relations \eqref{AntiCommDimless} and \eqref{JnOscillator}, we can compute the commutation relation of $J$ with the fermionic oscillators:
 \es{CommutJpsi}{
  [J_{ij}(n), B_{kl}(m)] &= \delta_{kj} B_{il} (n+m) - \delta_{il} B_{kj} (n+m) \,, \\
  [J_{ij}(n), D_{k\alpha}(m)] &= \delta_{kj} D_{i\alpha}(n+m)
   - \frac{1}{N}  \delta_{ij} D_{k \alpha} (n+m) \,,\\
   [J_{ij}(n), C_{\alpha k }(m)] &= -\delta_{ik} C_{\alpha j}(n+m) + \frac 1N \delta_{ij} C_{\alpha k}(n+m) \,.
 } 
From \eqref{CommutJpsi} and the expression \eqref{JnOscillator} for $J$ in terms of fermionic oscillators, we can derive the Kac-Moody (KM) current algebra
 \es{CommutJJ}{
  [J_{ij}(n), J_{kl}(m)]  = \delta_{kj} J_{il} (n+m) - \delta_{il} J_{kj} (n+m) + k_\text{KM} \frac{n\,  \delta_{n+m, 0}}{2} \left(\delta_{il} \delta_{kj}  
   - \frac 1N \delta_{ij} \delta_{kl} \right) \,,
 }
with level $k_\text{KM}=N + N_f$.\footnote{If instead of a single adjoint Majorana fermion we had $n_\text{adj}$ flavors of adjoint Majorana fermions, we would have obtained \eqref{CommutJJ} with $k_\text{KM} = n_\text{adj} N + N_f$.}  Note that \eqref{CommutJpsi} and the first two terms in \eqref{CommutJJ} follow from the $\SU(N)$ transformation properties of the operators being commuted with the current. The operators $J_{ij}(n)$ carry $P^+$ momentum equal to $-n / (2L)$, as can be seen from the commutation relation
 \es{JCommPp}{
  [P^+, J_{ij}(n)] = - \frac{n}{2L} J_{ij}(n) \,.
 }
Thus, when $J_{ij}(n)$ acts on a state with a definite value of $K = 2 L P^+$, it lowers $K$ by $n$ units.

For massless fermions, the expression for $P^-$ is
 \es{Pm}{
  P^- = -\frac{g^2}{2}  \int dx^- \tr \left( J^+ \frac{1}{\partial_-^2} J^+ \right)  \,.
 }
Plugging \eqref{JModes} into this expression gives 
 \es{PmFinal}{
  P^- = \frac{g^2 L }{\pi} \sum_{\text{even } n\neq 0}  \frac{ \tr \left[ J(-n)  J(n) \right]}{n^2} =   \frac{2g^2 L }{\pi} \sum_{\text{even } n>0}  \frac{ \tr \left[ J(-n)  J(n) \right]}{n^2}  \,,
 }
where there are no $n=0$ terms in these sums because $J(0)$ annihilates all gauge-invariant states, and where in the second equality we used the algebra \eqref{CommutJJ} to interchange $J(-n)$ with $J(n)$ for $n<0$ at the expense of a divergent $c$-number energy shift.  The requirement that $P^-$ annihilate the vacuum $\ket{0}$ fixes the regularized value of this divergent term to zero.  Unlike the expressions \eqref{PmOscillatorsDiscrete}--\eqref{TDef} which hold only to leading order in large $N$ when acting on single-trace states, the expression \eqref{PmFinal} is exact at finite $N$ as well.

As we will explain in more detail below, the advantage of the form \eqref{PmFinal} is to make it manifest that $P^-$ only mixes states that belong to the same representation of the current algebra \eqref{CommutJJ} \cite{Kutasov:1994xq}.  Representations of the Kac-Moody current algebra \eqref{CommutJJ} can be constructed in the free theory of $\psi_{ij}$, $\quark_{i \alpha}$, and $v^\dagger_{\alpha i}$, and they each contain both gauge-invariant states (i.e.,~annihilated by $J_{ij}(0)$) and non-gauge-invariant states.  Because $P^-$ commutes with $J_{ij}(0)$, it follows that $P^-$ maps gauge-invariant states to gauge-invariant states, so after constructing a given KM representation we can restrict our attention to its gauge-invariant subspace.

A representation of the KM algebra (also referred to below as a current block or KM block) starts with a KM primary state $|\chi \rangle_I$ (generally not gauge invariant with all $\SU(N)$ indices grouped together into the multi-index $I$) that is annihilated by all $J_{ij}(n)$ with $n>0$.  The other states in the representation are KM descendants obtained by acting with $J_{ij}(n)$ with $n<0$:\footnote{Acting with $J_{ij}(0)$ on the KM primary or one of its descendants can be reduced to a linear combination of the KM primary and the descendants because $J_{ij}(0)$ acts as the corresponding $\SU(N)$ generator.}
 \es{GenericState}{
  J_{i_1 j_1}(-n_1)  J_{i_2 j_2}(-n_2)  \cdots  J_{i_p j_p}(-n_p) \ket{\chi}_I
 }
with $n_i > 0$.  For such a state to be gauge-invariant, all $\SU(N)$ indices of the $J$'s and of $\ket{\chi}$ must be fully contracted.  Because $J_{ij}(n)$ carries $P^+$ momentum equal to $-n/2L$ (see \eqref{JCommPp}), the states of a KM representation are graded by the value of $K = 2 L P^+$, with the KM primary having the smallest value of $K$.

When acting with $P^-$ on a state of the form \eqref{GenericState}, we can use the algebra \eqref{CommutJJ} to commute the $J(n)$ with $n>0$ from \eqref{PmFinal} all the way to the right, where it annihilates $\ket{\chi}_I$.  Any $J(n')$ with $n'>0$ that was generated in this process from the commutators can also be commuted all the way to the right, and the process can be repeated until we are left with a linear combination of states of the form \eqref{GenericState} that do not contain any $J(n)$ with $n>0$.  Thus, $P^-$ takes the states \eqref{GenericState} to linear combinations of states of the same form, and thus the diagonalization of $P^-$ can be done independently for each current block.

As explained in \cite{Kutasov:1994xq}, this construction not only shows that $P^-$ is diagonal in each current block, but also that the eigenvalues of $P^-$ within the gauge-invariant subspace of a given current block depend {\em only} on the level $k_\text{KM}$ of the current algebra and on the $\SU(N)$ representation of the KM primary.  Indeed, the structure of the gauge-invariant states does depend on the representation of the KM primary, and when commuting the $J(n)$'s with $n>0$ all the way to the right, we use \eqref{CommutJJ} which depends only on the gauge group and the level $k_\text{KM}$.  It follows that for any two current blocks whose KM primaries transform in the same representation of $\SU(N)$, either belonging to the same theory or to two different theories with the same value of $k_\text{KM}$, the eigenvalues of $P^-$ will be the same.  This universality of the massive spectrum, first noticed in \cite{Kutasov:1994xq}, will be very important in explaining some of the degeneracies we observe numerically in the spectrum of $P^-$ in the following sections.

 For a current block to contain singlets it is necessary that the KM primary transform in a representation of $\SU(N)$ with $N$-ality $0$, because the $N$-ality is conserved under tensor products and $J(-n)$, being an $\SU(N)$ adjoint, has $N$-ality zero.  (All other KM representations where the KM primary does not have $N$-ality $0$ does not contain any gauge-invariant states.)   The representations of $N$-ality $0$ can be obtained by taking tensor products of the adjoint representation.  We will denote by $\mathfrak{n}$ the smallest number of adjoint factors in such tensor products that are needed to obtain a given representation.  See Table~\ref{RepTable} for the $\SU(N)$ representations obtained for up to $\mathfrak{n} =3$.  For each representation, we stated in the last column whether or not it belongs to the totally anti-symmetric product of $\mathfrak{n}$ adjoint factors.  This fact will be important in the next subsection. 
\begin{table}[htp]
\begin{center}
\begin{tabular}{c|c|c}
 $\mathfrak{n}$ & $\SU(N)$ irreps & anti-symmetric product? \\
 \hline \hline
 $0$ & $\mathfrak{R}_0 = [00\ldots 0]$ (identity) & yes \\
 \hline
 $1$ & $\mathfrak{R}_1 =[10\ldots 01]$ (adjoint) & yes \\ \hline
  & \\[\dimexpr-\normalbaselineskip+2pt]
 $2$ & $\mathfrak{R}_2 =[20\ldots 010]$, $\overline{\mathfrak{R}}_2 = [010\ldots 02]$ & yes \\
 & $\mathfrak{R}_2' = [20\ldots 02]$, $\mathfrak{R}_2''= [01\ldots 10]$ & no \\
 \hline
   & \\[\dimexpr-\normalbaselineskip+2pt]
 $3$ & $\mathfrak{R}_3 = [30\ldots 0100]$, $\overline{\mathfrak{R}}_3 \oplus [0010\ldots 03]$ & yes \\
   & $\mathfrak{R}_3' = [110\ldots 011]$ & yes \\
   & $\mathfrak{R}_3'' = [30\ldots 03]$,  $\mathfrak{R}_3''' = [0010\ldots 0100]$, etc. & no \\
 \hline
 etc. & 
\end{tabular}
\end{center}
\caption{The Dynkin labels of various irreducible $\SU(N)$ representations for various values of $\mathfrak{n}$. The last column states whether the representations are contained in the anti-symmetric product of $\mathfrak{n}$ adjoint representations.  All irreps are real except for the complex representations $\mathfrak{R}_2$ and $\mathfrak{R}_3$ and their conjugates $\overline{\mathfrak{R}}_2$ and $\overline{\mathfrak{R}}_3$.}
\label{RepTable}
\end{table}%

\subsection{Gluinoball spectrum at finite $N$ when $N_f = 0$}
\label{sec:GLUINO}

Before tackling the theory with one Majorana adjoint and $N_f$ fundamental fermions, let us first review current block construction when $N_f = 0$, following \cite{Kutasov:1994xq}.  Labeling the current blocks as above by $\mathfrak{n}$, we have the following primaries:\footnote{For the $n=3$ state, the traces that need to be subtracted are: 
 \es{Tracesn3}{
  & \frac 1N \biggl[ \delta_{kj} J_{il}(-2) B^\dagger_{nm}(1) - \delta_{il} J_{kj}(-2) B^\dagger_{nm}(1)
   -  \delta_{mj} J_{in}(-2) B^\dagger_{lk}(1) + \delta_{in} J_{mj} (-2) B^\dagger_{lk} (1) \\
    &{}- \delta_{kn} J_{ml} (-2) B^\dagger_{ji}(1) + \delta_{ml} J_{kn} (-2) B^\dagger_{ji}(1)  \biggr] \,. 
 } }
 \es{PrimariesPureAdjoint}{
  \mathfrak{n} &= 0: \qquad |0 \rangle \,, \\ 
  \mathfrak{n} &= 1: \qquad B^\dagger_{ji}(1) |0 \rangle \,, \\
  \mathfrak{n} &= 2: \qquad  \left( B^\dagger_{ji}(1) B^\dagger_{lk}(1) -  \frac 1N \delta_{kj} J_{il}(-2)  +  \frac 1N \delta_{il} J_{kj} (-2)  \right) |0 \rangle \,, \\
  \mathfrak{n} &= 3:  \qquad  \left( B^\dagger_{ji}(1)B^\dagger_{lk}(1) B^\dagger_{nm}(1)  -  \text{traces}  \right) |0 \rangle  \,, \\
  &\text{etc.}
 }
They all involve products of $\mathfrak{n}$ factors of $B^\dagger(1)$ acting on the Fock vacuum, with all the $\SU(N)$ traces removed.
  Even though they are all constructed from $B_{ji}^\dagger(1)$, one can obtain descendants where no $B_{ji}^\dagger(1)$'s appear because the currents can contain $B_{ji}(1)$'s which may annihilate the $B_{ji}^\dagger(1)$'s.  It is straightforward to check that these states are annihilated by all $J_{ij}(n)$ with $n<0$.  Note that the states of even $\mathfrak{n}$ are bosons while those of odd $\mathfrak{n}$ are fermions. 
  
For $\mathfrak{n}>1$, the primaries in \eqref{PrimariesPureAdjoint} transform in reducible representations of $\SU(N)$.  To form $\SU(N)$ irreps, we should further symmetrize and/or anti-symmetrize in the fundamental indices $i$, $k$, $m$, \ldots, and then because the $B^\dagger$'s anti-commute, the states in \eqref{PrimariesPureAdjoint} will have the opposite symmetry properties in the anti-fundamental indices $j$, $l$, $n$, etc.  The number of $\SU(N)$ irreps one can construct is equal to the number of Young diagrams with $\mathfrak{n}$ boxes, with the first few representations being the ones marked as appearing in the anti-symmetric product of $\mathfrak{n}$ adjoints in Table~\ref{RepTable}.

At finite $N$, each block can be diagonalized separately, and one expects no relation between the eigenvalues of blocks whose primaries are in different $\SU(N)$ representations, with one exception.  Because the adjoint representation is real, each time a complex representation appears in the product of several adjoint representations, its conjugate representation must appear too.  For example, at $\mathfrak{n}=2$, we have $\mathfrak{R}_2$ and $\overline{\mathfrak{R}}_2$, and at $\mathfrak{n}=3$ we have $\mathfrak{R}_3$ and $\overline{\mathfrak{R}}_3$, etc.  Since our gauge theory has charge conjugation symmetry, defined in \eqref{eq:charge_conjugation}, the eigenvalues of $P^-$ for a current block with primary in representation $\mathfrak{R}$ must be equal to those of the conjugate current block whose primary is in $\overline{\mathfrak{R}}$.  Thus, one expects two-fold degeneracies for these states in the finite $N$ spectrum of $P^-$.  Such degeneracies were noticed in the numerical study of \cite{Antonuccio:1998uz}, but no clear explanation was presented.  In particular, the double degeneracies in Table~4 of  \cite{Antonuccio:1998uz} correspond to the $\mathfrak{n}=2$ degeneracies between the $\mathfrak{R}_2$ and $\overline{\mathfrak{R}}_2$ blocks and the double degeneracy discussed after Eq.~(12) of \cite{Antonuccio:1998uz} correspond to the $\mathfrak{n}=3$ degeneracy between states in the $\mathfrak{R}_3$ and $\overline{\mathfrak{R}}_3$ blocks.

\subsection{Gluinoball degeneracies at large $N$ when $N_f =0$}
\label{sec:GLUINODEG}

At large $N$, the states in each block further split into single-trace and multi-trace sectors, and more exact degeneracies appear between single-trace and multi-trace states, as we now explain.  In particular, at leading order in large $N$, the single-trace eigenvalues from a sector labeled by $\mathfrak{n}_*>1$ are in one-to-one correspondence with the sums of $\mathfrak{n}_*$ eigenvalues of single-trace states from the $\mathfrak{n}=1$ sector.  Thus, in the current block labeled by $\mathfrak{n}_*$ there will be multi-trace states that are exactly degenerate with single-trace ones.

The proof of this fact relies on the fact that the theory with one Majorana adjoint and no fundamentals has the same value of the KM level $k_\text{KM} = N$ as the theory of $N_f = N$ fundamental fermions with no adjoints.  For brevity, let us denote the former theory as $\cT_\text{adj}$ and the latter as ${\cal T}_\text{fund}$.  As mentioned above, this means that the spectrum of $P^-$ in current blocks whose primaries transform in the same representation of $\SU(N)$ must be identical in the two theories.  The two theories are not equivalent, and indeed ${\cal T}_\text{fund}$ has more current blocks than $\cT_\text{adj}$ for a given $\mathfrak{n}$,  but the description in terms of fundamental fermions in ${\cal T}_\text{fund}$ will make it easier to determine the eigenvalues in the $\mathfrak{n}>1$ sectors at leading order in $1/N$.

In particular, instead of computing the $P^-$ eigenvalues in the current blocks in \eqref{PrimariesPureAdjoint} in $\cT_\text{adj}$, we can equivalently compute the $P^-$ eigenvalues in the following blocks of $\cT_\text{fund}$ with primaries given by\footnote{For a given $\mathfrak{n}$, these are not the only KM primaries of the $\cT_\text{fund}$ theory.}
 \es{PrimariesFund}{
  \mathfrak{n} &= 0: \qquad |0 \rangle \,,  \\ 
  \mathfrak{n} &= 1: \qquad \left( C^\dagger_{\alpha i}(1) D^\dagger_{j\beta}(1) - \text{flavor and gauge trace} \right) |0 \rangle \,, \\
  \mathfrak{n} &=2: \qquad \left( C^\dagger_{\alpha i}(1) D^\dagger_{j \beta}(1) C^\dagger_{\gamma k}(1) D^\dagger_{l \delta}(1) - \text{flavor and gauge traces} \right) |0 \rangle \,, \\
  &\text{etc.}
 }
We will set the flavor indices to be all distinct, in which case there is no flavor trace subtraction necessary.

At finite $N$ the two problems are equivalent, but at large $N$ there are two simplifications that occur.  The first is that the $\SU(N)$ trace subtractions are subleading in $1/N$ both in \eqref{PrimariesFund} and \eqref{PrimariesPureAdjoint}, so one can ignore the corresponding terms.  The second is that any commutators that break any of the strings are also subleading.  In the $\mathfrak{n}=0$ block, we have the gauge-invariant single-trace states 
 \es{SingleTracen0}{
  \mathfrak{n}=0: \qquad  \frac{1}{N^p} \tr ( J(-n_1) \cdots J(-n_p) ) |0 \rangle 
 }
in both the ${\cal T}_\text{adj}$ and ${\cal T}_\text{fund}$ theories.   For $\mathfrak{n}=1$, we have the gauge-invariant single-trace states
  \es{SingleTracen1}{
  \mathfrak{n}=1: \qquad  &\frac{1}{N^{p+\frac 12}} ( J(-n_1) \cdots J(-n_p) B(-1)) |0 \rangle \\
    &\cong \frac{1}{N^{p+\frac 12}} (  J(-n_1) \cdots J(-n_p))_{ji} C^\dagger_{\alpha i}(1) D^\dagger_{j \beta}(1) |0 \rangle \,,
 }
for any $\alpha \neq \beta$.  The ``$\cong$'' sign means that as far as the matrix elements and spectrum of $P^-$ are concerned, we can identify (up to normalization) the states on the left in the ${\cal T}_\text{adj}$ theory with the states on the right in the $\cT_\text{fund}$ theory.   While the ``single-trace'' terminology is appropriate in the ${\cal T}_\text{adj}$ theory, in ${\cal T}_\text{fund}$ the state \eqref{SingleTracen1} would be better referred to as ``single-string,'' because it consists of a sequence of $C^\dagger$'s and $D^\dagger$'s that alternate between having color indices and flavor indices contracted except at the endpoints of the string.

For $\mathfrak{n}=2$, we have the gauge-invariant single-trace states
 \es{SingleTracen2}{
  \mathfrak{n}=2: \qquad &\frac{[ J(-n_1) \cdots J(-n_{p_1}) ]_{jk} [ J(-m_1) \cdots J (-m_{p_2}) ]_{li} }{N^{p_1 + p_2+1}} 
    B^\dagger_{ji}(1) B^\dagger_{lk}(1) |0 \rangle \\
   &\hspace{-.5in}\cong \frac{[ J(-n_1) \cdots J(-n_{p_1}) ]_{jk} [ J(-m_1) \cdots J (-m_{p_2}) ]_{li} }{N^{p_1 + p_2+1}} 
    C^\dagger_{\alpha i}(1) D^\dagger_{j \beta }(1) C^\dagger_{\gamma k}(1) D^\dagger_{l \delta }(1) |0 \rangle \,,
 }
for any pairwise distinct $\alpha, \beta, \gamma, \delta$.  Again, these states are single-trace only in ${\cal T}_\text{adj}$;  in ${\cal T}_\text{fund}$ they are ``double-string'' states.   A similar construction holds for $\mathfrak{n}>2$:
\es{SingleTraceGeneraln}{
  &\frac{ \left(  \prod_{k=1}^{p_1} J(-n_{1, k})  \right)_{i_2 i_3} \cdots  \left(  \prod_{k=1}^{p_\mathfrak{n}} J(-n_{\mathfrak{n}, k})  \right)_{i_{2 \mathfrak{n}} i_1}}{N^{p_1 + \cdots + p_\mathfrak{n}+\frac{\mathfrak{n}}{2}}}  
    B^\dagger_{i_2 i_1}(1) \cdots B^\dagger_{i_{2 \mathfrak{n}} i_{2\mathfrak{n}-1} }(1) |0 \rangle \\
   &\cong \frac{ \left(  \prod_{k=1}^{p_1} J(-n_{1, k})  \right)_{i_2 i_3} \cdots  \left(  \prod_{k=1}^{p_\mathfrak{n}} J(-n_{\mathfrak{n}, k})  \right)_{i_{2 \mathfrak{n}} i_1}}{N^{p_1 + \cdots + p_\mathfrak{n}+\frac{\mathfrak{n}}{2}}}  
    C^\dagger_{\alpha_1 i_1}(1) D^\dagger_{i_2 \beta_1}(1) \cdots 
    C^\dagger_{\alpha_\mathfrak{n} i_{2\mathfrak{n}-1}}(1) D^\dagger_{i_{2\mathfrak{n}} \beta_\mathfrak{n}}(1) 
     |0 \rangle \,,
 }
with $\{ \alpha_i, \beta_i \}$ all distinct.  Neglecting terms suppressed in $1/N$, the state on the RHS of \eqref{SingleTraceGeneraln} can also be written as
 \es{SingleTraceGeneralnAgain}{
  \frac{  \left(  \prod_{k=1}^{p_1} J(-n_{1, k})  \right)_{i_2 i_3}  
   C^\dagger_{\alpha_2 i_3 }(1) D^\dagger_{i_2 \beta_1}(1)}{N^{p_1 +\frac{1}{2}}}  
  \cdots 
    \frac{  \left(  \prod_{k=1}^{p_\mathfrak{n}} J(-n_{1, k})  \right)_{i_{2\mathfrak{n}} i_1}  
   C^\dagger_{\alpha_1 i_1 }(1) D^\dagger_{i_{2 \mathfrak{n}} \beta_\mathfrak{n}}(1)}{N^{p_\mathfrak{n} +\frac{1}{2}}} 
    \ket{0} \,,
 }
which has the form of a multi-string state, with $\mathfrak{n}$ factors as in \eqref{SingleTracen1}.

Thus, single-trace states of the current block labeled by $\mathfrak{n}$ in the $\cT_\text{adj}$ theory are in fact multi-string states in the $\cT_\text{fund}$ theory, with the single-string factors coming from the $\mathfrak{n} = 1$ block.  Because  for multi-string states large $N$ factorization works similarly as for multi-trace states, it follows that the eigenvalues of single-trace states in the $\mathfrak{n}$th block of the $\cT_\text{adj}$ theory are in one-to-one correspondence sums of $\mathfrak{n}$ eigenvalues from the 1st current block.

The formula \eqref{SingleTraceGeneraln} explains how to construct the eigenstates.  Suppose in the $\mathfrak{n}=1$ sector the eigenstate of $P^-$ eigenvalue $P^- = E_a$ is
 \es{n1Estates}{
  \sum_{\{n_1, \ldots, n_p\}} c^a_{n_1, \ldots, n_p}  \tr ( J(-n_1) \cdots J(-n_p) B(-1) ) |0 \rangle \,,
 }
then the state in the $\mathfrak{n}$th block with eigenvalue $E_{a_1} + \cdots + E_{a_\mathfrak{n}}$ is
  \es{nEstates}{
  \sum_{\{n_{i, 1}, \ldots, n_{i, p}\}} c^{a_1}_{n_{1, 1}, \ldots, n_{1, p}}  \cdots
   c^{a_\mathfrak{n}}_{n_{\mathfrak{n}, 1}, \ldots, n_{\mathfrak{n}, p}} 
   &( J(-n_{1, 1}) \cdots J(-n_{1, p_1}) )_{i_2 i_3}  \cdots 
    ( J(-n_{\mathfrak{n}, 1}) \cdots J(-n_{\mathfrak{n}, p_\mathfrak{n}}) )_{i_{2\mathfrak{n}} i_1}   \\
    &{}\times
    B_{i_1 i_2}(-1) \cdots B_{i_{2\mathfrak{n}-1} i_{2\mathfrak{n}}}(-1)  |0 \rangle \,.
 }
One can obtain a multi-trace state in ${\cal T}_\text{adj}$  with the same $P^-$ eigenvalue $E_{a_1} + \cdots + E_{a_\mathfrak{n}}$ by contracting the indices in the first line with those in the second line of \eqref{nEstates} in a different way.  Of course, some of these states may vanish because of Fermi statistics.

The discussion above is restricted to the large $N$ spectrum of gluinoballs in $\cT_\text{adj}$.  The discussion also applies to the gluinoball spectrum of the theory with a Majorana adjoint and $N_f$ fundamental fermions that we are interested in here, in the limit $N \to \infty$ at fixed $N_f$.

\subsection{Current blocks and degeneracies at finite $N$ for $N_f > 0$}
\label{sec:BLOCKS}

Let us now go back to the $\SU(N)$ gauge theory with a Majorana adjoint fermion and $N_f$ Dirac fermions of primary interest in this paper, and let us list the first few KM blocks.   

Unlike in the pure adjoint theory whose primaries are listed in \eqref{PrimariesPureAdjoint}, we now have several current blocks for each $\mathfrak{n}$.  For $\mathfrak{n}=0$ ($\SU(N)$ singlets), we have the Fock vacuum $\ket{0}$, the equal linear combinations of two-bit mesons of the form $C_{\alpha i}^\dagger(n_1) D_{i \beta}^\dagger(n_2) \ket{0}$ with any total even value of $K \geq  2$,\footnote{The state  $\ket{\zeta_{\alpha \beta, K}}$ is  the unique exactly massless state of the 't Hooft model \cite{tHooft:1974pnl}.} the equal linear combination of three-bit mesons of the form $C_{\alpha i}^\dagger(n_1) B^\dagger_{ij}(n_2) D_{j \beta}^\dagger(n_3) \ket{0}$ with any total odd value of $K \geq 3$, as well as infinitely many other more complicated singlet KM primaries with more than three-bit components for which we do not have a concise expression:
 \es{SingletPrimaries}{
  \mathfrak{n} = 0: \qquad &\ket{0} \,, \qquad
   \ket{\zeta_{\alpha \beta, K}} = \sum_{\substack{m_1 + m_2 = K\\m_{1, 2}>0 \text{ odd}}}  C^\dagger_{\alpha i}(m_1) D^\dagger_{i \beta}(m_2)  \ket{0} \,, \\
   &\ket{\xi_{\alpha \beta, K}} = \sum_{\substack{m_1 + m_2 + m_3 = K\\m_{1, 2, 3}>0 \text{ odd}}} C^\dagger_{\alpha i}(m_1) B^\dagger_{ij}(m_2) D^\dagger_{j \beta}(m_3)   \ket{0} \,, \ldots 
 }
As per the discussion above, each such block has the exact same $P^-$ spectrum, so every $P^-$ eigenvalue in the $\mathfrak{n}=0$ sector is highly degenerate.  In particular, all KM primaries in the $\mathfrak{n}=0$ sector have $P^- = 0$ just like the Fock vacuum $\ket{0}$.  As we will explain in the next section, this theory has a finite but growing number of $P^- = 0$ states at each $K$, and their presence is due to the fact that the infrared is controlled by a nontrivial CFT\@.  Each $P^- =0$ state generates its own $\mathfrak{n}=0$ current block.\footnote{The expression for $P^-$ in \eqref{PmFinal} makes manifest that $P^-$ is a non-negative-definite operator and that $P^- \ket{\chi} = 0$ if and only if $J_{ij}(n) \ket{\chi}= 0$ for all $n>0$.}  Note that as emphasized in \cite{Kutasov:1994xq}, in light-cone quantization one can see all massive states and the right-moving massless ones, but not the left-moving massless states.  Thus, the $P^- = 0$ states seen in DLCQ do not provide a complete description of the massless sector.

For $\mathfrak{n}=1$ ($\SU(N)$ adjoints), we have the KM primary from the pure adjoint theory in \eqref{PrimariesPureAdjoint}, as well as primaries constructed using the fundamental fermions: 
 \es{AdjointPrimaries}{
   &\mathfrak{n}=1:  \quad B^\dagger_{ji}(1) |0 \rangle \,,  \\
  &\left( D^\dagger_{j \beta }(1) C^\dagger_{\alpha i}(1)   -  \frac {\delta_{\alpha \beta}}{N} B^\dagger_{jk}(1) B^\dagger_{ki}(1)  -  \text{gauge traces} \right) |0 \rangle \,, \\
  &\left(D^\dagger_{j \beta }(1) [C^\dagger (1) B^\dagger(1)]_{\alpha i} 
    -  \frac {\delta_{\alpha \beta}}{2N} [ B^\dagger(1)  B^\dagger(1) B^\dagger(1) ]_{ji} - \frac {\delta_{\alpha \beta}}2  B^\dagger_{ji}(3)  -  \text{gauge traces} \right) |0 \rangle\,, \\
   &\left([B^\dagger(1) D^\dagger(1)]_{j \beta }(1) C^\dagger_{\alpha i} (1) 
    -  \frac {\delta_{\alpha \beta}}{2N} [ B^\dagger(1)  B^\dagger(1) B^\dagger(1) ]_{ji} + \frac {\delta_{\alpha \beta}}2  B^\dagger_{ji}(3)  -  \text{gauge traces} \right) |0 \rangle\,,  \\
      &\ldots
 }
As we can see, the state in the first line has $K=1$; the state on the second line has $K=2$; and the states on the third and fourth lines have $K=3$.  The pattern continues:  we can find $p$ KM primaries with $K = p+1$ that start with 
 \es{OtherMesonPrimaries}{
   \left(   [B^\dagger(1)^q D^\dagger(1)]_{j\beta} [C^\dagger(1) B^\dagger(1)^{p-1-q}]_{\alpha i} + \cdots \right) \ket{0} \,, \qquad
    q = 0, 1, \ldots, p-1 \,.
 }  
Thus, for every eigenstate of $P^-$ with some value of $K$ belonging to the block on the first line of \eqref{AdjointPrimaries}, we will have a whole family of states degenerate with it: a $K+1$ state from the block on the second line, two $K+2$ states from the blocks on the third and fourth lines, etc.  In general, at resolution parameter $K+p+1$ we will have $p$ eigenstates of $P^-$ degenerate with the eigenstate of $P^-$ belonging to the first block in \eqref{AdjointPrimaries} at resolution parameter $K$.

For $\mathfrak{n}=2$, we can construct KM primaries from the $\U(N_f)$ singlet $B^\dagger_{ji}(1)$ and the $\U(N_f)$ adjoint $D^\dagger_{j\beta}(1) C^\dagger_{\alpha i}(1)$, as well as factors such as the first term in \eqref{OtherMesonPrimaries}.  Unlike the $N_f = 0$ case, we can now obtain all $\SU(N)$ representations that  appear in the product of two adjoints, namely $\mathfrak{R}_2$, $\overline{\mathfrak{R}}_2$ and also $\mathfrak{R}_2'$ and $\mathfrak{R}_2''$---see Table~\ref{RepTable}.  While we leave a full analysis of these representations to future work, let us point out that the first couple $\mathfrak{n}=2$ primaries in $\mathfrak{R}_2 \oplus \overline{\mathfrak{R}}_2$ are
 \es{n2PrimariesMesons}{
   \mathfrak{n}=2& \text{ primaries in $\mathfrak{R}_2 \oplus \overline{\mathfrak{R}}_2$}:  \\
  & \left( B^\dagger_{ji}(1) B^\dagger_{lk}(1) -  \text{traces}  \right) |0 \rangle \,,  \\
  &  \left( B^\dagger_{ji}(1) D^\dagger_{l\beta}(1) C^\dagger_{\alpha k}(1) 
   + \frac{\delta_{\alpha \beta}}{N} B^\dagger_{ji}(1) B^\dagger_{lm}(1) B^\dagger_{mk}(1) -   \text{traces}  \right) |0 \rangle \,,  \\
   &\ldots
 }
where the first line is as in \eqref{PrimariesPureAdjoint}.  As in the $\mathfrak{n}=1$ sector, we see that for every $P^-$ eigenstate at $K$ coming from the first block we will have a $P^-$ eigenstate at $K+1$, and so on.

\subsection{Degeneracies at large $N$ for $N_f > 0$}
\label{sec:DEGENERACIES}

As discussed in Section~\ref{sec:GLUINO}, at large $N$ the states in each current block split into single-trace and multi-trace, and there are additional degeneracies between the single-trace and multi-trace ones.

Of the $\mathfrak{n}=0$ current blocks in \eqref{SingletPrimaries}, the block whose primary is the Fock vacuum $\ket{0}$ contains single-trace gluinoball states of the form \eqref{SingleTracen0} (as well as multi-trace gluinoball states) that are in one-to-one correspondence with and have the exact same $P^-$ spectrum as the $\mathfrak{n}=0$ sector of the pure adjoint theory $\cT_\text{adj}$ at leading order in $1/N$.  Each of the other $\mathfrak{n}=0$ blocks in \eqref{SingletPrimaries} whose KM primaries are gauge-invariant single-trace states such as $\ket{\zeta_{\alpha \beta, K}}$ or $\ket{\xi_{\alpha \beta, K}}$ contains only one single-trace state (namely the KM primary itself) as well as multi-trace states.

Of the $\mathfrak{n}=1$ current blocks in \eqref{AdjointPrimaries}, the block whose primary is $B^\dagger_{ji}(1) \ket{0}$ contains single-trace gluinoball states of the same form as on the LHS of \eqref{SingleTracen1} (as well as multi-trace gluinoball states) that are in one-to-one correspondence with and have the exact same $P^-$ spectrum as the $\mathfrak{n}=1$ sector of the pure adjoint theory.  The other $\mathfrak{n}=1$ blocks in \eqref{AdjointPrimaries}--\eqref{OtherMesonPrimaries} contain single-trace mesons with the exact same $P^-$ spectrum as the $\mathfrak{n}=1$ gluinoballs (at leading order in large $N$ we should only keep the first terms in \eqref{AdjointPrimaries}--\eqref{OtherMesonPrimaries} when constructing gauge-invariant states because the remaining terms are subleading), with a degeneracy pattern described after \eqref{OtherMesonPrimaries}:  for every single-trace fermionic $\mathfrak{n}=1$ gluinoball at $K$, we will have $p$ single-trace meson at $K+p$, with $p=1, 2, 3, \ldots$.  In Section~\ref{sec:symmetry} we will also ``explain'' this degeneracy pattern of single-trace mesons through the existence of an $\mathfrak{osp}(1|4)$ symmetry.

The structure of states in the $\mathfrak{n}=2$ sector is similar to that in the $\mathfrak{n}=1$ sector.  In particular, we have the block whose primary is $\left( B^\dagger_{ji}(1) B^\dagger_{lk}(1) -  \text{traces}  \right) |0 \rangle$, which contains single-trace gluinoball states with the same $P^-$ spectrum at large $N$ as the $\mathfrak{n}=2$ single-trace gluinoball of the pure adjoint theory.  For each such gluinoball state, there is then a tower of single-trace meson states, namely for every gluinoball state at $K$ there are $p$ single-trace meson states at $K+p$ that are degenerate with it.  For $p=1$, this state belongs to the block on the second line of \eqref{n2PrimariesMesons}.  We did not analyze the case $\mathfrak{n}>2$, but we expect a similar structure of single-trace states for these sectors too.

\section{Numerical results for gluinoball spectrum}\label{sec:glue_spectrum}

Let us now study numerically the spectrum of single-trace gluinoballs and mesons in the large $N$ limit.  The two decouple at leading order in large $N$, so in this section we will focus only on the single-trace gluinoballs.  As already mentioned, in the limit where $N_f$ is kept fixed while $N$ is taken to infinity, the spectrum of gluinoballs is exactly the same as in the $N_f = 0$ case.  The gluinoball spectrum was studied  using the DLCQ procedure in \cite{Bhanot:1993xp,Gross:1997mx}.  As was done in \cite{Bhanot:1993xp,Gross:1997mx}, working at fixed $K$ and considering the basis of gluinoball states of the form \eqref{GlueballStates}, the operators $M^2$ and $P^-$ can be written as finite-dimensional matrices that can be diagonalized numerically.  

\subsection{Massless adjoint fermion}

For $y_\text{adj} = 0$, Refs.~\cite{Bhanot:1993xp,Gross:1997mx} obtained the spectrum of $M^2$ up to $K = 25$. With modern computers, one can do much better.  We redid this analysis, and using the ``Scalable Library for Eigenvalue Problem Computations'' (SLEPc)  \cite{petsc-user-ref,petsc-efficient,slepc-toms,slepc-manual}, we are able to obtain the masses of the lowest-lying states up to $K = 41$, which represents more than a thousandfold increase in the number of states in the discretized basis. Table \ref{tab:gluinoball_degeneracies} gives the numbers of states at our highest values of $K$.

\begingroup
\renewcommand{\arraystretch}{1.2}
\begin{table}
	\centering
	\begin{tabular}{|c|c|c|c|c|c|c|c|}
	\hline
	$K$ & 35 & 36 & 37 & 38 & 39 & 40 & 41 \\
	\hline
	Dim & \num{5.9e5} & \num{9.3e5} & \num{1.5e6} & \num{2.3e6} & \num{3.6e6} & \num{5.7e6} & \num{9.0e6} \\
	\hline
	\end{tabular}
	\caption{The number of gluinoball states in the discretized basis at the highest values of $K$ we reach.}
	\label{tab:gluinoball_degeneracies}
\end{table}
\endgroup

\begin{figure}
    \centering
    \begin{subfigure}{.75\linewidth}%
	    \centering
    	\includegraphics[width=\linewidth]{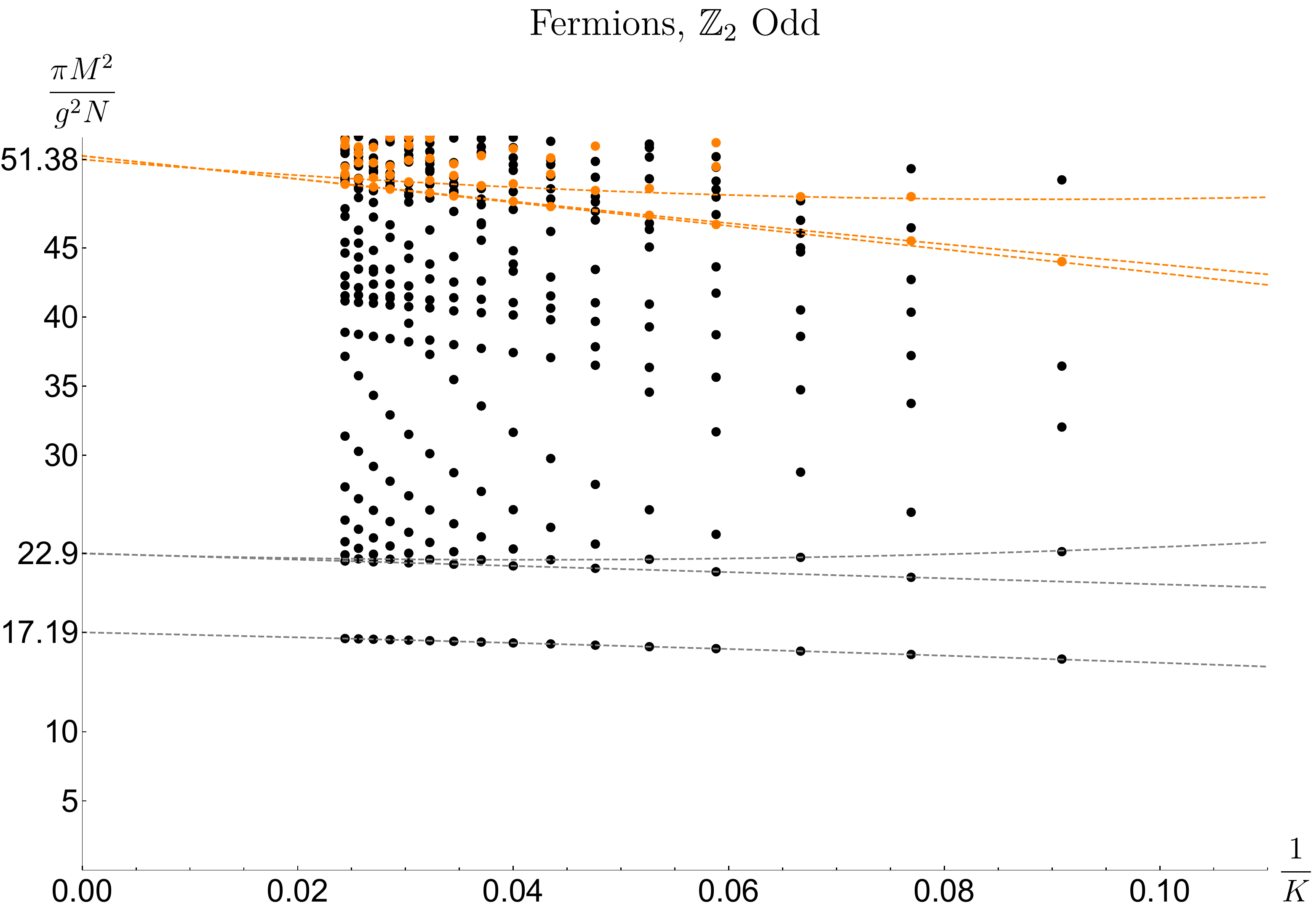}
    	\caption{The squared masses of fermionic gluinoball states with odd $\mathbb{Z}_2$ parity. A low-lying trajectory appears to converge towards $M_{F1}^2 \approx 17.2 g^2 N / \pi$.  }
    	\label{fig:glue_spectrum_fermi_odd}
    \end{subfigure}\\
    \vspace{2em}%
    \begin{subfigure}{.75\linewidth}%
	    \centering
    	\includegraphics[width=\linewidth]{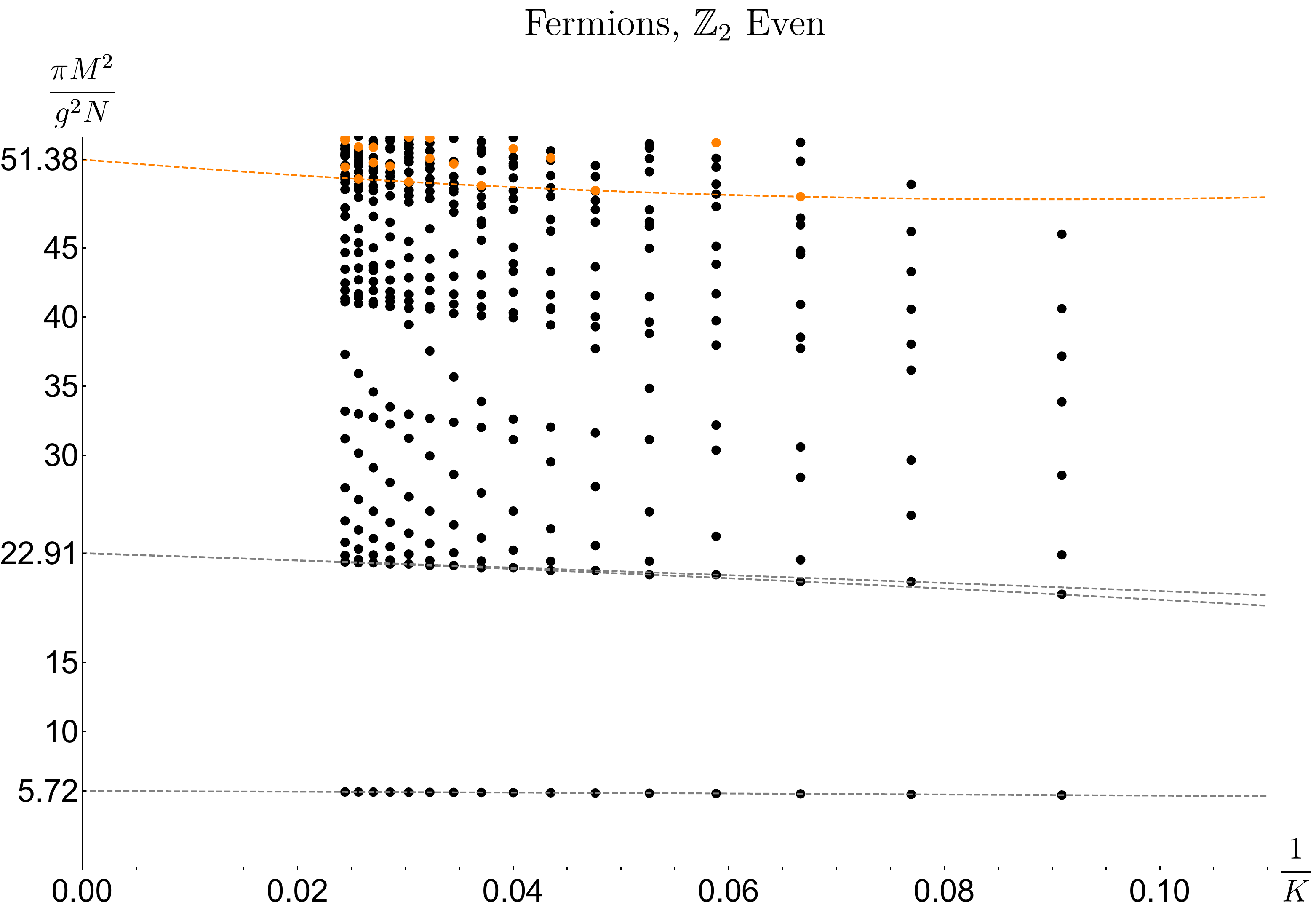}
    	\caption{The squared masses of fermionic gluinoball states with even $\mathbb{Z}_2$ parity. A low-lying trajectory appears to converge towards $M_{F0}^2 \approx 5.72 g^2 N / \pi$.  }
    	\label{fig:glue_spectrum_fermi_even}
    \end{subfigure}\\
    \phantomcaption
\end{figure}
\begin{figure}
	\centering
	\ContinuedFloat
    \begin{subfigure}{.75\linewidth}%
	    \centering
    	\includegraphics[width=\linewidth]{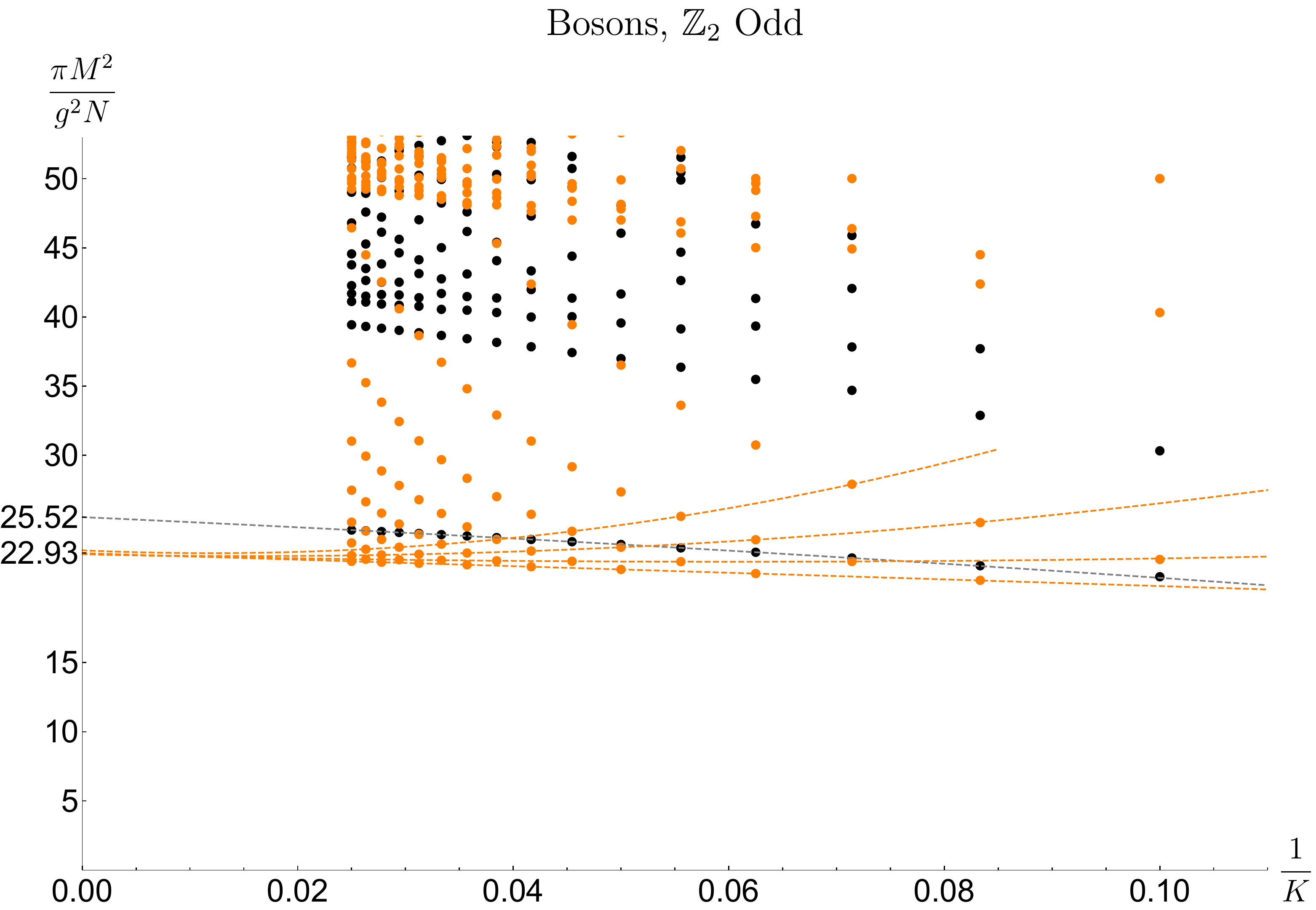}
    	\caption{The squared masses of bosonic gluinoball states with odd $\mathbb{Z}_2$ parity. The lowest single-particle state appears to converge towards $M_{B1}^2 \approx 25.5 g^2N/\pi$.  }
    	\label{fig:glue_spectrum_bose_odd}
    \end{subfigure}\\
    \vspace{2em}%
    \begin{subfigure}{.75\linewidth}%
	    \centering
    	\includegraphics[width=\linewidth]{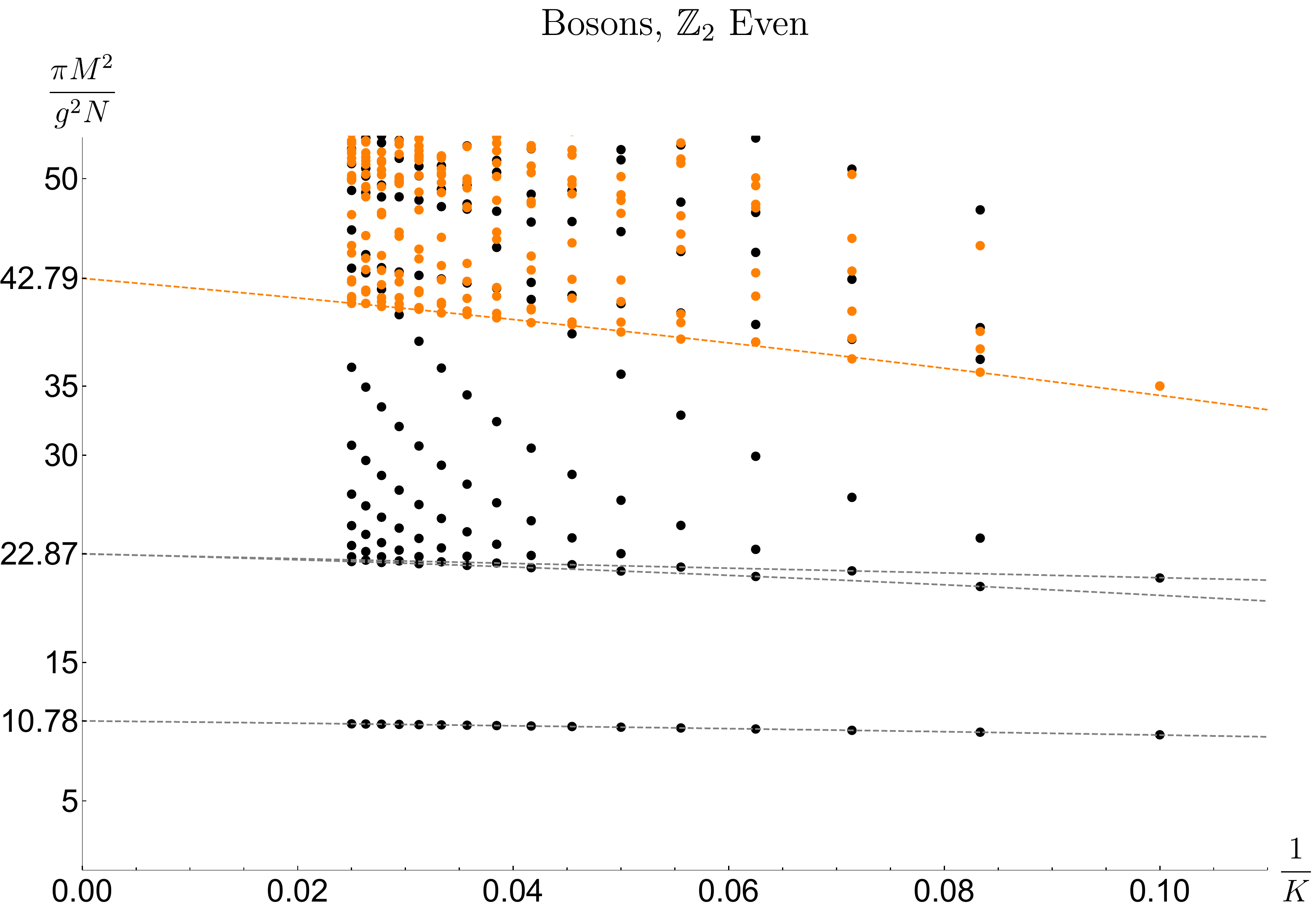}
    	\caption{The squared masses of bosonic gluinoball states with even $\mathbb{Z}_2$ parity. A low-lying trajectory appears to converge towards $M_{B0}^2 \approx 10.8 g^2 N / \pi$.  }
    	\label{fig:glue_spectrum_bose_even}
    \end{subfigure}%
    \caption{The masses of gluinoball states in DLCQ with $m_\text{adj}=0$ as a function of $1/K$, up to $K = 41$.  The spectrum was first described in \cite{Bhanot:1993xp,Gross:1995bp} up to $K=25$.  The orange points are single-trace gluinoball states that are exactly degenerate with multi-trace states.}
    \label{fig:glue_spectrum}
\end{figure}

In Figure~\ref{fig:glue_spectrum}, we show the masses of gluinoball states as a function of $1/K$. We split the states according to their statistics (states with odd $K$ are fermions, and those with even $K$ are bosons), and their quantum numbers with respect to the $\mathbb{Z}_2$ charge conjugation symmetry \eqref{eq:charge_conjugation}. For the fermionic states, we find that the first 
two ``single-particle states" converge to $M_{F0}^2 \approx 5.72 g^2 N/\pi$ and $M_{F1}^2 \approx 17.2 g^2 N/\pi$ in the continuum limit $K \to \infty$.  At higher mass-squared values, we see what appears to be the beginning of a continuum in the spectrum, as was noticed in \cite{Gross:1997mx,Trittmann:2000uj,Trittmann:2001dk,Trittmann:2015oka}. Extrapolating the lowest masses in this region, we find that they converge to around $M^2 \approx 22.9 g^2 N / \pi$ in both the $\Z_2$-even and odd sectors.  Among the bosonic states, we find the first single-particle state converging to $M_{B0}^2 \approx 10.8 g^2 N / \pi$ in the continuum limit, as well as evidence for a continuous spectrum starting around $M^2 \approx 22.9 g^2 N / \pi$ in both the $\Z_2$-even and odd sectors.  That the values of $M^2$ where the continuum starts in both sectors is very close to $4M_{F0}^2 \approx 22.9 g^2 N / \pi$ suggests that the continuum is composed of two-particle states \cite{Gross:1997mx}.

A striking feature of these plots are the exact relations stating that some of the eigenvalues of $P^-(K)$ at some given $K$ are sums of $P^-$ eigenvalues for fermionic states at $K_i$ with $K = \sum_i K_i$, i.e.~there are states obeying $P^-(K) = \sum_{i} P^-(K_i)$.  Such relations were explained in Section~\ref{sec:GLUINODEG} as arising at large $N$ in the $\mathfrak{n}>1$ sectors.  In terms of the masses, the relation $P^-(K) = \sum_{i=1}^{\mathfrak{n}} P^-(K_i)$ implies
 \es{MassSum}{
  \frac{M^2(K)}{K} = \sum_{i=1}^{\mathfrak{n}} \frac{M^2(K_i)}{K_i}  \,.
 }
 The states whose masses obey these exact relations are shown in orange in Figure~\ref{fig:glue_spectrum}.  In particular, the orange dots at the top of the plots in the fermionic sectors are $\mathfrak{n}=3$ states with $P^-$ written as a sum of three fermionic eigenvalues.  For the bosonic states in Figure~\ref{fig:glue_spectrum}, the orange dots visible in the plots correspond to $\mathfrak{n}=2$ states.  States with $\mathfrak{n}>3$ have masses that are larger  than the upper cutoffs of the plots.

While the threshold at $M^2 \approx 22.9 g^2 N / \pi$ in the bosonic $\Z_2$-odd sector is marked by states exactly degenerate with the double-trace states, we observe similar thresholds in the other three sectors. It appears that, in the continuum limit $K\rightarrow \infty$, every sector therefore exhibits threshold bound states. The reason why the other three sectors, for example, the fermionic ones, do not exhibit exact thresholds is related to the fact that a pair of fermionic bound states cannot simply bind to form another fermionic state. To make the threshold single-trace state a fermion, it needs to contain another fermionic insertion, such as $B^\dagger_{ij} (1)$ (since we are using the anti-periodic boundary conditions, this is the lowest frequency mode). This insertion would explain the breaking of exact degeneracy, but would have negligible effect in the limit $K\rightarrow \infty$. As noted in \cite{Trittmann:2001dk}, this provides an argument for the existence of the approximate thresholds when the DLCQ boundary conditions are anti-periodic.

\begin{figure}
    \centering
    \begin{subfigure}{.8\linewidth}%
	    \centering
    	\includegraphics[width=\linewidth]{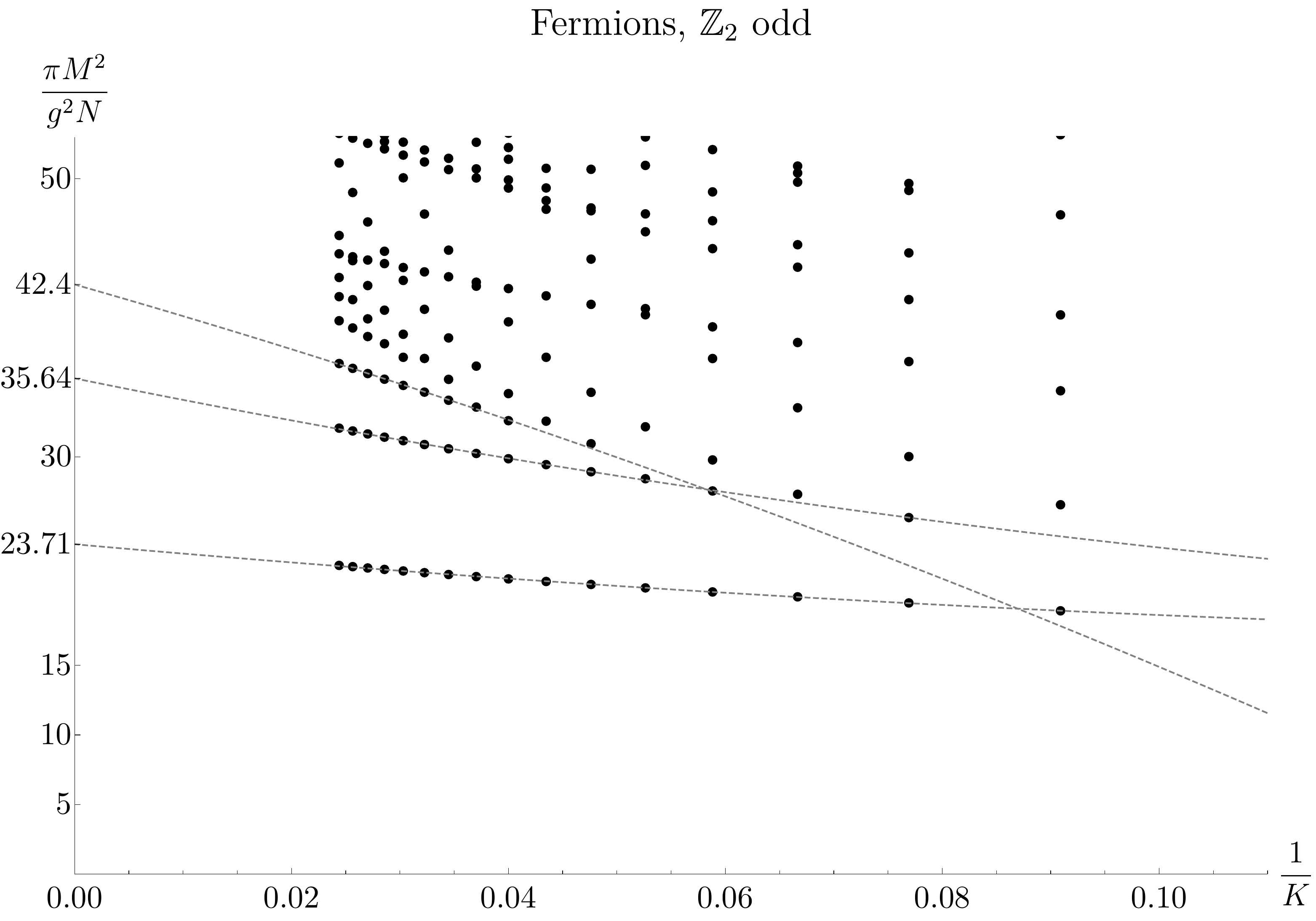}
    	\caption{}
    	\label{fig:glue_spectrum_massive_fermi_even}
    \end{subfigure}\\
    \vspace{2em}%
    \begin{subfigure}{.8\linewidth}%
	    \centering
    	\includegraphics[width=\linewidth]{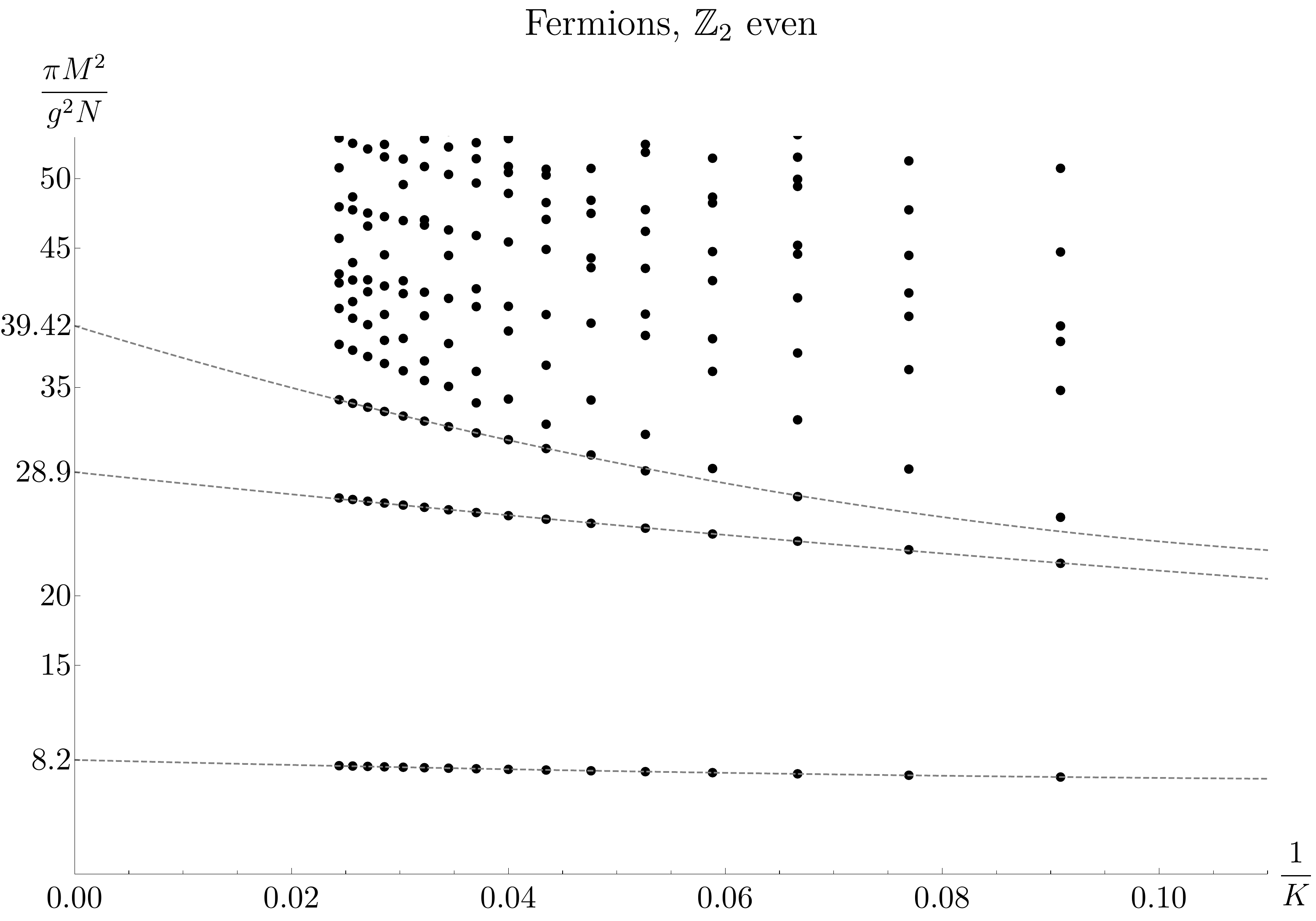}
    	\caption{}
    	\label{fig:glue_spectrum_massive_fermi_odd}
    \end{subfigure}\\
    \phantomcaption
\end{figure}
\begin{figure}
	\centering
	\ContinuedFloat
    \begin{subfigure}{.8\linewidth}%
	    \centering
    	\includegraphics[width=\linewidth]{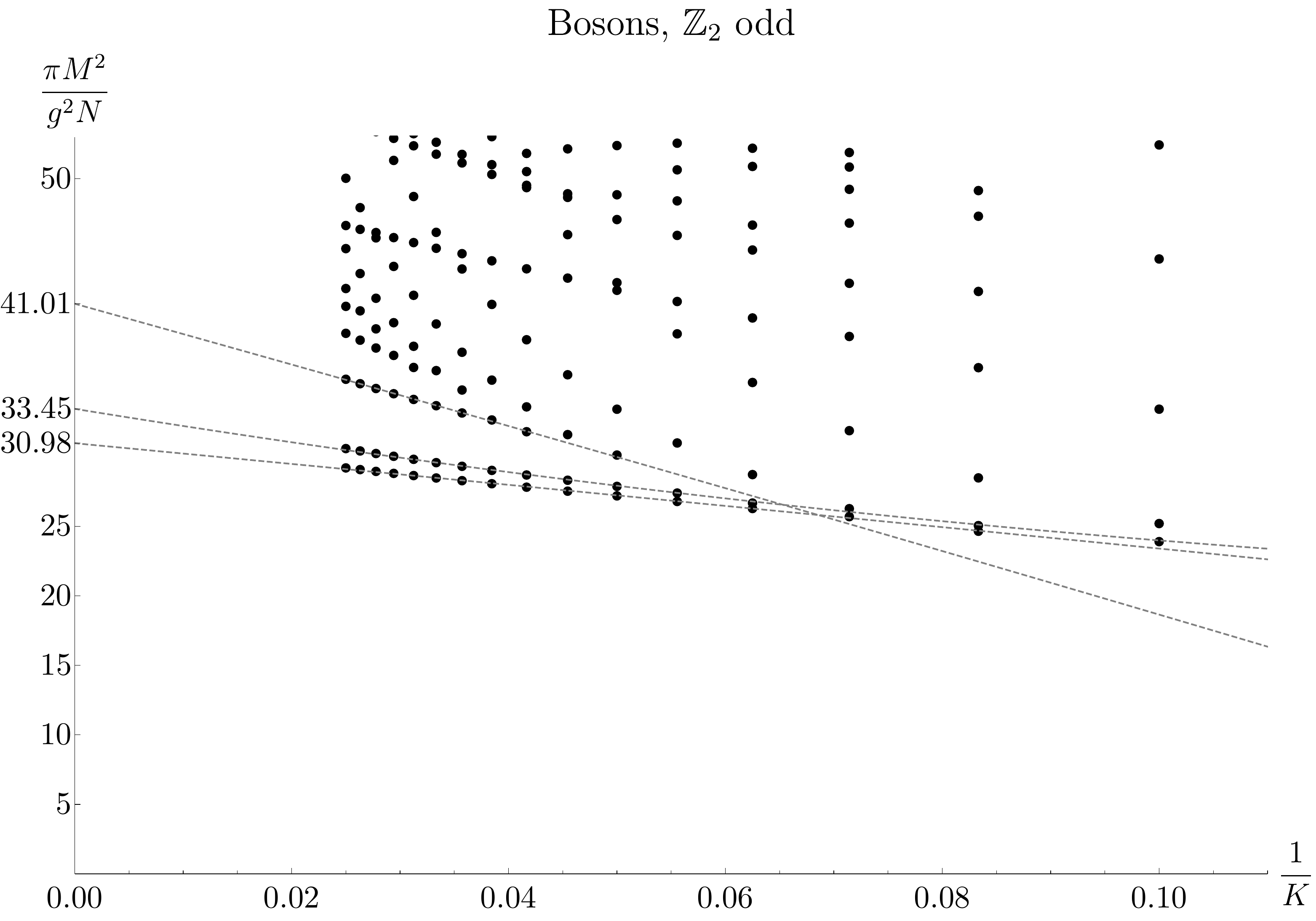}
    	\caption{}
    	\label{fig:glue_spectrum_massive_bose_even}
    \end{subfigure}\\
    \vspace{2em}%
    \begin{subfigure}{.8\linewidth}%
	    \centering
    	\includegraphics[width=\linewidth]{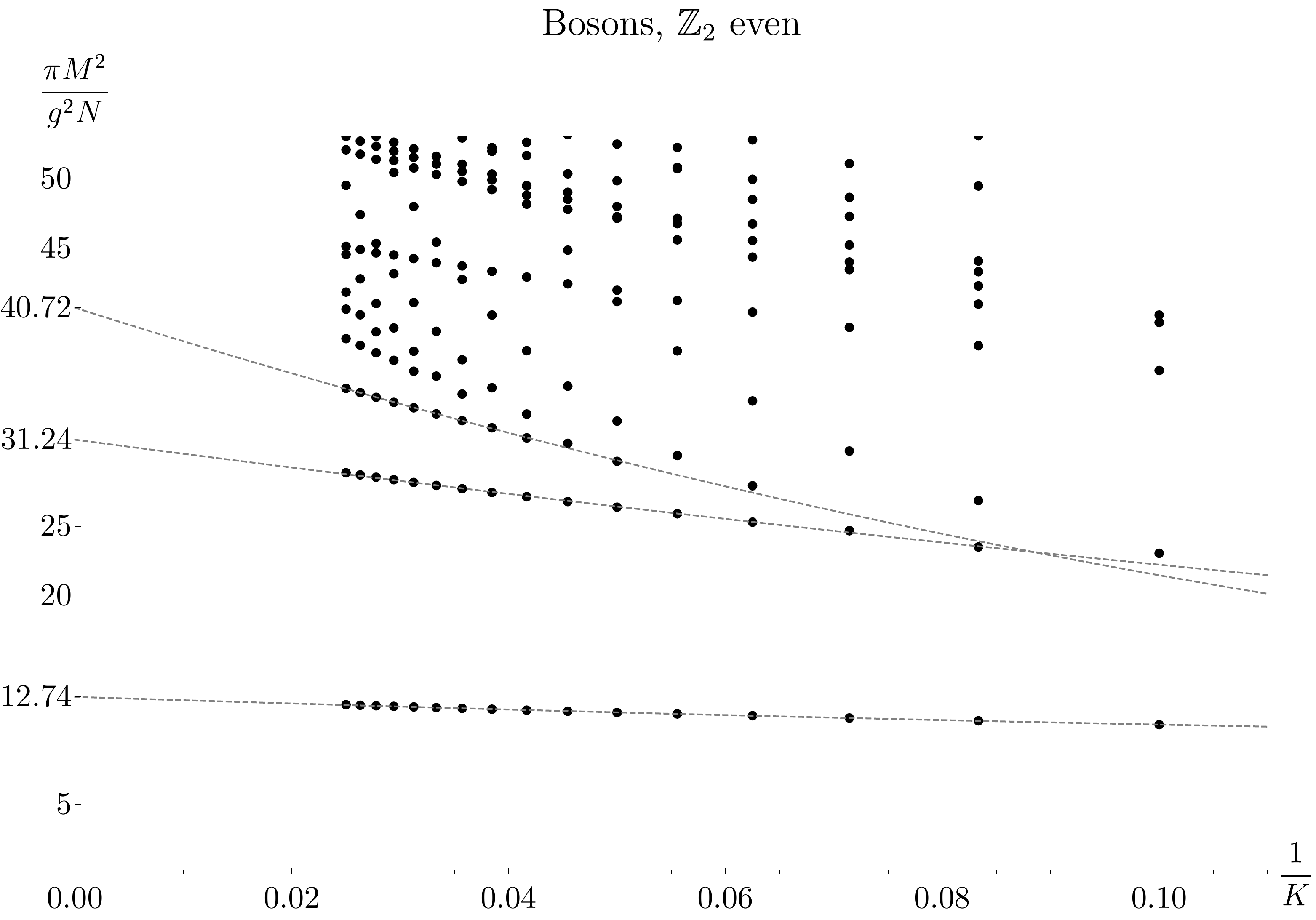}
    	\caption{}
    	\label{fig:glue_spectrum_massive_bose_odd}
    \end{subfigure}%
    \caption{The masses of gluinoball states in DLCQ as a function of $1/K$, up to $K = 41$, with the adjoint mass parameter $y_\text{adj} = \frac{m^2_\text{adj} \pi}{g^2 N} = 0.1$.}
    \label{fig:glue_spectrum_massive}
\end{figure}

Note that in the continuum limit $K \to \infty$, the relation \eqref{MassSum} implies that for any group of $\mathfrak{n}$ trajectories whose masses asymptote to $\{ M_{i} \}_{1 \leq i \leq \mathfrak{n}} $ as $K \to \infty$, we will have a continuum of states with masses starting at 
 \es{MCont}{
  M_\text{threshold} = \sum_{i = 1}^\mathfrak{n} M_{i} \,.
 }
Indeed, at large enough $K$, we can approximate $M(K_i) \approx M_i$, and we can effectively vary continuously the momentum fractions $x_i \equiv K_i / K$.  Thus, \eqref{MassSum} becomes, at large $K$, 
 \es{MSQApprox}{
  M^2 \approx \sum_{i=1}^\mathfrak{n} \frac{M_i^2}{x_i} \,, \qquad \sum_{i=1}^\mathfrak{n} x_i = 1 \,.
 }
We see from \eqref{MSQApprox} that $M^2$ can be made arbitrarily large by taking (at least) one of the $x_i$ to be small.  The lowest value of $M^2$ is attained when $x_i = M_i / \sum_{j=1}^\mathfrak{n} M_j$ and it is given by $M_\text{threshold}^2$.  Since the $x_i$ can be varied continuously, we thus find a continuum of states with $M \geq M_\text{threshold}$ as $K \to \infty$.  For any finite mass window $\Delta M^2$ above the threshold, the number of states that approximate the continuum grows as $K^{\mathfrak{n}-1}$ at large but finite $K$.  At finite $K$, we will in general not find states with masses given by $M_\text{threshold}$, partly because $M(K_i)$ varies slightly with $K$ for single particle states, and partly because we will not find rational $x_i = K_i / K$ that come arbitrarily close to the optimal value $x_i = M_i / \sum_{j=1}^\mathfrak{n} M_j$.

As we will see shortly, the gluinoball states in Figure~\ref{fig:glue_spectrum} are related to the spectrum of single-trace meson states.

\subsection{Massive adjoint fermion}

For a fixed $m_\text{adj}>0$, we expect the continua in the gluinoball spectrum at large $K$ to disappear and be replaced by a discrete spectrum. Since we have diagonalized the discretized $P^-$ matrix up to higher values of $K$ than was achieved in previous work, we present here some additional evidence for this phenomenon.

Figure~\ref{fig:glue_spectrum_massive} shows the squared masses of DLCQ states up to $K = 41$ with the adjoint mass parameter fixed at $y_\text{adj} = 0.1$. Indeed, the apparent continua in Figure \ref{fig:glue_spectrum} are absent here. In each sector, we extrapolate to estimate the masses of the lowest few states in the continuum.

\section{Numerical results for meson spectrum}
\label{sec:MESON}

Let us now proceed to a similar analysis for the single-trace meson spectrum.  In the theory with $N_f$ fundamental quarks, single-trace meson states transform in the adjoint representation of the $\U(N_f)$ flavor symmetry and thus have degeneracy $N_f^2$.   At leading order in large $N$ with fixed $N_f$, the meson masses are independent of $N_f$.  Thus, in the numerical computations that follow we set $N_f = 1$ without loss of generality.

In this section we restrict to the case $y_\text{adj} = y_\text{fund} =0$ where the fermions are massless; we examine the case where the fundamental fermions are massive in Section~\ref{sec:MASSIVE}.   As in the gluinoball case, to find the single-trace meson spectrum we can again write $M^2$ and $P^-$ as finite-dimensional matrices if we work in the basis of states \eqref{MesonStates} at fixed $K$.  The basis states \eqref{MesonStates} are in one-to-one correspondence with the ordered partitions of the integer $K$ into any number of odd integers.  It can be shown that the number of meson states at given $K$ equals $F_K$ if $K$ is even and $F_K - 1$ if $K$ is odd, where $\{F_n\}_{n=1}^\infty$ is the Fibonacci sequence with $F_1 = F_2 = 1$---see Table~\ref{MesonicStates} for the first few examples.  As written, the states in \eqref{MesonStates} are orthonormal at leading order in large $N$. 
\begin{table}[htp]
\begin{center}
\begin{tabular}{c|c}
 $K$ & \# of mesonic states  \\
 \hline 
 $2$ & $1$ \\
 $3$ & $1$ \\
 $4$ & $3$ \\
 $5$ & $4$ \\
 $6$ & $8$ \\
 $7$ & $12$ \\
 $8$ & $21$ \\
\end{tabular}
\end{center}
\caption{The number of single-trace mesonic states as a function of $K$.} \label{MesonicStates}
\end{table}%

\subsection{Massless meson states}

Diagonalizing $M^2$ and $P^-$ numerically as in the gluinoball case, we find that for $y_\text{fund} = y_\text{adj} = 0$, the spectra of the discretized $M^2$ and $P^-$ operators are highly degenerate.  First, unlike in the gluinoball case, we now find meson states with $M^2=0$ whose number grows with $K$, leading to an infinite number of such states in the continuum limit.  The first few massless states were presented around Eq.~\eqref{SingletPrimaries}, where it was pointed out that they are necessarily the Kac-Moody primaries of their respective current blocks.  

The presence of an infinite number of massless states in the continuum limit is not unexpected, because in the deep IR, our theory is governed by a non-trivial CFT\@.  To support this conclusion analytically, let us calculate the central charge of the gauged WZW model which describes the far infrared limit of the theory.   Since the free UV theory has $N^2-1+ 2N N_f$ Majorana fermion fields, it may be bosonized \cite{Witten:1983ar} into the $\SO(N^2-1+2N N_f)_1$ WZW model with central charge $c_{\rm UV}= NN_f+ \frac{1}{2}(N^2-1)$. As explained in \eqref{CommutJJ}, the $\SU(N)$ currents satisfy the Kac-Moody algebra at level $k_\text{KM} = N+N_f$. Therefore, the IR limit of the gauge theory is described by the
\begin{equation}
{  \SO(N^2-1+2N N_f)_1\over \SU(N)_{N+ N_f} }
\label{coset}
\end{equation}
coset model. The central charge of the $\SU(N)_k$ WZW model is well-known \cite{Knizhnik:1984nr} to be $(N^2-1)\frac{k}{N+k}$. Therefore, we find that the coset model has central charge
\begin{equation}
c_{\rm IR}= c_{\rm UV}- (N^2-1)\frac{N+N_f}{2N+N_f}= 
\frac{N_f( 3N^2 +2N_f N +1)}{2(2N+N_f)}\ .
\label{IRcentral}
\end{equation}
For large $N$ this grows as $3N N_f/4$ and explains the proliferation of massless meson bound states we observe in DLCQ\@. We should emphasize again that the DLCQ description of the massless bound states is not complete, since for $y_\text{fund} = y_\text{adj} = 0$ there is no justification for discarding the components of the fermions moving along $x^-$. The precise description of the far IR limit of the theory is provided by the coset model \eqref{coset}, but we will not discuss it further in this paper.

\subsection{Massive meson states}

\begin{figure}
    \centering
    \includegraphics[width=.9\linewidth]{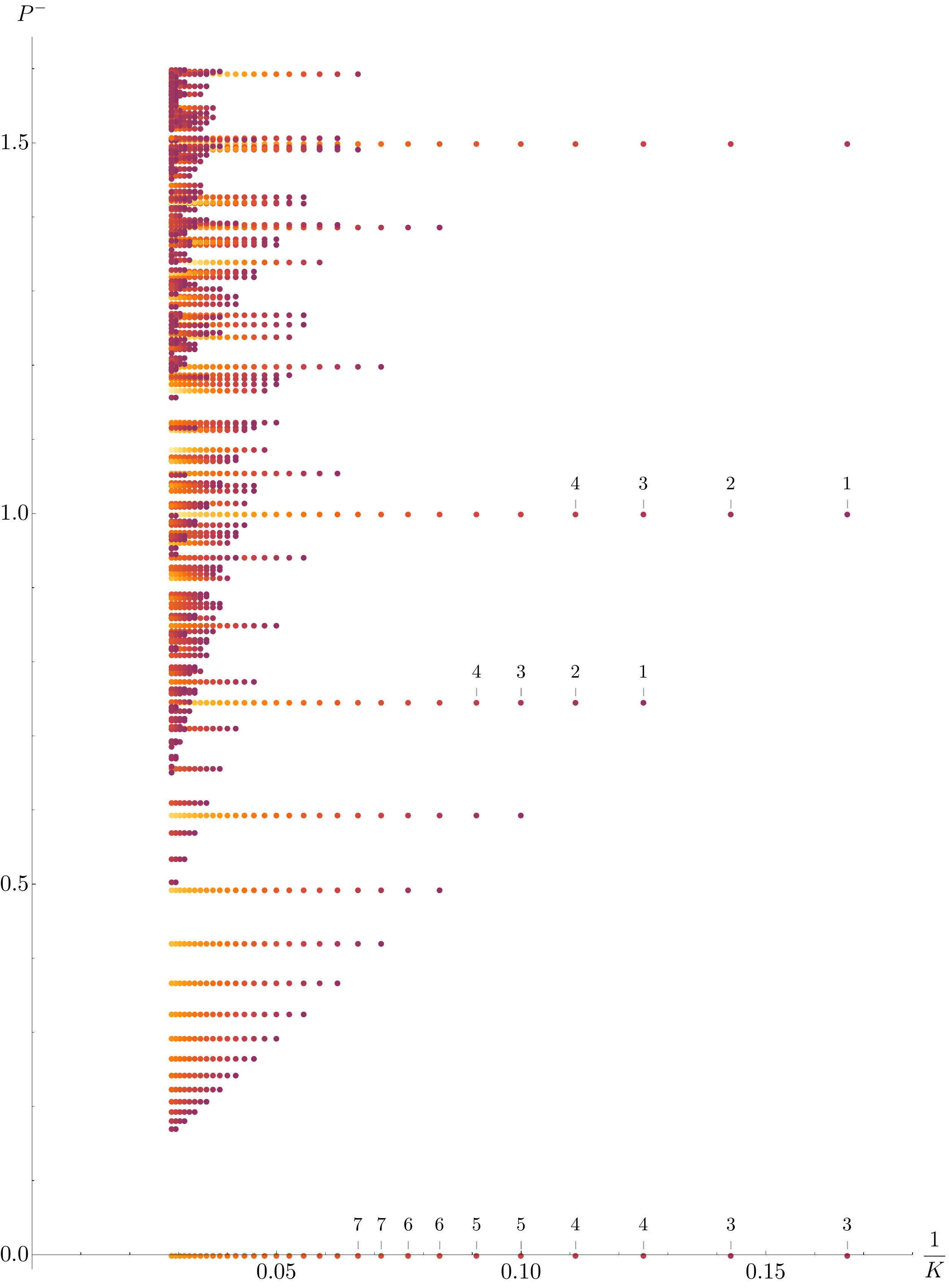}
    \caption{The eigenvalues $P^-$ up to $K = 35$. States are colored according to their degeneracies, with the darkest states being singlets. Along the horizontal trajectories, the degeneracies increase in steps of 1 moving from right to left, except for the series of massless states, which have degeneracy $\lfloor K/2\rfloor$.}
    \label{fig:pminus_degeneracies}
\end{figure}

For the massive states, the spectrum of $P^-$ has degeneracies that are even more striking than in the gluinoball case. The eigenvalues of $P^-$ for mesonic states are shown up to $K=35$ in Figure~\ref{fig:pminus_degeneracies}, where at each value of $K$ we also colored the eigenvalues according to their degeneracies.   We notice that all the $P^-$ eigenvalues we encounter at some $K$ are also $P^-$ eigenvalues at $K+1$.  
More precisely, we find that an eigenvalue $\lambda > 0$ which has degeneracy $n$ at $K$ will have degeneracy at least $n+1$ at $K+1$ (and usually exactly $n+1$, except in rare cases to be described later).  Moreover, if the eigenvalue $\lambda$ first appears in the spectrum at $K = K_\lambda$ (usually with degeneracy $1$ but sometimes with higher degeneracy), we find that, in most cases, at $K = K_\lambda - 1$, $\lambda$ is a $P^-$ eigenvalue in the single-trace gluinoball spectrum.  

Let us first comment on the pattern of degeneracies between the mesonic states, and discuss the degeneracy between mesons at $K_\lambda$ and gluinoballs at $K_\lambda-1$ in Section~\ref{sec:glueball_degeneracies}.  We have two explanations for the degeneracies between the massive mesons.  The first was already provided in Section~\ref{KMA} and it relies on the fact that the $P^-$ spectrum agrees for any two (or more) KM blocks whose KM primaries transform in the same $\SU(N)$ representation.  Indeed, for the $\mathfrak{n}=1$ blocks, we found in \eqref{AdjointPrimaries}--\eqref{OtherMesonPrimaries} a number of KM blocks that increases with $K$ in precisely the same way as the pattern of degeneracies we observed above.  While we have not constructed all the Kac-Moody primaries explicitly, we expect that a similar explanation would hold for $\mathfrak{n}>1$.  This construction would provide a complete explanation of the degeneracies in the single-trace massive meson spectrum at large $N$, because all massive states belong to $\mathfrak{n} \geq 1$ blocks.

In Section~\ref{sec:symmetry}, we will provide a different explanation that applies to all single-trace mesonic states.  This explanation relies on the existence of a $\mathfrak{osp}(1|4)$ symmetry enjoyed by the discretized theory at large $N$.   In particular, the massive spectrum splits into infinite-dimensional unitary irreducible representations of $\mathfrak{osp}(1|4)$.  As we explain in more detail in Section~\ref{sec:symmetry}, for an eigenvalue $\lambda$, the meson states at $K = K_\lambda$ where that eigenvalue first appears are referred to as ``$\mathfrak{osp}(1|4)$ primary states,'' whereas the states with the same $\lambda$ at $K > K_\lambda$ are $\mathfrak{osp}(1|4)$ descendants.

\begin{figure}
    \centering
    \begin{subfigure}{\linewidth}
    	\includegraphics[width=\linewidth]{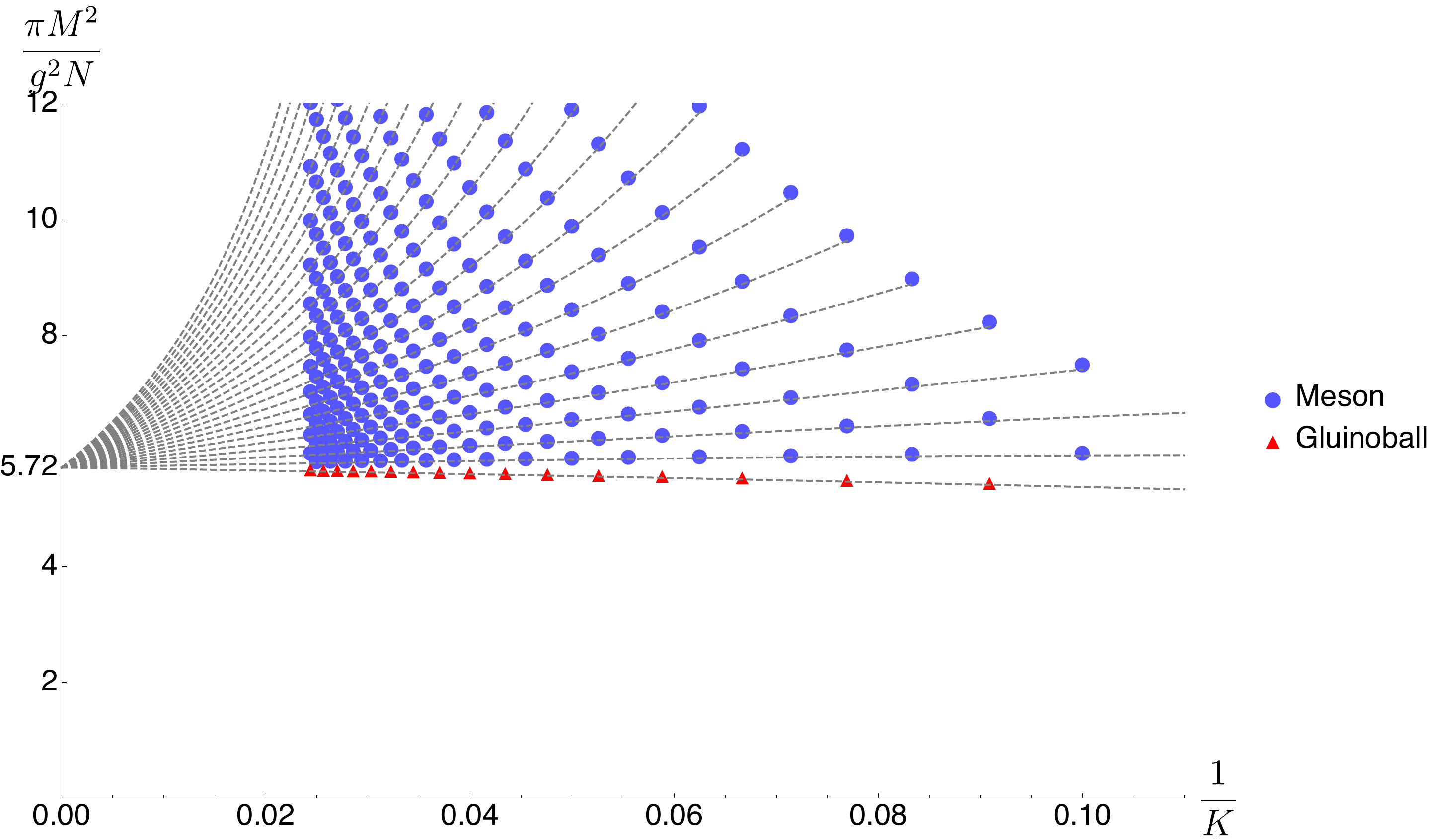}
    	\caption{The states approaching $M_0^2 \approx 5.72 g^2N / \pi$.}
	    \label{fig:m2_convergence_1}
    \end{subfigure}\\
    \begin{subfigure}{\linewidth}
    	\includegraphics[width=\linewidth]{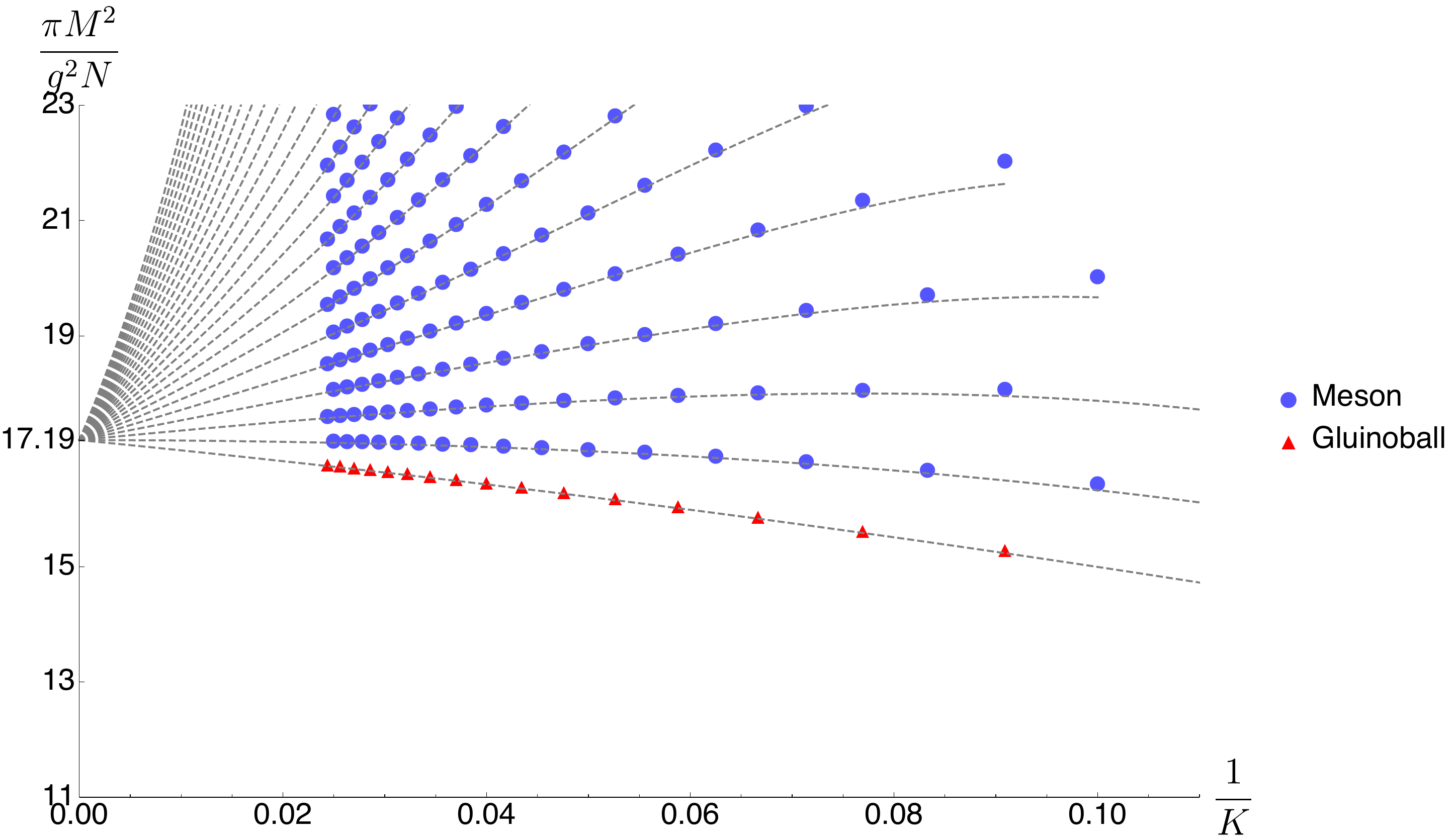}
    	\caption{The states approaching $M_1^2 \approx 17.2 g^2 N / \pi$.}
	    \label{fig:m2_convergence_2}
    \end{subfigure}
    \caption{The masses of the states in Figure \ref{fig:pminus_degeneracies}, along with the gluinoball states described in Section \ref{sec:glueball_degeneracies}, and the trajectories along which they approach their $K\to\infty$ values.}
    \label{fig:m2_convergence}
\end{figure}

From the $P^-$ eigenvalues, we can immediately recover the $M^2 = 2P^+ P^-$ eigenvalues.  The latter organize into different groups of multiple trajectories of consecutive degeneracies that converge to the same $M^2$ as $K \to \infty$, as a consequence of the degeneracy of the $P^-$ eigenvalues at different values of $K$.  Indeed, if $M^2_0(K)$ are the squared masses of a trajectory of singlet $M^2$ eigenvalues (arising from distinct $P^-$ eigenvalues at each $K$), then there exist trajectories of degeneracy $2j+1$ with $j= 0, \frac 12, 1, \frac 32, \cdots$.  As a consequence of the $\mathfrak{osp}(1|4)$ symmetry, the squared masses on the trajectory with degeneracy $2j+1$ are related to those in the singlet trajectory through $M^2_j(K) = \frac{K}{K-2j}M^2_0(K-2j)$.  This relation implies that as we take $K \to \infty$, we find that all trajectories converge to the same $M^2$:
\begin{equation}\label{eq:continuum_degeneracy}
    \lim_{K\to\infty} M^2_j(K) = \lim_{K\to\infty} M^2_0(K) \,.
\end{equation}
We show the lowest two groups of such trajectories in Figure~\ref{fig:m2_convergence}.  The lowest one converges to $M_0^2 \approx 5.72 g^2N / \pi$ as $K \to \infty$ and the next one converges to $M_1^2 \approx 17.2 g^2 N / \pi$.  These are precisely the lowest two squared masses in the fermionic gluinoball spectrum shown in the top panel of Figure~\ref{fig:glue_spectrum}, which is a consequence of the relation between single-trace meson eigenvalues and gluinoball ones mentioned above.

\subsection{Relations between meson and gluinoball spectra}\label{sec:glueball_degeneracies}

Let us now discuss the relation between the $P^-$ eigenvalues for mesons and gluinoballs in more detail.  As already mentioned, a given eigenvalue $\lambda$ for a $\mathfrak{osp}(1|4)$ primary meson at $K_\lambda$ also appears, in most cases, in the gluinoball spectrum at $K_\lambda-1$.  The converse is, however, not true:  there are (many) bosonic gluinoball eigenvalues that do not appear in the fermionic meson spectrum at larger $K$.  In particular, looking at the lowest 50 $P^-$ eigenvalues of meson $\mathfrak{osp}(1|4)$ primary states up to $K = 35$, we find the following pattern of degeneracies between mesons at $K$ and gluinoballs at $K-1$:
\begin{itemize}
	\item The set of $P^-$ eigenvalues of single-trace {\em bosonic} primary mesons at any fixed even $K$ is identical to the set of $P^-$ eigenvalues of single-trace {\em fermionic} gluinoballs at $K-1$.
	\item The set of $P^-$ eigenvalues of single-trace {\em fermionic} primary mesons at any fixed odd $K$ is identical to the set of sums of two single-trace {\em fermionic} gluinoball eigenvalues at odd $K_1$ and odd $K_2$, with $K-1 = K_1 + K_2$.  In most cases, the sum of fermionic gluinoball eigenvalues at $K_1$ and $K_2$ is also a bosonic gluinoball eigenvalue at $K_1 + K_2$.
\end{itemize}

For a concrete example of the pattern in the first bullet point, we refer to Table \ref{tab:meson8_gluinoball7}. Here, we give all the eigenvalues of $P^-$ for mesons at $K = 8$ and for gluinoballs at $K = 7$. All of the eigenvalues at $K = 8$ are either $\mathfrak{osp}(1|4)$ descendant states  of $\mathfrak{osp}(1|4)$ primary mesons at lower $K$, or are $\mathfrak{osp}(1|4)$ primary states and occur in the gluinoball spectrum at $K = 7$.
\begingroup
\renewcommand{\arraystretch}{1.5}
\begin{table}
\begin{tabularx}{\linewidth}{YY|YY}
	\hline
	\multicolumn{2}{c|}{Mesons, $K = 8$} & \multicolumn{2}{c}{Gluinoballs, $K = 7$} \\
	$P^-$ & Degeneracy & $P^-$ & Degeneracy \\
	\hline
	\rowcolor{LightCyan}
	$\frac{7}{2}$ & 1 & $\frac{7}{2}$ & 1 \\
	\rowcolor{Gray}
	3 & 2 & ~ & ~ \\
	\rowcolor{LightCyan}
	$\frac{1}{36}\left(59+\sqrt{1033}\right)$ & 1 & $\frac{1}{36}\left(59+\sqrt{1033}\right)$ & 1 \\
	\rowcolor{Gray}
	$\frac{5}{2}$ & 3 & ~ & ~ \\
	\rowcolor{LightCyan}
	2 & 1 & 2 & 1 \\
	\rowcolor{Gray}
	$\frac{3}{2}$ & 5 & ~ & ~ \\
	\rowcolor{Gray}
	1 & 3 & ~ & ~ \\
	\rowcolor{LightCyan}
	$\frac{1}{36}\left(59-\sqrt{1033}\right)$ & 1 & $\frac{1}{36}\left(59-\sqrt{1033}\right)$ & 1 \\
	\rowcolor{Gray}
	0 & 4 & ~ & ~ \\
	\hline
\end{tabularx}
\caption{All of the eigenvalues of $P^-$ in the meson sector at $K = 8$ and in the gluinoball sector at $K = 7$. Cyan rows are meson $\mathfrak{osp}(1|4)$ primary states, and gray rows are descendant states. The set of primary state eigenvalues of mesons at $K = 8$ is identical to the set of eigenvalues of gluinoballs at $K = 7$.}
\label{tab:meson8_gluinoball7}
\end{table}
\endgroup
For a concrete example of the pattern in the second bullet point above, see Table~\ref{tab:meson11_gluinoball10}.   Here, we first tabulate the $P^-$ eigenvalues of $\mathfrak{osp}(1|4)$ primary mesons at $K=11$.  We then write each of these eigenvalues as sums of gluinoball eigenvalues at $(K_1, K_2) = (3, 7)$ and $(K_1, K_2) = (5, 5)$.
\begingroup
\renewcommand{\arraystretch}{1.5}
\begin{table}
\centering
\begin{tabular}{C{0.2\linewidth}C{0.2\linewidth}|C{0.25\linewidth}|C{0.25\linewidth}}
	\hline
	\multicolumn{2}{c|}{$\mathfrak{osp}(1|4)$ primary mesons, $K = 11$} & Gluinoballs at $(K_1, K_2) = (3, 7)$ & Gluinoballs at $(K_1, K_2) = (5, 5)$ \\
	$P^-$ & Degeneracy & $P^-$ & $P^-$ \\
	\hline
	5 & 3 & $\frac 32 + \frac 72$ & $\frac 52 + \frac 52$ \\
	$\frac{1}{36}\left(113 + \sqrt{1033}\right)$ & 2 & $ \frac 32 + \frac{1}{36}\left(59 + \sqrt{1033}\right)$ & ~ \\
	$\frac{7}{2}$ & 4 & $\frac 32 + 2$ & $1 + \frac 52$ \\
	$\frac{1}{36}\left(113 - \sqrt{1033}\right)$ & 2 & $ \frac 32 + \frac{1}{36}\left(59 - \sqrt{1033}\right)$ & ~ \\
	2 & 1 & ~ & $1+ 1$ \\
	\hline	
\end{tabular}
\caption{The $P^-$ eigenvalues of $\mathfrak{osp}(1|4)$ primary meson states at $K = 11$.  They agree with sums of eigenvalues of fermionic gluinoballs at $K_1$ and $K_2$ with $K_1 + K_2 = K-1 = 10$.  In particular, the gluinoball eigenvalues are:  $\left\{ \frac 32 \right\} $ at $K=3$, $\left\{ 1, \frac 52 \right\} $ at $K=5$, and $\left\{\frac{1}{36}\left(59 - \sqrt{1033}\right), 2, \frac{1}{36}\left(59 + \sqrt{1033}\right) , \frac 72 \right\} $ at $K=7$.}
\label{tab:meson11_gluinoball10}
\end{table}
\endgroup

These two facts in the bullet points above can be explained in a straightforward way in the Kac-Moody approach of Section~\ref{KMA}.  In particular, the first bullet point is explained by the fact that for every odd $\mathfrak{n}$ Kac-Moody block that contains single-trace gluinoball states (for instance the first line of \eqref{AdjointPrimaries} in the $\mathfrak{n}=1$ case), one can also construct Kac-Moody blocks in the same representation of $\SU(N)$ that contain single-trace mesons and that have one extra unit of $K$ relative to the gluinoball blocks (for instance the second of \eqref{AdjointPrimaries} in the $\mathfrak{n}=1$ case).   Similarly, the second bullet point is explained by the fact that for every even $\mathfrak{n} \geq 2$ Kac-Moody block that contains single-trace gluinoball states (for instance the first line of \eqref{n2PrimariesMesons} in the $\mathfrak{n}=2$ case), one can also construct Kac-Moody blocks in the same representation of $\SU(N)$ that contain single-trace mesons and that have one extra unit of $K$ relative to the gluinoball blocks (for instance the second of \eqref{n2PrimariesMesons} in the $\mathfrak{n}=2$ case).

The degeneracies between mesons and gluinoballs at finite $K$ give rise to degeneracies between the continuum spectra of mesons and gluinoballs. For instance, at every odd value of $K$ there is a state in the gluinoball spectrum on a trajectory approaching $M_0^2 \approx 5.72 \frac{g^2 N}{\pi}$. Each of these fermionic gluinoballs corresponds to a bosonic meson at one higher $K$, with the same eigenvalue of $P^-$. Their mass-squared values are thus related by
\begin{equation}
	\frac{M^2_\text{meson}(K+1)}{K+1} = \frac{M^2_\text{gluinoball}(K)}{K},
\end{equation}
It follows that there is a series of meson primary states also approaching $M_0^2\approx 5.72 \frac{g^2 N}{\pi}$. These states form the lowest meson trajectory in Figure \ref{fig:m2_convergence_1}. The other meson trajectories approaching the same value are explained by the pattern of degeneracies among meson states which we have already described.

\section{Exact symmetry at large $N$}\label{sec:symmetry}

Let us now provide another explanation for the degeneracy in the $P^-$ spectrum observed in Figure~\ref{fig:pminus_degeneracies}.  As already mentioned, this degeneracy can be traced to the existence of an $\mathfrak{osp}(1|4)$ symmetry algebra that commutes with $P^-$.

\subsection{$\mathfrak{osp}(1|4)$ algebra}

The $\mathfrak{osp}(1|4)$ generators can be constructed from the four basic ``supercharges'' of the discretized model\footnote{The quantities denoted by $q$ in this section are unrelated to the fundamental quarks $q_{i \alpha}$ appearing in \eqref{STotal}.}
 \es{qLRDef}{
    q^L_{\pm} &=  \frac{1}{\sqrt{2N}}  \sum_{n_1,n_2} \left(C^\dagger_j(n_1+n_2\pm 1) B_{ij}(n_1) C_i(n_2) + C^\dagger_j(n_1) B^\dagger_{ji}(n_2) C_i(n_1+n_2\mp 1)\right),\\
    q^R_{\pm} &=  \frac{1}{\sqrt{2N}}  \sum_{n_1,n_2} \left(D^\dagger_i(n_1+n_2\pm 1) B_{ij}(n_1) D_j(n_2) + D^\dagger_i(n_1) B^\dagger_{ji}(n_2) D_j(n_1+n_2\mp 1)\right) \,.
 }
The first two supercharges, $q^L_{\pm}$, act on the left end of the mesonic string, with $q^L_+$ raising the value of $K$ by one unit and $q^L_-$ lowering it by one unit.   Similarly, $q^R_{\pm}$ act on the right end of the mesonic string, and $q^R_+$ raises $K$ by one unit while $q^R_-$ lowers it by the same amount.  That $K$ changes in this fashion under the action of \eqref{qLRDef} means that
 \es{CommutP}{
  [P^+, q^L_\pm] = \pm \frac{1}{2L} q^L_\pm \,, \qquad  [P^+, q^R_\pm] = \pm \frac{1}{2L} q^R_\pm \,.
 }

The left supercharges and the right supercharges each separately generate an $\mathfrak{osp}(1|2)$ subalgebra, and, as we will show, together they generate an $\mathfrak{osp}(1|4)$ algebra. The $\mathfrak{osp}(1|2)$ subalgebras follow from the commutation relations
 \es{CommutqLqR}{
	[q^L_{-}, q^L_{+}] = [q^R_{-}, q^R_{+}] = \frac{1}{2} \,.
 }
To see this, let us first compute $[q^L_{-}, q^L_{+}]$ by plugging in the definitions \eqref{qLRDef}:
\es{Commutator}{
	[q^L_{-}, q^L_{+}] &= \frac{1}{2N}\sum_{\substack{n_1,n_2\\p_1,p_2}} \Biggl(
	\left\lbrack C^\dagger_j(n_1+n_2-1) B_{ij}(n_1) C_i(n_2), C^\dagger_l(p_1) B^\dagger_{lk}(p_2) C_k(p_1+p_2-1)\right\rbrack \\
	+ &\left\lbrack C^\dagger_j(n_1) B^\dagger_{ji}(n_2) C_i(n_1+n_2+1), C^\dagger_l(p_1+p_2+1) B_{kl}(p_1) C_k(p_2)\right\rbrack\Biggr) \,.
}
There are several terms in the commutators on the right hand side, but most of them are either suppressed by factors of $1/N$ or annihilate all single-trace mesonic states.  From the first commutator, the terms that survive in the large $N$ limit in the summand are
 \es{FirstCommutator}{
	&N \delta_{n_1, p_2} \delta_{n_2, p_1} C^\dagger_i(n_1+n_2-1)C_i(p_1+p_2-1) 
	+ \delta_{n_1+n_2, p_1+p_2} C^\dagger_i(p_1)B^\dagger_{ij}(p_2)C_k(n_2)B_{kj}(n_1) \\
	&{}-N \delta_{n_1, p_2}\delta_{n_2, p_1} C^\dagger_i(p_1+p_2+1)C_i(n_1+n_2+1) 
	- \delta_{n_1 + n_2, p_1 + p_2} C^\dagger_i(n_1)B^\dagger_{ij}(n_2)C_k(p_2)b_{kj}(p_1)
 }
where the first line comes from the first commutator in \eqref{Commutator} while the second line comes from the second commutator in \eqref{Commutator}. The $C^\dagger B^\dagger CB$ terms cancel under the summation. The $C^\dagger C$ terms telescope, and so the full commutator \eqref{Commutator} at leading order in large $N$ is
\begin{equation}
	[q^L_{-}, q^L_{+}] = \frac{1}{2}\sum_n C^\dagger_i(k)C_i(k) = \frac{1}{2} \,,
\end{equation}
where the second equality holds only on the single-trace mesonic states that involve only one $C^\dagger$ operator.  An analogous computation shows that $[q^R_{-}, q^R_{+}] = \frac 12$ when acting on single-trace mesonic states at leading order in $1/N$, thus concluding the proof of \eqref{CommutqLqR}.

Having established \eqref{CommutqLqR}, it is straightforward to show that $q^L_{\pm}$ generate an $\mathfrak{osp}(1|2)$ algebra and similarly for $q^R_{\pm}$.  In particular, focusing on $q^L_\pm$, let
\begin{equation}
	Q_1 \equiv q^L_{-}\,, \qquad Q_2 \equiv q^L_{+} \,,
\end{equation}
and define the symplectic form $\omega^{(2)}_{AB}  \equiv  (i \sigma_2)_{AB} = \begin{psmallmatrix} 0 & 1 \\ -1 & 0 \end{psmallmatrix}$, with $A, B = 1, 2$.  The $\mathfrak{osp}(1|2)$ is generated by the $Q_A$ as well as the $\mathfrak{sl}(2, \R)$  generators 
 \es{MDef}{
	M_{AB} \equiv \{Q_A, Q_B\} \,.
 }
Eq.~\eqref{CommutqLqR} can be written as $[Q_A, Q_B] = \frac{1}{2}\omega^{(2)}_{AB}$, which can be used to show that
 \es{eq:qm}{
	[Q_A, M_{BC}] &= [Q_A, Q_B Q_C] + [Q_A, Q_C Q_B] 
	= \omega^{(2)}_{AB} Q_C + \omega^{(2)}_{AC} Q_B \,,
 }
as well as
 \es{eq:mm}{
	[M_{AB}, M_{CD}] = \omega^{(2)}_{AD}M_{BC} + \omega^{(2)}_{BC}M_{AD} + \omega^{(2)}_{AC} M_{BD} + \omega^{(2)}_{BD} M_{AC} \,.
 }
Eqs.~\eqref{MDef}--\eqref{eq:mm} are the defining equations of the $\mathfrak{osp}(1|2)$ algebra.  We will denote this algebra generated by  $\{Q_1, Q_2, M_{11}, M_{12}, M_{22}\}$ by $\mathfrak{osp}(1|2)_L$  because it acts at the left end of the mesonic string.  Similarly, defining 
 \es{Q34Def}{
  Q_3 \equiv (-1)^F q^R_{+} \,, \qquad Q_4 \equiv  (-1)^F q^R_{-} \,,
 }
where $F$ is the fermion number operator, we find that $\{Q_3, Q_4, M_{33}, M_{34}, M_{44}\}$ (with $M_{AB}$ defined as in \eqref{MDef}) also obey the commutation relations of an  $\mathfrak{osp}(1|2)$ algebra that we denote by $\mathfrak{osp}(1|2)_R$  because it acts at the right end of the mesonic string.\footnote{We could've defined $Q_3 = q^R_-$ and $Q_4 = q^R_+$ and then $\{Q_3, Q_4, M_{33}, M_{34}, M_{44}\}$ would've also defined an $\mathfrak{osp}(1|2)$ algebra.  We used the definition \eqref{Q34Def} instead because with this definition we can extend the symmetry algebra to $\mathfrak{osp}(1|4)$ when acting on the massive states.}$^{,}$\footnote{Note that the existence of an $\mathfrak{osp}(1|2)_L$ algebra acting on the left end of the string and an $\mathfrak{osp}(1|2)_R$ algebra acting on the right end of the string does not imply that there is a $\mathfrak{osp}(1|2)_L \oplus \mathfrak{osp}(1|2)_R$ symmetry algebra acting on the single-trace mesonic states in the large $N$ limit.  We did not show, for instance, that $M_{AB}$, with $A, B = 1, 2$ commutes with $M_{CD}$, with $C, D = 3, 4$, so the two $\mathfrak{osp}(1|2)$ algebras may not be independent.  We will show shortly that the two $\mathfrak{osp}(1|2)$ act independently only on the massive states.}

Furthermore, as we now show, the $\{ Q_A, M_{AB} \}$ where now $A, B = 1, \ldots, 4$ generate an $\mathfrak{osp}(1|4)$ algebra when acting on the massive states.  To begin, let us compute the commutator between $Q_1, Q_2$ and $Q_3, Q_4$.  With $\alpha, \beta = \pm$, we have
  \es{CommutQ12Q34}{
   [q^L_\alpha, (-1)^F q^R_\beta] = - (-1)^F \{ q^L_\alpha,  q^R_\beta \} \,,
  }
so let us then compute the anticommutator of $q^L_{\alpha}$ with $q^R_{\beta}$:
\begin{equation}
\begin{split}
	\left\{q^L_{\alpha}, q^R_{\beta}\right\} = \frac{1}{2N}\sum_{\substack{n_1,n_2\\p_1,p_2}} \Bigg\lbrack &\left\{C^\dagger_j(n_1+n_2+\alpha) B_{ij}(n_1) C_i(n_2), D^\dagger_k(p_1) B^\dagger_{lk}(p_2) D_l(p_1+p_2-\beta)\right\} \\
	+ &\left\{C^\dagger_j(n_1) B^\dagger_{ji}(n_2) C_i(n_1+n_2-\alpha), D^\dagger_k(p_1+p_2+\beta) B_{kl}(p_1) D_l(p_2)\right\}\Bigg\rbrack \,.
\end{split}
\end{equation}
We only have to anti-commute the $B$ and $B^\dagger$ operators, which gives
\begin{equation}
\begin{split}
	\left\{q^L_{\alpha}, q^R_{\beta}\right\} = -&\frac{1}{2N}\sum_{n_1,n_2,n_3,n_4}\delta_{n_1+n_2, n_3+n_4+\alpha+\beta}\\
	&{}\times\left(C^\dagger_i(n_1)D^\dagger_i(n_2)C_j(n_3)D_j(n_4) - \frac{1}{N}C^\dagger_i(n_1)D^\dagger_j(n_2)C_i(n_3)D_j(n_4)\right) \,.
\end{split}
\end{equation}
The matrix elements of $C^\dagger_i D^\dagger_i C_j D_j$ are only of order $N$ between two 2-bit states, and the matrix elements of $C^\dagger_i D^\dagger_j C_i D_j$ are always of order 1 or smaller. Thus, for two single-trace meson states $\ket{\psi}$ and $\ket{\chi}$, the matrix elements are  
\begin{equation}
	\braket{\chi|\left\{q^L_{\alpha}, q^R_{\beta}\right\}|\psi} = -\frac{\braket{\chi | \psi}}{2} \times \begin{cases} 1 & \text{if both $\ket{\psi}$ and $\ket{\chi}$ are 2-bit} \\ 0 & \text{otherwise} \end{cases} \,,
\end{equation}
at leading order in large $N$.  This relation means that $\{q^L_\alpha,q^R_\beta\}$ is a rank one operator.  It annihilates all the states except for the equal linear combination of all two-bit states of a given $K$ (if $K$ is even) and it outputs an equal linear combination of two-bit states with $K + \alpha + \beta$ units of $P^+$ momentum.  These states are massless and were defined in \eqref{SingletPrimaries}.   Thus, we can write
\begin{equation}
	\left\{q^L_{\alpha}, q^R_{\beta}\right\} = -\sum_{K\text{ even}}\frac{K}{4}\sqrt{\frac{K+\alpha+\beta}{K}}\ket{\zeta_{K+\alpha+\beta}}\bra{\zeta_K} \,.
\end{equation}
Importantly, if we restrict ourselves to the massive sector, we simply have $\left\{q^L_{\alpha}, q^R_{\beta}\right\} = 0$ at leading order in $1/N$.

From now on, let us restrict to the massive states only.  Since $\left\{q^L_{\alpha}, q^R_{\beta}\right\} = 0$, we have from \eqref{CommutQ12Q34} that $Q_{1, 2}$ commute with $Q_{3, 4}$, so 
\begin{equation}\label{eq:comm2}
	[Q_A, Q_B] = \frac{1}{2}\omega^{(4)}_{AB} \,,
\end{equation}
where $\omega^{(4)}_{AB} = ({\bf 1}  \otimes \omega^{(2)})_{AB}$ is a $4 \times 4$ symplectic form and ${\bf 1}$ denotes the $2 \times 2$ identity matrix.  With the definition \eqref{MDef}, we then immediately see that \eqref{eq:comm2} implies
 \es{osp14Algebra}{
	[Q_A, M_{BC}] 
	&= \omega^{(4)}_{AB} Q_C + \omega^{(4)}_{AC} Q_B \,, \\
	[M_{AB}, M_{CD}] &= \omega^{(4)}_{AD}M_{BC} + \omega^{(4)}_{BC}M_{AD} + \omega^{(4)}_{AC} M_{BD} + \omega^{(4)}_{BD} M_{AC} \,,
 } 
which are the commutation relations defining the $\mathfrak{osp}(1|4)$ algebra.  Thus, $\{ Q_A, M_{AB} \}$ generate an $\mathfrak{osp}(1|4)$ algebra when acting on the massive states.

When restricted to the massive sector, the $\mathfrak{osp}(1|4)$ generators commute with $P^-$ at leading order in $1/N$.  While we do not have a proof of this fact in full generality, we checked it numerically for fixed $K$ up to $K = 35$.

\subsection{$\mathfrak{osp}(1|4)$ representations of $P^-$ eigenstates}

Let us now connect the discussion of $\mathfrak{osp}(1|4)$ with the degeneracies in the spectrum of mesons presented in Figure~\ref{fig:pminus_degeneracies}.  The massive states are acted on by $\mathfrak{osp}(1|4)$, so they must transform in unitary representations of $\mathfrak{osp}(1|4)$.  Such representations may be familiar to the reader from the study of 3d superconformal field theories with ${\cal N} = 1$ supersymmetry, because $\mathfrak{osp}(1|4)$ is the ${\cal N} =1$ superconformal algebra in three dimensions.  Given that the bosonic part of $\mathfrak{osp}(1|4)$ is $\mathfrak{sp}(4, \R) \cong \mathfrak{so}(3, 2)$, which is non-compact, such unitary representations are necessarily infinite dimensional.

Let us determine the irreducible representations under which the states in Figure~\ref{fig:pminus_degeneracies} transform in more detail.  The $P^-$ degeneracies occur both between states at the same $P^+$ and between states at different values of $P^+$, so the first question we should ask is which $\mathfrak{osp}(1|4)$ generators commute with $P^+$.  As already seen in  \eqref{CommutP}, the $Q_A$ do not commute with $P^+$, but some of the bosonic $M_{AB}$ generators, namely those that are anti-commutators of $Q_A$'s of opposite $P^+$ eigenvalues, do commute with $P^+$.   These are $\{M_{12}, M_{34}, M_{13}, M_{24}\}$, and they form an $\mathfrak{su}(2)\times\mathfrak{u}(1)$ algebra.   We can exhibit this algebra more clearly by defining
 \es{su2u1}{
	J_1 &\equiv \frac{M_{24}+M_{13}}{2} \,, \qquad
	 J_2 \equiv i  \frac{M_{12}+M_{34}}{2}\, \qquad J_3 \equiv \frac{M_{24}-M_{13}}{2} \,, \\
	D &\equiv \frac{M_{12}-M_{34}}{2}  \,.
 }
They satisfy
 \es{su2u1Algebra}{
  [J_i, J_j] = i \epsilon_{ijk} J_k \,, \qquad [D, J_i] = 0 \,,
 }
so the $J_i$ generate $\mathfrak{su}(2)$ while $D$ generates the $\mathfrak{u}(1)$.  In the language of 3d CFTs, the generator $D$ measures the scaling dimension while $J^2 = j(j+1)$ measures the spin $j$.

Because $J_i$ and $D$ commute with $P^+$, each multiplet represented by a dot in Figure~\ref{fig:pminus_degeneracies} forms a representation of this $\mathfrak{su}(2)\times\mathfrak{u}(1)$. For instance, the $P^- = 3/2$ triplet at $K = 6$ is an $\mathfrak{su}(2)$ triplet that can be split into three $J_3$ eigenstates (where $(n_1, n_2, \ldots, n_N)$ is a shortcut notation for the state $C^\dagger_i(n_1) B^\dagger_{ij}(k_2)\cdots D^\dagger_j(k_N)\ket{0}$):
\begin{center}
\renewcommand{\arraystretch}{1.5}
\begin{tabular}{c|c|c|c}
    Eigenvector & $P^-$ & $D$ & $J_3$ \\
    \hline
    $-(1, 5) - (5, 1) + 2(3, 3) - 3(1, 1, 3, 1) + 3(1, 3, 1, 1)$ & $\frac{3}{2}$ & $\frac 32$ & -1 \\
    $(1, 5) - (5, 1) + 2 (1, 1, 1, 3) - (1, 1, 3, 1) - (1, 3, 1, 1) + 2 (3, 1, 1, 1)$ & $\frac{3}{2}$ & $\frac 32$ & 0 \\
    $-(1, 5) - (5, 1) + 2(3, 3) + (1, 1, 3, 1) - (1, 3, 1, 1) + 4 (1, 1, 1, 1, 1, 1)$ & $\frac{3}{2}$ & $\frac 32$ & 1
\end{tabular}
\renewcommand{\arraystretch}{1}
\end{center}

The unitary irreps of $\mathfrak{osp}(1|4)$ that are familiar from ${\cal N} = 1$ SCFTs are lowest-weight representations that have the following tree-like structure.  There is a unique ``(superconformal) primary'' state with lowest $D$ eigenvalue $\Delta$ and some value of $j$.  All the other states in the irrep are generated by acting with a string of $Q$'s on the primary state.  The action of each $Q$ increases $\Delta$ by $1/2$.

Quite nicely, this tree-like structure is very explicit in Figure~\ref{fig:pminus_degeneracies} because, as we will show, for a given irrep increasing $\Delta$ by $1/2$ is equivalent to increasing $K$ by $1$.  Indeed, from the commutator between $Q_A$ and $M_{BC}$ in \eqref{eq:qm}, one can see that $D/L$ has the same commutation relations with $Q_A$ as $P^+$.  It follows that $P^+ - D/L$ commutes with the $\mathfrak{osp}(1|4)$ algebra, and consequently for a given $\mathfrak{osp}(1|4)$ representation, $P^+ - D/L$ is a constant.  This implies that increasing $\Delta$ by $1/2$ is equivalent to increasing $K = 2 L P^+$ by $1$.  Thus, the primary state in each irrep of a given $P^-$ eigenvalue must be the state with lowest $K$.\footnote{We can give a more refined description of the state counting in terms of the scaling dimension. One can show that the number of $\mathfrak{osp}(1|4)$ primary states at a given $K$ is $F_{K-4} + (-1)^K$, where $\{F_n\}_{n=1}^\infty$ is the Fibonacci sequence with $F_1 = F_2 = 1$. Using the tree-like structure described above, this implies that at a given $K$ there are $2\Delta \left(F_{K-3-2\Delta} + (-1)^{K-3-2\Delta}\right)$ massive states of scaling dimension $\Delta$. Summing over $\Delta$ and adding the $\lfloor K/2\rfloor$ massless states, we recover the total of $F_K$ states when $K$ is even and $F_K - 1$ when $K$ is odd.}

Explicitly acting with $D$ on the states of smallest $K$ for the various $P^-$ eigenvalues of massive states in Figure~\ref{fig:pminus_degeneracies} reveals that these states have $\Delta = 1/2$, and they also have $\mathfrak{su}(2)$ spin $j=0$.\footnote{Most of these states are singlets, and for them it is obvious that $j=0$.  We checked explicitly that even in cases in which the first time a $P^-$ eigenvalue appears in the spectrum as a doublet, the $J^2$ eigenvalue still vanishes.}  Such $\mathfrak{osp}(1|4)$ irreducible representations for which the primary has $\Delta = 1/2$ and $j=0$ are usually referred to as ``$\mathfrak{osp}(1|4)$ singletons'' and are shorter than generic irreps.  In SCFT language, a singleton correspond to the irrep consisting of a free massless scalar and a free massless Majorana fermion, together with their superconformal descendants.    Under the bosonic $\mathfrak{so}(3, 2)$ algebra, an $\mathfrak{osp}(1|4)$ singleton decomposes into a direct sum of two irreps, namely a scalar singleton (which corresponds to the free real scalar with $\Delta= 1/2$ and $j=0$) and a fermionic singleton (which corresponds to the free Majorana fermion with $\Delta= 1$ and $j=1/2$).

One can also confirm that all the massive states belong to $\mathfrak{osp}(1|4)$ singleton representations by computing the eigenvalues of the $\mathfrak{so}(3, 2)$ and $\mathfrak{osp}(1|4)$ quadratic Casimir.  The $\mathfrak{so}(3, 2)$ quadratic Casimir is 
 \es{so32Cas}{
   C_{\mathfrak{so}(3, 2)} = - \frac 14 \omega^{AC} \omega^{BD} M_{AB} M_{CD} \,,
 }
where $\omega^{AB} = - \omega_{AB}$, while the $\mathfrak{osp}(1|4)$ one is 
 \es{osp14Cas}{
   C_{\mathfrak{osp}(1|4)} = - \frac 14 \omega^{AC} \omega^{BD} M_{AB} M_{CD} - \frac 12 \omega^{AB} Q_A Q_B \,.
 }
(It is straightforward to check that $C_{\mathfrak{so}(3, 2)}$ commutes with all $M_{AB}$ and that $C_{\mathfrak{osp}(1|4))}$ commutes with the $Q_A$.) The eigenvalue of $C_{\mathfrak{so}(3, 2)}$ for an $\mathfrak{so}(3, 2)$ irrep whose primary state has $D = \Delta$ and $J^2 = j(j+1)$ is
 \es{lamso32}{
  \lambda_{\mathfrak{so}(3, 2)}(\Delta, j)  = \Delta( \Delta - 3) + j(j+1) \,, 
 }
while the eigenvalue of $C_{\mathfrak{osp}(1|4)}$ for an $\mathfrak{osp}(1|4)$ irrep whose primary state has $D = \Delta$ and $J^2 = j(j+1)$ is
  \es{lamosp14}{
  \lambda_{\mathfrak{osp}(1|4)}(\Delta, j) = \Delta( \Delta - 2) + j(j+1) \,.
 }
For the $\mathfrak{osp}(1|4)$ singleton representation we have $C_{\mathfrak{so}(3, 2)} =  \lambda_{\mathfrak{so}(3, 2)}(\frac 12, 0) =  \lambda_{\mathfrak{so}(3, 2)}(1, \frac 12) = 5/4$ on all the states and $C_{\mathfrak{osp}(1|4)} = \lambda_{\mathfrak{osp}(1|4)}(\frac 12, 0) = 3/4$.  We checked that we obtain these values when acting with \eqref{so32Cas} and \eqref{osp14Cas} explicitly on all the massive states.

While the $\mathfrak{osp}(1|4)$ symmetry explains the degeneracies in the meson spectrum, it does not explain the degeneracies pointed out in Section~\ref{sec:glueball_degeneracies} between the $\mathfrak{osp}(1|4)$ primary meson states and gluinoball states.  As described in Section~\ref{sec:glueball_degeneracies}, these latter degeneracies can be seen from the current algebra approach.  For a partial alternative explanation that involves operators related to the $\mathfrak{osp}(1|4)$ charges, see Appendix~\ref{sec:DEGMG}.

\section{Making the quarks massive} 
\label{sec:MASSIVE}

\subsection{Spectrum and degeneracies }

In order to further probe our model, we would like to study the meson spectrum as a function of the fundamental fermion mass.  If we make $y_\text{fund}>0$ while keeping $y_\text{adj} =0$, we expect the theory to retain some of the non-trivial dynamical properties of the massless adjoint QCD$_2$. 
We will note that some of the DLCQ degeneracies between the single-trace and multi-trace states are not lifted and that the meson spectrum is continuous above a certain threshold.
This provides new quantitative evidence, along the lines of Footnote~4 of~\cite{Gross:1995bp}, that the fundamental string tension vanishes in the massless adjoint QCD$_2$.

For $y_\text{fund}= y_\text{adj} =0$, we have found an infinite series of single-trace mesonic states with the same value of $P^-$ as some fermionic gluinoballs at an odd value of $K$
(see Figures \ref{fig:pminus_degeneracies} and \ref{fig:m2_convergence}). These include the bosonic mesons with resolution parameters $K+1, K+3, K+5, \ldots$ and the fermionic mesons with resolution parameters $K+2, K+4, K+4, \ldots$.
Since the spectrum of the model contains massless mesons, both bosonic and fermionic, this is consistent with the pattern of degeneracies between single-trace and multi-trace states.
In particular, some massive bosonic mesons are degenerate with double-trace states 
of a fermionic gluinoball and a massless fermionic meson, and some massive fermionic mesons are degenerate with double-trace states 
of a fermionic gluinoball and a massless bosonic meson.   As explained in Section \ref{sec:glueball_degeneracies}, when a meson is degenerate with a bosonic gluinoball, it is because the bosonic gluinoball is in turn degenerate with a multi-trace state formed from fermionic gluinoballs. This follows from the Kac-Moody approach reviewed in Section~\ref{KMA}. We may thus think of the degeneracy as being between a fermionic meson and a triple-trace state formed from a massless fermionic meson and two fermionic gluinoballs.

After we make the fundamental fermions massive, there are no more massless mesons in the spectrum; yet, some of the degeneracies survive. 
We continue to find that some bosonic mesons are degenerate with double-trace states 
of the fermionic gluinoballs and massive fermionic mesons. These degeneracies hold for any $y_\text{fund}>0$, and an example at $y_\text{fund} = 1$ is shown in Figure \ref{fig:meson_spectrum_massive_bose_even} and \ref{fig:meson_spectrum_massive_bose_odd}, where the degenerate states are marked in orange. In fact, we find that every double-trace state formed from a fermionic meson and a fermionic gluinoball is degenerate with a bosonic meson.
\begin{figure}
    \centering
    \begin{subfigure}{.7\linewidth}%
	    \centering
    	\includegraphics[width=\linewidth]{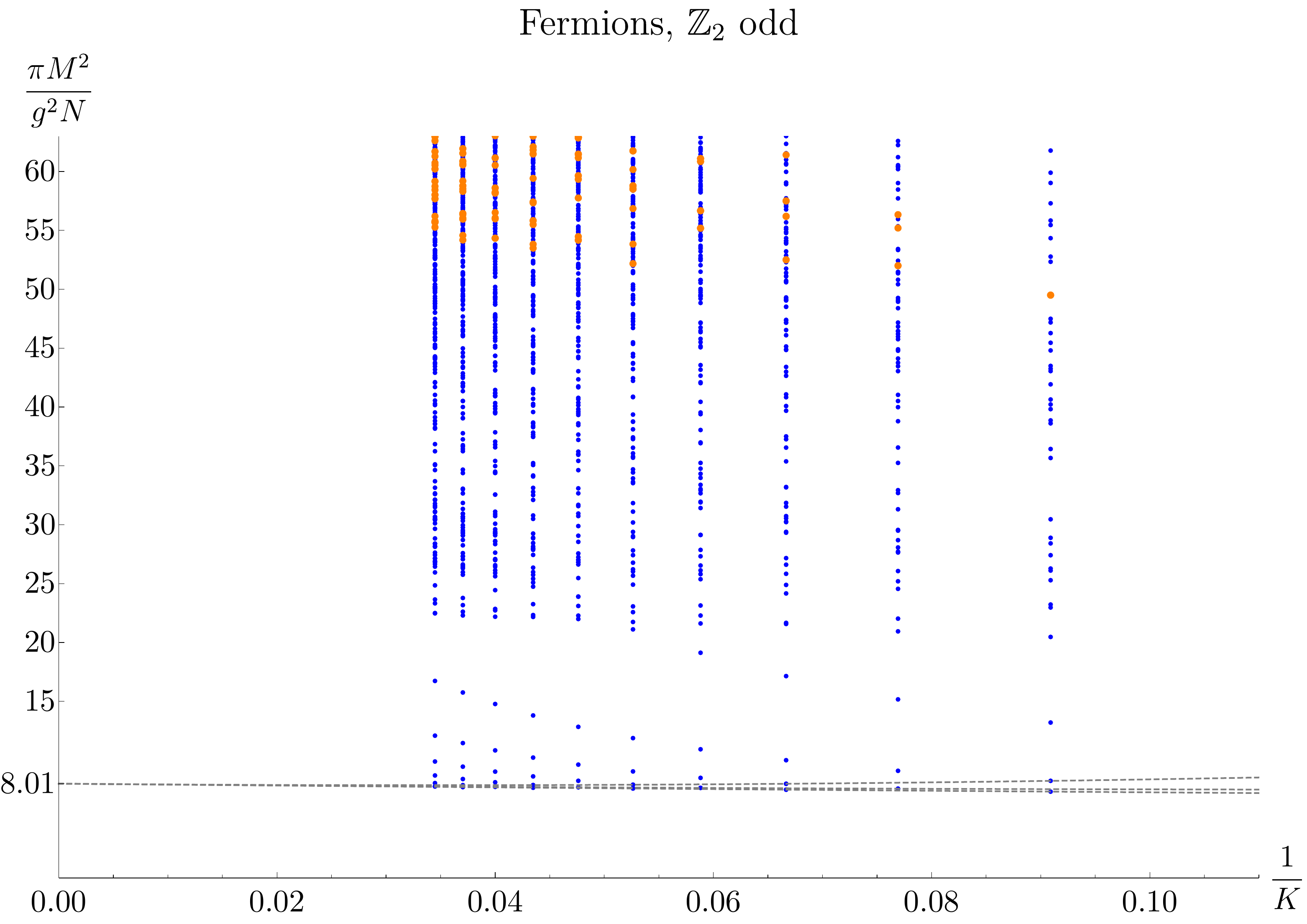}
    	\caption{}
    	\label{fig:meson_spectrum_massive_fermi_even}
    \end{subfigure}\\
    \vspace{2em}%
    \begin{subfigure}{.7\linewidth}%
	    \centering
    	\includegraphics[width=\linewidth]{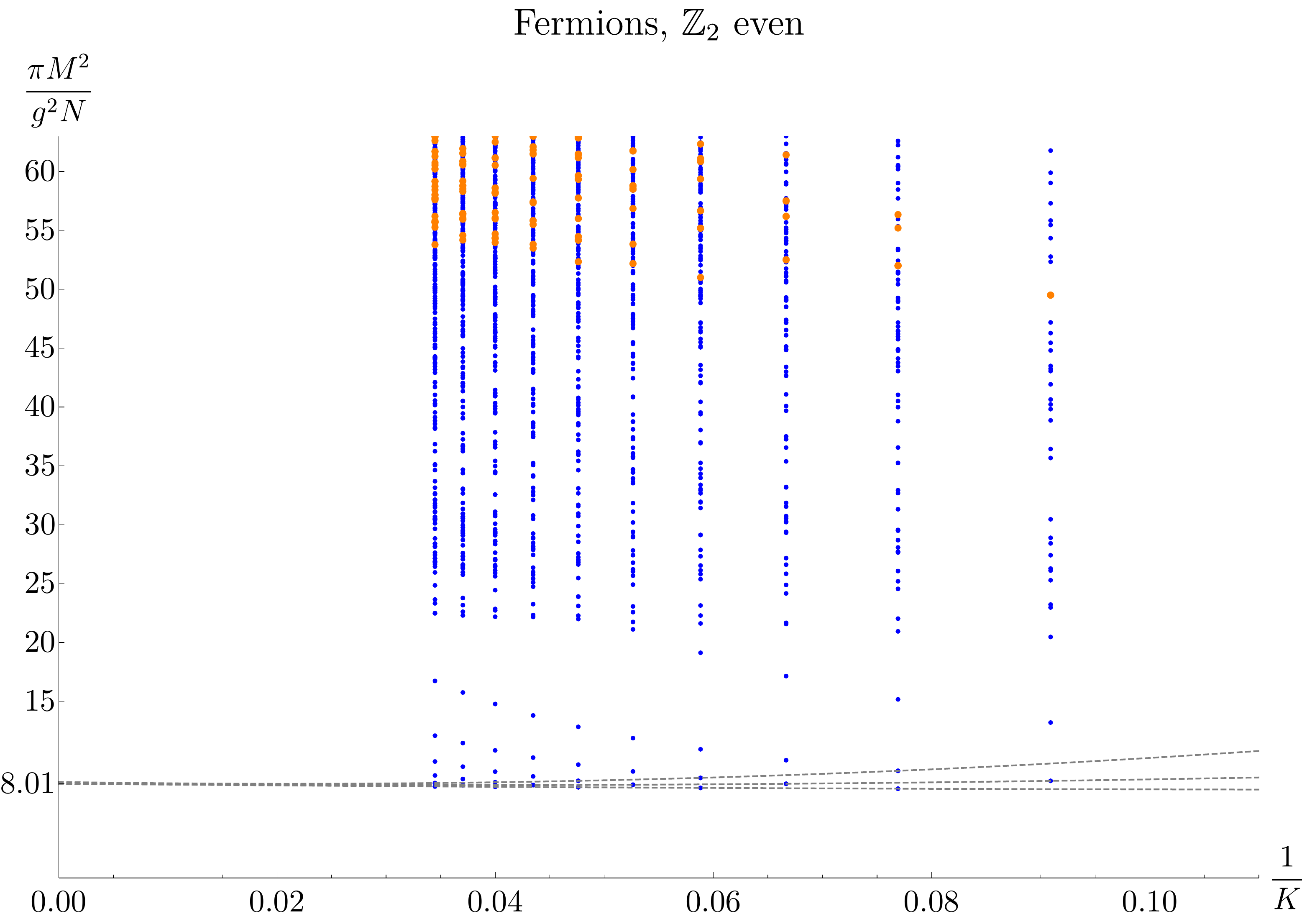}
    	\caption{}
    	\label{fig:meson_spectrum_massive_fermi_odd}
    \end{subfigure}\\
    \phantomcaption
\end{figure}
\begin{figure}
	\centering
	\ContinuedFloat
    \begin{subfigure}{.7\linewidth}%
	    \centering
    	\includegraphics[width=\linewidth]{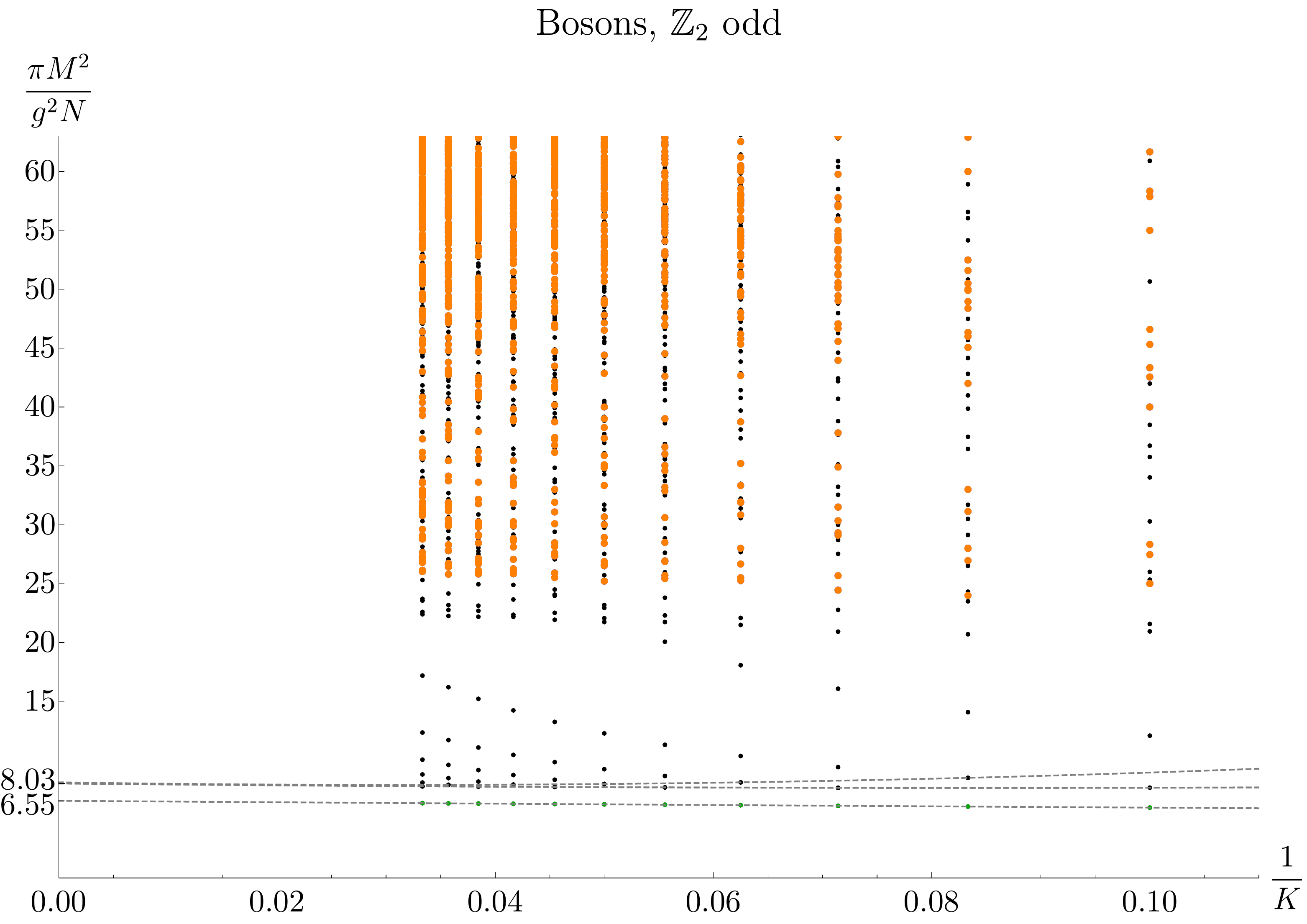}
    	\caption{}
    	\label{fig:meson_spectrum_massive_bose_even}
    \end{subfigure}\\
    \vspace{2em}%
    \begin{subfigure}{.7\linewidth}%
	    \centering
    	\includegraphics[width=\linewidth]{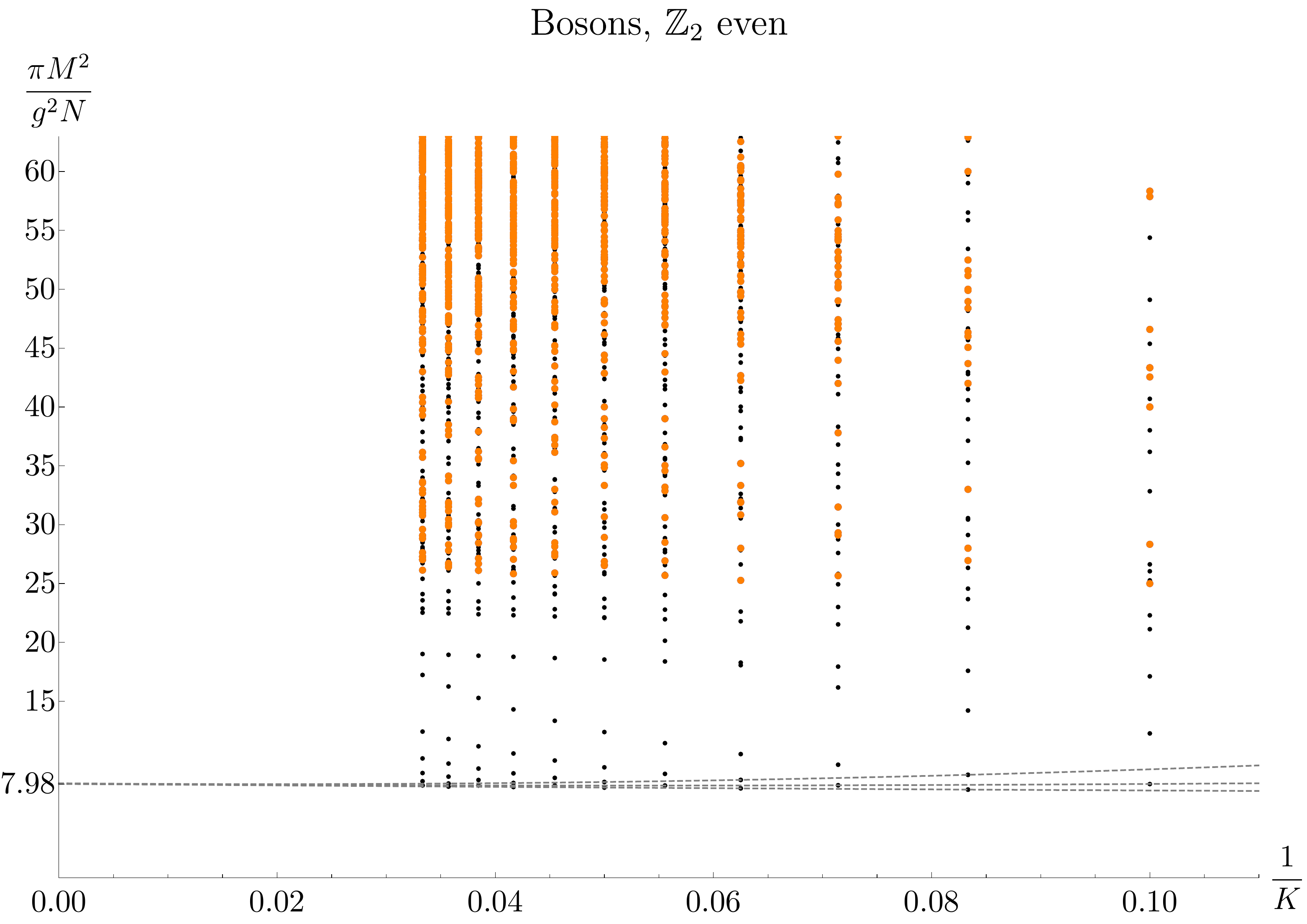}
    	\caption{}
    	\label{fig:meson_spectrum_massive_bose_odd}
    \end{subfigure}%
    \caption{The squared masses of single-trace meson states in theory ${\cal T}$ (the theory with an adjoint and a fundamental fermion)  with $y_\text{adj} = 0$ and $y_\text{fund} = 1$. The blue points are the states which are 
degenerate with multi-string states in  the theory $\mathcal{T}'$ defined in Section~\ref{sec:MORE}. The orange points are the states which additionally are degenerate with a multi-trace state formed from a meson and one or more gluinoballs in theory ${\cal T}$. The dashed lines show the threshold at $M^2\approx 8 \frac{g^2 N}{\pi}$ above which the extrapolated spectrum is continuous. There is a  
bosonic $\mathbb{Z}_2$-odd state shown with green dots
that lies below this threshold. Its extrapolated squared mass is $M^2 \approx 6.6\frac{g^2 N}{\pi}$. 
}
    \label{fig:meson_spectrum_massive}
\end{figure}

Likewise, we continue to find that some fermionic mesons are degenerate with triple-trace states built from a fermionic meson and two fermionic gluinoballs. Almost all of the triple-trace states of this form are degenerate with single-trace fermionic mesons with the same total $P^+$. Examples at $y_\text{fund} = 1$ are shown in Figure \ref{fig:meson_spectrum_massive_fermi_even} and \ref{fig:meson_spectrum_massive_fermi_odd}, with the degenerate states being marked in orange.

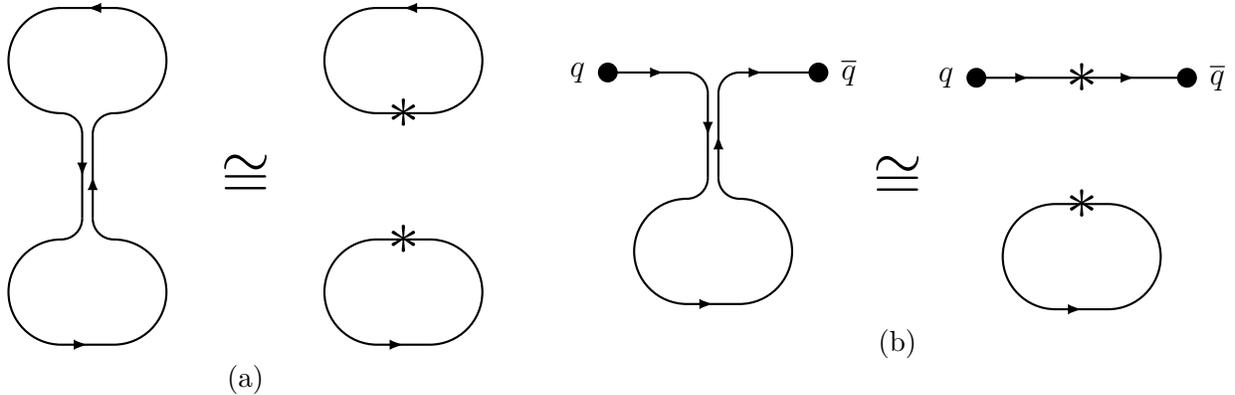
\begin{figure}
	\tikzset{
      mid arrow/.style={postaction={decorate,decoration={
            markings,
            mark=at position .5 with {\arrow[#1]{latex}}
          }}},          
      dot/.style={fill,circle,minimum size=5,inner sep=0}
    }
	\centering
	\begin{subfigure}{.4\linewidth}%
		\centering
		\begin{tikzpicture}[thick,scale=0.7]
            \begin{scope}[scale=-1,xshift=-4cm,yshift=2.4cm]
            \draw[mid arrow=black] (1.9, -2) arc(0:-90:.4) arc(90:270:1) -- ++(1, 0) arc(-90:90:1) arc(270:180:.4);
            \draw[mid arrow=black] (2.1,-2) -- (2.1,-.4);
            \end{scope}
            
            \draw[mid arrow=black] (1.9, -2) arc(0:-90:.4) arc(90:270:1) -- ++(1, 0) arc(-90:90:1) arc(270:180:.4);
            \draw[mid arrow=black] (2.1,-2) -- (2.1,-.4);
            
            \node[scale=2] at (5,-1.2) {$\cong$};
            
            \begin{scope}[xshift=6cm]
                \begin{scope}[scale=-1,xshift=-4cm,yshift=2.4cm]
                \draw[mid arrow=black] (2,-2.4) -- (1.5, -2.4) arc(90:270:1) -- ++(1, 0) arc(-90:90:1) -- cycle;
                \node[scale=2] at (2,-2.4) {$\ast$};
                \end{scope}
                \draw[mid arrow=black] (2,-2.4) -- (1.5, -2.4) arc(90:270:1) -- ++(1, 0) arc(-90:90:1) -- cycle;
                \node[scale=2] at (2,-2.4) {$\ast$};
            \end{scope}
        \end{tikzpicture}
        \caption{}
        \label{fig:gluinoball_sketch}
    \end{subfigure}%
    \hspace{.05\linewidth}%
	\begin{subfigure}{.55\linewidth}%
		\centering
		\begin{tikzpicture}[thick,scale=0.7]
            \draw[mid arrow=black] (0,0) node[dot,label=180:{$q$},scale=1.5] {} -- (1.5,0) arc(90:0:.4);
            \draw[mid arrow=black] (1.9,-.4) -- (1.9,-2);
            \draw[mid arrow=black] (1.9, -2) arc(0:-90:.4) arc(90:270:1) -- ++(1, 0) arc(-90:90:1) arc(270:180:.4);
            \draw[mid arrow=black] (2.1,-2) -- (2.1,-.4);
            \draw[mid arrow=black] (2.1,-.4) arc(180:90:.4) (2.5,0) -- (4,0) node[dot,label=0:{$\overline{q}$},scale=1.5] {};
            
            \node[scale=2] at (5.5,-2) {$\cong$};
            
            \begin{scope}[xshift=7cm,yshift=-0.1cm]
                \draw[mid arrow=black] (0,0) node[dot,label=180:{$q$},scale=1.5] {} -- (2,0);
                \draw[mid arrow=black] (2,0) -- (4,0) node[dot,label=0:{$\overline{q}$},scale=1.5] {};
                \node[scale=2] at (2,0) {$\ast$};
                \draw[mid arrow=black] (2,-2.4) -- (1.5, -2.4) arc(90:270:1) -- ++(1, 0) arc(-90:90:1) -- cycle;
                \node[scale=2] at (2,-2.4) {$\ast$};
            \end{scope}
        \end{tikzpicture}
        \caption{}
        \label{fig:meson_sketch}
    \end{subfigure}%
    \caption{The degeneracies among different states in theory $\cT$, which survive at $y_\text{fund} > 0$, can be heuristically interpreted in terms of an adjoint flux line of vanishing energy. In (a), pinching away an adjoint flux line in a single-trace gluinoball state yields a double-trace gluinoball state. In (b), pinching away an adjoint flux line in a single-trace meson state yields a double-trace state composed of a meson and a gluinoball.}
    \label{fig:degeneracy_sketches}
\end{figure}

Qualitatively, we may think of the degeneracies we have found in terms of adjoint flux lines with vanishing energy. Figure \ref{fig:degeneracy_sketches} shows schematically how an adjoint flux line contributing zero mass to a state could lead to degeneracies between a single-trace gluinoball and a double-trace gluinoball, or between a single-trace meson and a double-trace state composed of a meson and a gluinoball. Similar pictures could be drawn for multi-trace states with more components.

\subsection{More relations between eigenvalues and screening in adjoint QCD$_2$}
\label{sec:MORE}

The relations between the eigenvalues at $y_\text{fund} > 0$ do not end here.  To explain the relations noticed so far and to discover new ones, let us recall from Section~\ref{sec:GLUINODEG} that the fact that the $P^-$ eigenvalues of $\mathfrak{n}>1$ gluinoballs are degenerate with sums of $P^-$ eigenvalues of $\mathfrak{n}=1$ gluinoballs can be traced back to the fact that the massive spectrum of the adjoint QCD$_2$ theory ${\cal T}_\text{adj}$ is part of the massive spectrum of the theory ${\cal T}_\text{fund}$ whose matter content consists of $N$ massless Dirac fermions, as explained in \cite{Kutasov:1994xq}.  In particular, the $P^-$  eigenvalues of $\mathfrak{n}>1$ states in ${\cal T}_\text{adj}$ at resolution parameter $K$ correspond to ${\cal T}_\text{fund}$ states at resolution parameter $K+\mathfrak{n}-1$ that are manifestly ``multi-string'' states---see Eq.~\eqref{SingleTraceGeneralnAgain}.  Essentially, each  $B^\dagger_{ij}(1)$ in the construction of the states in ${\cal T}_\text{adj}$ is replaced by $C^\dagger_{\alpha i}(1) D^\dagger_{j \beta}(1)$ in ${\cal T}_\text{fund}$.  Intuitively, each $B^\dagger_{ij}(1)$ serves as a breaking point of the closed gluinoball string, and thus if $B^\dagger_{ij}(1)$ appears  $\mathfrak{n}-1$ times in ${\cal T}_\text{adj}$, then in  ${\cal T}_\text{fund}$ we end up with $\mathfrak{n}-1$ strings. 

Now let us couple the two theories ${\cal T}_\text{adj}$ and ${\cal T}_\text{fund}$ to a {\it massive} fundamental Dirac fermion and compare the two resulting theories. One of them, denoted by $\cT$, has a Dirac fermion with mass $m_q$ coupled to a massless adjoint and it is the theory we studied in the previous subsection; the other, denoted by $\cT'$, has $N+1$ Dirac fermions, the first $N$ of which are massless and the $(N+1)$st with mass $m_q$. The mass spectrum of the theory ${\cal T}$ is again part of the mass spectrum of the theory ${\cal T}'$, generalizing the result of \cite{Kutasov:1994xq}.  At large $N$, it is easier to see various relations between the $P^-$ eigenvalues in the theory ${\cal T}'$.

While we leave a careful analysis for the future, let us describe how the correspondence between the meson states in the theory of interest ${\cal T}$ and states in ${\cal T}'$ works at a qualitative level. We can construct the states by considering KM primaries only with respect to the adjoint contribution to the $SU(N)$ current $J_{ij}$.  Thus, in the current algebra construction, each meson state in ${\cal T}$ will have some number of $B_{ij}^\dagger(1)$'s which in ${\cal T}'$ will be replaced by $C^\dagger_{\alpha i}(1) D^\dagger_{j \beta}(1)$, thus increasing $K$ by one unit and breaking the string.

\begin{figure}
	\tikzset{
      mid arrow/.style={postaction={decorate,decoration={
            markings,
            mark=at position .5 with {\arrow[#1]{latex}}
          }}},
      dot/.style={fill,circle,minimum size=7,inner sep=0}
    }
	\centering
	\begin{subfigure}{.4\linewidth}%
	\centering
	\begin{tikzpicture}[thick]
		\draw[mid arrow] (-1,2) node[dot,label=180:{$q$}] {} -- (-1,-2) node[dot,label=180:{$\overline{q}$}] {};
        \node[scale=2] at (1,0) {$\cong$};
		\draw[mid arrow] (3,2) node[dot,label=180:{$q$}] {} -- (3,0.5) node[draw,cross out] {};
		\draw[mid arrow] (3,-0.5) node[draw,cross out] {} -- (3,-2) node[dot,label=180:{$\overline{q}$}] {};
	\end{tikzpicture}
	\caption{}
	\end{subfigure}%
	\hspace{0.1\linewidth}%
	\begin{subfigure}{.5\linewidth}%
	\centering
	\begin{tikzpicture}[thick]
		\draw[mid arrow] (-0.5,0) arc(0:180:1);
		\draw[mid arrow] (-2.5,0) arc(180:360:1);
        \node[scale=2] at (1,0) {$\cong$};
		\begin{scope}[xshift=3.5cm]
			\draw[mid arrow] (30:1) node[draw,rotate=-30,cross out] {} arc(30:150:1);
			\draw (150:1) arc(150:210:1);
			\draw[mid arrow] (210:1) arc(210:330:1) node[draw,rotate=-30,cross out] {};
		\end{scope}
	\end{tikzpicture}
	\caption{}
	\end{subfigure}
    \caption{The degeneracies among different states can be interpreted as splitting a string.  In (a), we show an $\mathfrak{m}=1$ meson in theory ${\cal T}$ which splits into a two-string state $ [\text{H $-$ L}]  [\text{L $-$ H}]$ in theory ${\cal T}'$.  In (b), we show an $\mathfrak{n}=1$ gluinoball in ${\cal T}$ which splits into a one-string state  $[\text{L $-$ L}]$ in theory ${\cal T}'$.  Diagram (a) illustrates the origin of the continuous spectrum of mesons in theory $\mathcal{T}$.}
    \label{fig:tprime_degeneracy_sketches}
\end{figure}
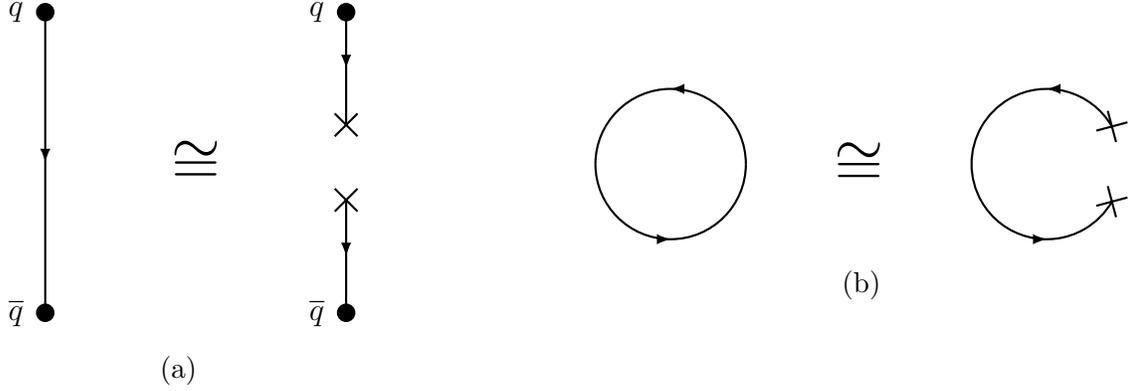

Thus,  a meson with $\mathfrak{m}$ $B^\dagger(1)$'s in ${\cal T}$ becomes an $\mathfrak{m}+1$-string state in ${\cal T}'$, which we can write schematically as
 \es{schematic}{
  \text{meson with $\mathfrak{m}$ $B^\dagger(1)$'s in ${\cal T}$}  \longleftrightarrow \underbrace{
   [\text{H $-$ L}] [\text{L $-$ L}] \cdots [\text{L $-$ L}] [\text{L $-$ H}]}_\text{$\mathfrak{m}+1$ factors} \text{ in ${\cal T}'$}  \,,
 }
where $[\text{L $-$ L}]$ denotes a string where the quarks at both ends are light (i.e.~massless), while $[\text{H $-$ L}]$ and $[\text{L $-$ H}]$ denote strings where the quark at the end marked with $H$ is heavy (i.e.~of mass $m_q$) while the one at the other end is light.  If the state on the LHS of \eqref{schematic} is at resolution parameter $K$ in ${\cal T}$, the state on the RHS is at resolution parameter $K + \mathfrak{m}$ in ${\cal T}'$.  Note that with the same notation, we have 
 \es{gluinoballschematic}{
  \text{$\mathfrak{n}=1$ gluinoball in ${\cal T}$}  \longleftrightarrow
    [\text{L $-$ L}]  \text{ in ${\cal T}'$} \,,
 }
where if the resolution parameters are $K$ and $K+1$ on the LHS and RHS, respectively.  See Figure~\ref{fig:tprime_degeneracy_sketches} for a diagramatic representation of the relations \eqref{schematic}--\eqref{gluinoballschematic}.

In \eqref{schematic}, the states in ${\cal T}$ with even $\mathfrak{m}$ are bosons and those with odd $\mathfrak{m}$ are fermions.  Because for multi-string states in ${\cal T}'$ the values of $P^-$ add, based on \eqref{schematic} and \eqref{gluinoballschematic} we expect the following relations between the eigenvalues of $P^-$ within theory ${\cal T}$:
 \begin{itemize}
  \item There are bosonic mesons (namely with $\mathfrak{m}= 0$) whose $P^-$ eigenvalues are unrelated to other states in theory ${\cal T}$.  These states are marked in black in Figures~\ref{fig:meson_spectrum_massive_bose_even} and \ref{fig:meson_spectrum_massive_bose_odd}, and they correspond to states of the form $[\text{H $-$ H}]$ in ${\cal T}'$.
  \item There are fermionic mesons (namely with $\mathfrak{m}= 1$) whose $P^-$ eigenvalues are unrelated to other states in theory ${\cal T}$.  They are marked in blue in Figures~\ref{fig:meson_spectrum_massive_fermi_even} and \ref{fig:meson_spectrum_massive_fermi_odd}, and they correspond to states of the form $[\text{H $-$ L}][\text{L $-$ H}]$ in~${\cal T}'$.
  \item The remaining bosonic mesons (namely with $\mathfrak{m}=2, 4, 6, \ldots$) are degenerate with a sum of $\mathfrak{n}= 1$ fermionic mesons and an odd number $\mathfrak{m}-1$ of $\mathfrak{n}=1$ gluinoballs, with the same total $K$.  They are marked in orange in Figures~\ref{fig:meson_spectrum_massive_bose_even} and \ref{fig:meson_spectrum_massive_bose_odd}.
\item The remaining fermionic mesons (namely with $\mathfrak{m} = 1, 3, 5, \ldots$) are degenerate with a sum of $\mathfrak{n}= 1$ fermionic mesons and an even number $\mathfrak{m}-1$ of  $\mathfrak{n}=1$ gluinoballs, with the same total $K$.   They are marked in orange in Figures~\ref{fig:meson_spectrum_massive_fermi_even} and \ref{fig:meson_spectrum_massive_fermi_odd}.
 \end{itemize}
The last two bullet points explain the degeneracies noticed in the previous section.

The second bullet point above has remarkable implications.  It states that the $P^-$ eigenvalues of all fermionic mesons with $\mathfrak{m}=1$ at resolution parameter $K$ in theory ${\cal T}$ can be written as sums of two eigenvalues of $[\text{H $-$ L}]$ states in theory ${\cal T}'$ with total resolution parameter $K+1$.  From this information, we can reconstruct the $[\text{H $-$ L}]$ spectrum of $P^-$ eigenvalues in theory ${\cal T}'$.  

\begin{figure}
	\centering
    \includegraphics[width=\linewidth]{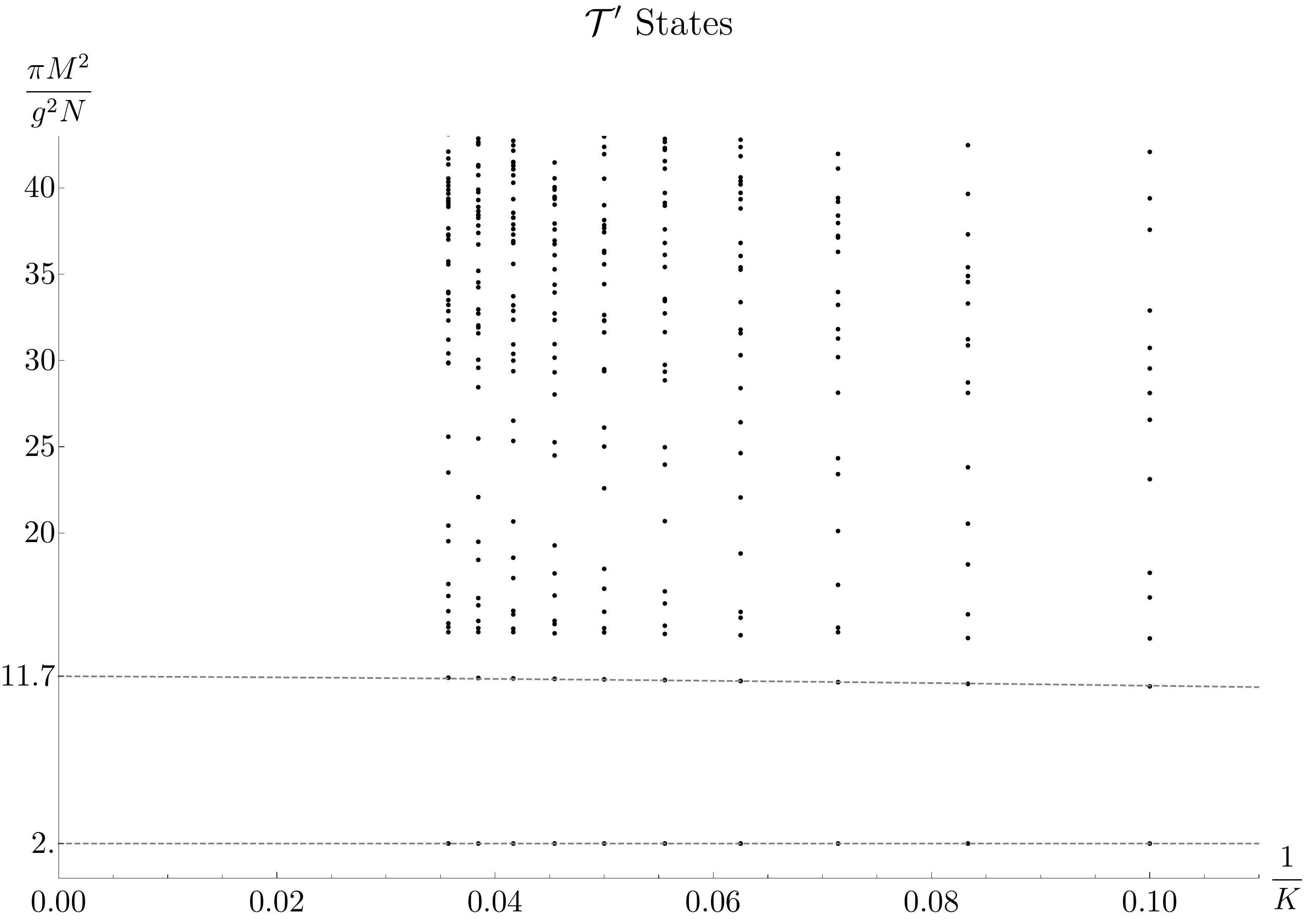}
    \caption{The inferred masses of bosonic single-string states in the theory $\mathcal{T}'$ which contain exactly one massive quark. The massive quark has $m^2_\text{fund} = \frac{g^2 N}{\pi}$, corresponding to $y_\text{fund} = 1$. }
    \label{fig:tprime_spectrum}
\end{figure}

For example, at $K=3$ in ${\cal T}$ we have only meson state, $C^\dagger(1) B^\dagger(1) D^\dagger(1)$ with $P^-$ eigenvalue $2 y_\text{fund}$.  The only possibility is that this is the sum of two $K=2$ $[\text{H $-$ L}]$ states in ${\cal T}'$ each with eigenvalue
 \es{p1}{
  p_1 =y_\text{fund} \,.
 }
At $K=5$, in ${\cal T}$ we have the following eigenvalues 
 \es{eval5}{
  \frac{9 + 20 y_\text{fund}  \pm \sqrt{81 + 24 y_\text{fund} + 16 y_\text{fund}^2}}{12} \,.
 }
These must be the sum of one $K=2$ eigenvalue and one $K=4$ eigenvalue in ${\cal T}'$.  Since we know that the only $K=2$ eigenvalue is given by $p_1$, we should subtract it from \eqref{eval5} to find the $K=4$ eigenvalues in ${\cal T}'$:
 \es{p23}{
  p_2 =  \frac{9 + 8 y_\text{fund}  + \sqrt{81 + 24 y_\text{fund} + 16 y_\text{fund}^2}}{12} \,, \quad
   p_3 = \frac{9 + 8 y_\text{fund}  - \sqrt{81 + 24 y_\text{fund} + 16 y_\text{fund}^2}}{12} \,.
 }
We can already have a check on these equations:  the quantities $2p_2$, $2p_3$, and $p_2 + p_3$ would be $P^-$ eigenvalues of two-string states in ${\cal T}'$ at $K=8$, so they must appear in the meson spectrum in ${\cal T}$ at $K=7$.  One can check that this is indeed the case.  Continuing to higher $K$ analytically is not feasible because the $P^-$ eigenvalues are roots of high order polynomials, but one can continue this procedure and reconstruct the spectrum of  $[\text{H $-$ L}]$ states in the ${\cal T}'$ theory numerically for given $y_\text{fund}$.  For $y_\text{fund}=1$, we show this reconstructed spectrum in Figure~\ref{fig:tprime_spectrum}.  It would be interesting to provide a confirmation of this plot by directly diagonalizing $P^-$ in the theory~${\cal T}'$.

An immediate implication of these results is that the spectrum of the fermionic mesons in ${\cal T}$ is continuous, and the threshold at which the continuum starts is 4 times the squared mass of the lightest
$[\text{H $-$ L}]$ state in ${\cal T}'$.  As shown in Figure \ref{fig:tprime_spectrum}, for $y_\text{fund}= 1$ the latter extrapolates to $\approx 2 \frac{g^2 N}{\pi}$ in the continuum limit. The lowest threshold at $M^2\approx 8 \frac{g^2 N}{\pi}$ is clearly seen not only in the fermionic meson sectors of ${\cal T}$, but also in the bosonic ones (see   
Figure~\ref{fig:meson_spectrum_massive}). Also, there is a bosonic $\mathbb{Z}_2$-odd meson with squared mass $\approx 6.6 \frac{g^2 N}{\pi}$, which lies below this threshold. 
The existence of this bound state suggests that there is attraction between the quark and antiquark at short distances; at long distances the attractive force is screened by the massless adjoint fermions. 

For $y_\text{fund} \gg 1$ we may interpret the massive quarks as heavy fundamental probes of the massless adjoint QCD$_2$ theory. One implication of the vanishing fundamental string tension in massless adjoint QCD$_2$ is that the spectrum of single-meson states at large $N$ in theory $\cT$ becomes continuous above a certain threshold \cite{Gross:1995bp}. Now we have established this result quantitatively for any $y_\text{fund}$ (and in particular for $y_\text{fund} \gg 1$), simply as a consequence of the fact that, at large $N$, the $P^-$ eigenvalues of every fermionic meson in $\cT$ are sums of eigenvalues of $[\text{H $-$ L}]$ states in ${\cal T}'$.  This provides a new confirmation of the screening phenomenon directly through the studies of meson spectra.

\section{Discussion}
\label{sec:DISCUSSION}

In this paper, we used DLCQ to study a 2d $\SU(N)$ gauge theory with a Majorana fermion in the adjoint representation of $\SU(N)$ and $N_f$ quarks in the fundamental representation.  With anti-periodic boundary conditions for the fermions, we diagonalized the light-cone components of the momentum $P^+$ and $P^-$ in the large $N$ limit, and extracted the mass spectrum of the single-trace gluinoball and meson states of this theory.  When the adjoint and fundamental fermions are massless, we observed an intricate pattern of exact degeneracies of the $P^-$ eigenvalues, both between mesons with different $P^+$ eigenvalues, between mesons and gluinoballs, and between single-trace and multi-trace states.  We provided two seemingly independent explanations for these degeneracies, one building on the Kac-Moody approach of \cite{Kutasov:1994xq} and the other based on an $\mathfrak{osp}(1|4)$ symmetry present in this model at large $N$. Under the $\mathfrak{osp}(1|4)$ symmetry, we found that the single-trace meson states transform in infinite-dimensional unitary representations referred to as $\mathfrak{osp}(1|4)$ singletons.  

Lastly, we noticed that when the fundamental quark mass parameter $y_\text{fund} > 0$ and the adjoint quark mass parameter $y_\text{adj} = 0$, some of the degeneracies between the single-trace mesons and double-trace states comprised of a single-trace gluinoball and a single-trace meson still survive.  These are thus non-trivial examples of threshold states in our 2d model for QCD\@.   Moreover, by relating the $P^-$ eigenvalues of meson states to eigenvalues of states in a theory with $N$ massless fundamental fermions and a massive one, we showed that the meson spectrum is continuous above a certain threshold, with the fermionic mesons having no discrete states below the bottom of the continuum. The presence of this continuum provides a direct confirmation of the screening of the fundamental flux line by the massless adjoint fermions.

There are various unanswered questions which we hope to return to in the future.   While our numerical analysis was limited to the large $N$ limit, it would be interesting to perform a similar analysis at finite $N$ generalizing the study of \cite{Antonuccio:1998uz}.  The Kac-Moody analysis of Section~\ref{KMA} indicates that the same pattern of degeneracies in the meson sector persists at finite $N$ as well.  On the other hand, the $\mathfrak{osp}(1|4)$ symmetry we observed was constructed only in the large $N$ limit.  Thus, it would be interesting to explore whether the $\mathfrak{osp}(1|4)$ symmetry persists at finite $N$ as well and, if so, construct the symmetry generators.  More generally, one can aim to combine the $\mathfrak{osp}(1|4)$ analysis with the Kac-Moody approach.

It would also be interesting to consider various generalizations of the model presented here.  For instance, one can consider quarks in different representations of the gauge group or gauge theories with different gauge groups.  Another generalization would be to consider periodic boundary conditions for the fermions in the light-cone direction parameterized by $x^-$.  The current algebra approach of Section~\ref{KMA} should generalize to this case, suggesting that a pattern of degeneracies similar to the one we noticed would still be present.

\section*{Acknowledgments}

We are grateful to Thomas Dumitrescu, Zohar Komargodski, Fedor Popov, and Andrei Smilga for useful discussions.  RD and SSP were supported in part by the Simons Foundation Grant No.~488653, and by the US NSF under Grant No.~PHY-1820651.  RD was also supported in part by an NSF Graduate Research Fellowship.   The research of IRK was supported in part by the US NSF under Grant No.~PHY-1914860.  

\appendix
\section{Degeneracy between mesons and gluinoballs}
\label{sec:DEGMG}

In this Appendix, we provide an alternative partial explanation for the degeneracies between gluinoballs and $\mathfrak{osp}(1|4)$ primary mesons observed in Section~\ref{sec:glueball_degeneracies}. As discussed there, the meson primary states have $P^-$ eigenvalues equal to a sum of one or two $P^-$ eigenvalues of fermionic gluinoballs with one less unit of total $P^+$. Clearly $q^L_\pm$ and $q^R_\pm$ cannot be used to construct the relevant gluinoball states from the meson primaries, since each term in both operators contains a fundamental creation operator and a fundamental annihilation operator. However, we note that the terms of $q^L_\pm + q^R_\pm$ can be found in the discretization of the continuum operator
\begin{equation}
	\tilde q_\pm = \frac{1}{\sqrt{N}}\int_0^{2\pi L} dx^-\,e^{\pm ix^-/(2L)} \quark^\dagger_i\psi_{ij}\quark_j \,.
\end{equation}
In addition, the discretization of this operator contains the terms
\begin{equation}
	\tilde q_\pm \supset -\frac{1}{(2\pi)^{3/2}\sqrt{N}}\sum_{n_1,n_2}\left(B^\dagger_{ij}(n_1+n_2\pm 1)C_i(n_1)D_j(n_2) + C^\dagger_i(n_1)D^\dagger_j(n_2)B_{ij}(n_1+n_2\mp 1)\right) \,,
\end{equation}
which can couple our mesonic states to the gluinoball states. We define the following string-closing and string-opening operators based on terms of $\tilde q_-$:
\begin{align}
  q_\text{close} &= \sum_{n_1,n_2} B^\dagger_{ij}(n_1 + n_2 - 1)C_i(n_1)D_j(n_2) \,, \\
  q_\text{open} &= \sum_{n_1, n_2} C^\dagger_i(n_1)D^\dagger_j(n_2) B_{ij}(n_1+n_2-1) \,.
\end{align}

The operator $q_\text{close}$ converts a meson at $K$ to a single-trace gluinoball at $K-1$. In fact, if we act on a bosonic meson singlet state at $K$ with $\tilde q_-$, we obtain the corresponding single-trace fermionic gluinoball at $K-1$ with the same value of $P^-$. For instance, at $K = 6$, there is a meson primary state with $P^- = 1$,
\begin{equation}
  \ket{\psi} = \frac{1}{2\sqrt{N}}\left(\!C^\dagger_i(1)D^\dagger_i(5) - C^\dagger_i(5)D^\dagger_i(1) + \frac{1}{N}C^\dagger_i(1) \left(B^\dagger_{ij}(1) B^\dagger_{jk}(3) + B^\dagger_{ij}(3)B^\dagger_{jk}(1)\right)D^\dagger_k(1)\!\right)\!\ket{0} \,.
\end{equation}
Acting with $q_\text{close}$, we find
\begin{equation}
  q_\text{close}\ket{\psi} \propto \frac{1}{\sqrt{N}}B^\dagger_{ij}(1)B^\dagger_{jk}(1)B^\dagger_{ki}(3)\ket{0} \,, 
\end{equation}
which is a gluinoball eigenstate with $P^- = 1$.

For fermionic meson primaries which are degenerate with double-trace gluinoball states, acting with $q_\text{close}$ gives a single-trace bosonic gluinoball state with the same $P^-$ eigenvalue. For instance, at $K = 9$, there are two meson primary states with $P^- = 4$. These are both degenerate with a double-trace gluinoball state at $K = 8$,
\begin{equation}
  \Tr\left(B^\dagger(1)^5\right)\Tr\left(B^\dagger(1)^3\right)\ket{0} \,.
\end{equation}
This double-trace state is in turn degenerate with two single-trace bosonic gluinoballs at $K = 8$. We can choose a basis for the two meson primaries at $K = 9$ and these two bosonic gluinoballs at $K = 8$ such that one of the meson primaries is annihilated by $q_\text{close}$, and the other gives one of the two degenerate bosonic gluinoballs.

Finally, fermionic meson primaries which are not degenerate with double-trace gluinoball states are annihilated by $q_\text{close}$. The $P^-$ eigenvalues of these meson primaries are equal to double the $P^-$ eigenvalues of a fermionic gluinoball, and thus the corresponding double-trace state would be null by the fermion statistics.

In summary, we see that the commutator $[q_\text{close}, P^-]$ annihilates all meson primary states. If we project onto meson primary states before acting with $q_\text{close}$, then we will have an operator which commutes with $P^-$. We can construct such an operator using the observation that all meson primaries have $\Delta = \frac{1}{2}$. Since the eigenvalues of $D$ are all half-integers,
\begin{equation}
	{\cal O} \equiv \prod_{n=1}^\infty \left( 1- \frac{4 (D-1/2)^2}{n^2} \right) = \frac{\sin (2 \pi (D - 1/2) )}{2 \pi (D-1/2)}
\end{equation}
projects onto the space of meson primary states. We thus have
\begin{equation}\label{eq:qclose}
	\left\lbrack q_\text{close} {\cal O}, P^-\right\rbrack = 0 \,.
\end{equation}
We have checked this numerically up to $K = 20$.\footnote{Since the operator ${\cal O}$ is a dense matrix in our basis, we cannot use the fast linear algebra methods which allow us to reach up to higher $K$ for other numerical checks.}

Naturally we expect for $q_\text{open}$ to act on gluinoball $P^-$ eigenstates and give corresponding meson $P^-$ primary eigenstates. However, the codomain of $q_\text{open}$ acting upon the gluinoball states at $K - 1$ does not coincide with the space of meson primary states at $K$. Instead, we must act with the projection operator ${\cal O}$. Analogously with \eqref{eq:qclose}, we have
\begin{equation}
  \left\lbrack {\cal O} q_\text{open}, P^-\right\rbrack = 0 \,.
\end{equation}
We have also checked this relation up to $K = 20$.

\bibliographystyle{ssg}
\bibliography{DLCQ}

\end{document}